\documentclass[11pt]{article}
\usepackage{jheppub} % for details on the use of the package, please
\usepackage{braket,slashed,bm}
\usepackage{array,multirow}
\usepackage[normalem]{ulem}
\usepackage{xcolor,cancel}

\usepackage[T1]{fontenc} % if needed

\usepackage{mathrsfs}

\usepackage{tikz}
\usetikzlibrary{arrows,decorations.pathmorphing,backgrounds,positioning,fit,petri,automata,shadows,calendar,mindmap,
decorations.markings,calc}

\def\nn{\nonumber\\ }
\def\rd{{\rm d}}
\def\abs#1{\left| #1 \right| }
\def\lsix{ \mathcal{L}^{(6)}}

\def\hyp{\mathsf{y}}
\def\tr{{\rm Tr}\,}

\def\zero#1{\hbox{\xout{$#1$}}}
\def\lsix{ \mathcal{L}^{(6)}}
\def\gcb{{\overline g_{1}}}
\def\gcw{{\overline g_{2}}}
\def\gcg{{\overline g_{3}}}
\def\gcZ{{\overline g_{Z}}}

\def\tc{{\overline \theta}}
\def\ec{{\overline e}}
\def\sc{{\overline s}}

\def\ckin{c_{H,\text{kin}}}

\title{
Renormalization Group Evolution of the Standard Model Dimension Six Operators\\
III: Gauge Coupling Dependence and Phenomenology}

\author[a]{Rodrigo Alonso,}

\author[a]{Elizabeth E.~Jenkins,}

\author[a]{Aneesh V.~Manohar,}

\author[b,1]{Michael Trott}\note{Corresponding author.}

\affiliation[a]{Department of Physics, University of California at San Diego, 9500 Gilman Drive,\\ La Jolla, CA 92093-0319, USA}
\affiliation[b]{Theory Division, Physics Department, CERN, CH-1211 Geneva 23, Switzerland}

\emailAdd{ralonsod@ucsd.edu}
\emailAdd{ejenkins@ucsd.edu}
\emailAdd{amanohar@ucsd.edu}
\emailAdd{michael.trott@cern.ch }

\abstract{
We calculate the gauge  terms of the  one-loop anomalous dimension matrix for  the dimension-six operators of the Standard Model effective field theory (SM EFT). Combining these results with our previous results for the $\lambda$ and Yukawa coupling terms completes the calculation of the one-loop anomalous dimension matrix for the dimension-six operators.  There are 1350 $CP$-even and $1149$ $CP$-odd parameters in the dimension-six Lagrangian for 3 generations,  and our results give the entire $2499 \times 2499$ anomalous dimension matrix. We discuss how the  renormalization of the dimension-six operators, and the additional renormalization of the dimension $d \le 4$ terms of the SM Lagrangian due to dimension-six operators, lays the groundwork for future precision studies of the SM EFT aimed at constraining the effects of new physics through precision measurements at the electroweak scale. As some sample applications, we discuss some aspects of the full RGE improved result for essential processes such as $gg \to h$, $h \to \gamma \gamma$ and $h \to Z \gamma$, for Higgs couplings to fermions,  for the precision electroweak parameters $S$ and $T$, and for the operators that modify important processes in precision electroweak phenomenology, such as the three-body Higgs boson decay $h \rightarrow Z \, \ell^+ \, \ell^-$ and triple gauge boson couplings.  We  discuss how the renormalization group improved results can be used to study the flavor problem in the SM EFT, and to test the minimal flavor violation (MFV) hypothesis.  We briefly discuss the renormalization effects on the dipole coefficient $C_{e\gamma}$ which contributes to $\mu \to e \gamma$ and to the muon and electron magnetic and electric dipole moments. 
}

\begin{document} 
\maketitle

\section{Introduction}\label{sec:intro}

The LHC experiments have recently found strong evidence for a scalar particle with mass 126~GeV, and properties consistent with the  Higgs boson of the Standard Model (SM)~\cite{Aad:2012gk,Chatrchyan:2012gu}.  The absence of any clear evidence of new particles at energies up to several times the scalar boson mass allows one to parametrize the effects of arbitrary new physics residing at energies $\Lambda \gg v$ on physical observables at the electroweak scale in terms of higher dimension operators built out of SM fields.  Experimental measurements of the properties of the scalar boson and other observables at the electroweak scale can then be used to constrain or determine the coefficients of the higher dimension operators, and hence the effects of arbitrary beyond-the-standard-model (BSM) theories with characteristic energy scale $\Lambda$ in a model independent way.

In this paper, we adopt the assumption that the scalar boson observed at LHC is the SM Higgs boson, and that the Higgs mechanism generates the mass of the SM gauge fields and fermions.  Specifically, we assume that the observed scalar boson $h$ is part of a $SU(2)_L$ doublet $H$ with hypercharge $\hyp_h = \frac12$, and that the electroweak $SU(2)_L \times U(1)_Y$ gauge symmetry is a linearly realized symmetry in the scalar sector which is spontaneously broken by the vacuum expectation value of $H$.  These assumptions yield the simplest and most direct interpretation of the LHC data, and the related experimental observations from LEP and the Tevatron.\footnote{There are other alternatives being investigated, such as a nonlinearly realized $SU(2)_L \times U(1)_Y$ gauge symmetry in the scalar sector with a light scalar $h$; see~\cite{Grinstein:2007iv,Contino:2010mh,Alonso:2012px,Buchalla:2013rka,Brivio:2013pma} and references therein}. The SM effective field theory (SM EFT) based on these assumptions consists of the SM Lagrangian plus all possible higher dimension operators.

The leading higher dimension operators built out of SM fields that preserve baryon and lepton number are 59 dimension-six operators~\cite{Buchmuller:1985jz,Grzadkowski:2010es}.  It is important to keep in mind that many of these operators have flavor (generation) indices.   For $n_g=3$ generations, the dimension-six Lagrangian has 1350 $CP$-even and 1149 $CP$-odd couplings, for a total of 2499 hermitian operators and real parameters. The flavor indices obviously cannot be neglected --- there is no reason in general, for example, why the new physics contribution to $\mu \to e \gamma$ should be the same as the new physics contribution to the muon magnetic moment.  Despite the large number of operators, it is important to realize that the SM equations of motion (EOM) have been used extensively in reducing the operator basis. As a result, the coefficient of a removed operator is distributed among the remaining operators.

In this work, we complete the full calculation of the $2499 \times 2499$ one-loop anomalous dimension matrix of the 59 dimension-six operators in the operator basis of 
Ref.~\cite{Buchmuller:1985jz,Grzadkowski:2010es}, including flavor indices for an arbitrary number of generations $n_g$. We present the gauge coupling terms in the one-loop anomalous dimension matrix in this paper.  Combined with our past results \cite{Jenkins:2013wua,Jenkins:2013sda,Jenkins:2013zja,Grojean:2013kd}, this gives the full one-loop renormalization group evolution (RGE) of the dimension-six operators of the SM EFT.  Having the full one-loop RGE of an independent set of dimension-six operators in the SM EFT has the advantage that all physical effects are included, and there can be no cancellation of terms between independent operators.

To precisely interpret any pattern of deviations of SM processes using higher dimensional operators, one has to map the pattern of deviations observed at the electroweak scale back to the scale $\Lambda$, where the BSM physics was integrated out of the effective field theory.  Due to operator mixing, the pattern of Wilson coefficients that are observed at the low scale $\sim m_H$ is not identical to the pattern of Wilson coefficients at the matching scale $\Lambda$. Our RG calculation determines all of the logarithmically enhanced terms in observables at the renormalization group scale $\mu=m_H$ due to RG running from the high-energy scale of new physics  $\mu=\Lambda$.  

There are also other contributions from the finite parts of one-loop graphs at the low scale $\mu \sim m_H$, which we have not computed.  For $\Lambda \sim 1$\,TeV, $\ln(\Lambda^2/m_H^2) \sim 4$, so there is a modest enhancement of the log terms over the finite terms.  As experiments get more precise, and the scale $\Lambda$ is pushed higher, the log terms become even more important relative to the finite terms.  Nevertheless, the calculation of finite terms is  important, and these terms will eventually be required for a precise comparison of data with the SM EFT. The anomalous dimensions can also be viewed as computing the $\ln \Lambda/m_H$ enhanced finite terms. The anomalous dimension computation is easier because it can be done in the unbroken theory, whereas the computation of finite terms needs to be done in the broken theory.

An important application of the SM EFT is to test the hypothesis of minimal flavor violation~\cite{Chivukula:1987py,DAmbrosio:2002ex}. The dimension-six operators can have arbitrary flavor structure, and the renormalization group equations derived in Refs.~\cite{Jenkins:2013wua,Jenkins:2013sda,Jenkins:2013zja,Grojean:2013kd}  and in this paper give non-trivial mixing between  different particle sectors. MFV assumes that the only sources of $U(3)^5$ flavor symmetry violation are the Yukawa coupling matrices $Y_e$, $Y_u$ and $Y_d$. The SM respects MFV by definition. Since MFV is formulated in terms of symmetries, it is preserved by the RG evolution. If the dimension-six Lagrangian respects MFV, then the RG evolution preserves this property.

The general dimension-six  Lagrangian does not have to respect MFV, and RG evolution then feeds non-minimal flavor violation into different operator sectors. By constraining the parameters of the SM EFT, one can experimentally test the MFV hypothesis taking this RG running into account. It is important to test MFV directly in a model-independent way.  The SM EFT provides a model-independent formalism to test the MFV hypothesis.

The outline of this paper is as follows.  In Section~\ref{sec:anomdim}, we discuss our notation,  and the gauge coupling constant terms reported in this work. Some generalities about the structure of the anomalous dimension matrix are given in Sec.~\ref{sec:structure}.  Some interesting cancellations are pointed out in 
Sec.~\ref{sec:cancel}. A detailed presentation of the gauge coupling constant terms in the RG equations of the dimension-six operator coefficients is relegated to Appendix~\ref{app:rge}.  Section~\ref{sec:silh} compares the standard operator basis of Refs.~\cite{Buchmuller:1985jz,Grzadkowski:2010es} with SILH operators~\cite{Giudice:2007fh}.  A brief discussion of MFV and its implications is given in Section~\ref{sec:mfv}.
Section~\ref{pheno} presents the main applications of the SM EFT to phenomenology. We discuss the SM parameters at tree level, and how their values are modified by the SM EFT dimension-six operators. In particular, we discuss the modifications to the Higgs mass and couplings, and to the gauge boson masses.  We also discuss the scale dependence of the dimension-six operators, and how the dimension-six operators contribute to the running of the $d \le 4$ parameters of the SM Lagrangian. The complete expressions for the running of the $gg \to h$, $h \to \gamma \gamma$ and $h \to \gamma Z$ amplitudes are given in Secs.~\ref{sec:ggh}, \ref{sec:gg}, and \ref{sec:gZ}, respectively.  In Secs.~\ref{sec:EWPD} and~\ref{sec:tgc}, we  discuss  the operators corresponding to the electroweak precision data (EWPD) parameters $S$ 
and $T$, and operators modifying critical processes for precision electroweak phenomenology, such as triple gauge boson couplings and  the three-body decay $h \rightarrow Z \, \ell^+ \, \ell^-$.  In Sec.~\ref{sec:mag}, we discuss the dipole coefficients $C_{e\gamma}$ which contribute to the decay $\mu \to e \gamma$ and to the muon and electron magnetic and electric dipole moments.  We present our conclusions in Section~\ref{conclusion}. The counting of parameters in $\lsix$ is summarized in Appendix~\ref{app:counting}, and the conversion of SILH operators to the standard basis is given in Appendix~\ref{sec:convert}.

\section{The anomalous dimension matrix}\label{sec:anomdim}

The complete list of 59 independent dimension-six operators is given in Table~\ref{op59}.  The operators are divided into eight classes by field content and number of covariant derivatives.  The eight operator classes are $1:X^3$, $2:H^6$, $3:H^4 D^2$, $4:X^2 H^2$, $5:\psi^2 H^3$, $6:\psi^2 X H$, $7:\psi^2 H^2 D$ and $8:\psi^4$, where $X=G^A_{\mu \nu},W^I_{\mu \nu}, B_{\mu \nu}$ represents a gauge field strength, $H$ denotes the Higgs doublet scalar field,  $\psi$ is a fermion field $\psi=q,u,d,l,e$, and $D$ is a covariant derivative. The dimension-six Lagrangian is 
\begin{align}
\lsix &= \sum_i C_i Q_i
\label{lsix}
\end{align}
where the $Q_i$ are the dimension-six operators of Table~\ref{op59} and the operator coefficients $C_i$ have dimensions of $1/\Lambda^2$.  The one-loop anomalous dimension matrix $\gamma_{ij}$ is defined by the RG equation of the operator coefficients
\begin{align} 
\dot C_i &\equiv 16 \pi^2 \mu \frac{\rd C_i}{\rd \mu} = \gamma_{ij} C_j .
\end{align}
The explicit RG equations are given in  Appendix~\ref{app:rge} as differential equations, rather than as elements of the matrix $\gamma$. We will use $\gamma_{ij}$ to represent the $8 \times 8$ block  form of the anomalous dimension matrix, where the subscripts on $\gamma$ refer to the eight operator classes $i, j = 1, \ldots, 8$.  For example, $\gamma_{35}$ is the $2 \times 3$ anomalous dimension submatrix which mixes the 3 independent class 5 operator coefficients into the 2 independent class 3 operator coefficients (see Table~\ref{op59}).

Although there are 59 independent operators, many of them have flavor indices which take on $n_g=3$ values. Table~\ref{number} gives the number of $CP$-even and $CP$-odd coefficients for each operator class. For $n_g=3$, there are $(107 n_g^4+2n_g^3+213 n_g^2+30n_g+72) /8=1350$ $CP$-even coefficients and  $(107 n_g^4+2n_g^3+57 n_g^2-30n_g+48)/8=1149$ $CP$-odd parameters, for a total of 2499 parameters which need to be constrained by experiment. The counting of parameters is summarized in Appendix~\ref{app:counting}.

Such a large number of terms makes the calculation of the complete anomalous dimension matrix a formidable task. In  Ref.~\cite{Grojean:2013kd}, we began by computing the $8\times8$ one-loop anomalous dimension matrix $\gamma_{44}$ for the class-4 Higgs-gauge operators $X^2 H^2$, since these operators contribute directly to the experimentally interesting Higgs production and decay channels $gg \to h$, $h \to \gamma \gamma$, and $h \to \gamma Z$, which first occur at one loop in the SM.  The $8 \times 8$ submatrix  $\gamma_{44}$ has been subsequently verified by several independent calculations (e.g.\ Ref.~\cite{Chen:2013kfa}). In Ref.~\cite{Jenkins:2013zja}, we calculated the $\lambda$-dependent terms of the full anomalous dimension matrix for vanishing gauge coupling constants, as well as the complete running of the SM $d \le 4$ parameters due to the dimension-six operators. The running of the SM parameters resulting from the dimension-six operators is of order $m_H^2/\Lambda^2$, which is of the same order as the tree-level contribution of dimension-six operators. The Yukawa-dependent terms of the  anomalous dimension matrix for vanishing gauge couplings were computed in Ref.~\cite{Jenkins:2013wua}.  In this paper, we complete the full calculation of the one-loop anomalous dimension matrix of the dimension-six operators by computing the gauge coupling terms.

The one-loop anomalous dimension matrix has the usual $1/(16\pi^2)$ suppression of a one-loop calculation. However, there are several anomalous dimensions with large numerical factors. In Ref.~\cite{Jenkins:2013zja}, for example, we found that 
\begin{align}
 16 \pi^2 \mu \frac{\rd}{\rd  \mu} C_H = 108 \, \lambda \, C_H + \ldots \,.
\end{align}
Since $m_H^2=2\lambda v^2$, the anomalous dimension coefficient is  $108 \lambda = 54m_H^2/v^2\simeq 14$, independent of the normalization convention for the quartic coupling $\lambda$. In the study of the Yukawa coupling terms of Ref.~\cite{Jenkins:2013wua}, the numerical factors were generally $\mathcal{O}(1)$. These Yukawa terms give interesting nontrivial flavor mixing between the various operators.  The gauge terms calculated in this paper also contain several large coefficients.  For example, the mixing of the class 4 operators $X^2 H^2$ into the class 2 operator $H^6$ gives
\begin{align}
 16 \pi^2 \mu \frac{\rd}{\rd  \mu} C_H = - (48 g_1^4 \, \hyp_h^4 + 12 g_1^2 g_2^2 \hyp_h^2) C_{HB} \ldots
\end{align}

\begin{figure}
\centering
\begin{tikzpicture}[
decoration={
	markings,
	mark=at position 0.55 with {\arrow[scale=1.5]{stealth'}};
}]

\draw[postaction=decorate] (90:1) arc (90:270:1) ;
\draw[postaction=decorate] (-90:1) arc (-90:90:1) ;

\filldraw (-0.1,0.9) rectangle (0.1,1.1);

\filldraw (270:1) circle (0.075);

% top
\draw[postaction=decorate] (90:1) -- +(50:1.5) ;
\draw[postaction=decorate] (90:1)+(130:1.5) -- (90:1) ;

% 1
\draw[decorate,decoration={snake}]  (270:1) -- +(270:1.5);

\end{tikzpicture}
\caption{\label{fig:penguin} 
A penguin diagram. The solid square is a $\psi^4$ vertex from $\lsix$, and the dot is a SM gauge coupling.}
\end{figure}
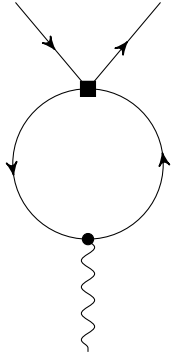

The lengthiest contributions to gauge coupling constant terms come from the well-known penguin graph Fig.~\ref{fig:penguin}.  The penguin graph itself is simple to compute.  However, there are 25 possible $\psi^4$ operators in the $\lsix$ Lagrangian, and the penguin graph is proportional to $D_\mu X^{\mu \nu}$, which is replaced by a gauge current summed over all fermion and scalar fields. The resulting  four-fermion and fermion-scalar operators then have to be Fierzed to the canonical operator basis, resulting in the bulk of the terms given in Appendix~\ref{app:rge}.

One finds a substantial amount of operator mixing in the SM EFT, and such mixing affects observables measured at the electroweak scale in a manner which must be unraveled to understand BSM theories. One of the consequences of this mixing is the propagation of $CP$ violation through different sectors of the Lagrangian. For instance, dipole operators receive contributions from $CP$ violating class 4 operators (that enter, e.g., $h\rightarrow \gamma Z$ at tree level), the latter are therefore subject to electric dipole moment constraints, see Sec.~\ref{sec:mag}.
On the other hand, it is already known~\cite{Grojean:2013kd} that mixing effects are relevant for studies of $h \to \gamma \gamma$.

\subsection{The structure of $\gamma_{ij}$}\label{sec:structure}

The complication of dealing with a large operator basis naturally leads to the desire to simplify the calculation, or to look for hidden structure in the anomalous dimension matrix to more easily understand the physics of the one-loop RGE flow. In Ref.~\cite{Jenkins:2013zja}, we showed that the structure of the anomalous dimension matrix can be understood using Naive Dimensional Analysis (NDA) \cite{Manohar:1983md}. The argument is simplest using rescaled operators $\widehat Q_i$. The rescaled operators $\widehat Q_i$ are given by $g^2 X^3$, $H^6$, $H^4D^2$, $g^2 X^2 H^2$, $y\psi^2 H^3$, $gy \psi^2 X H$, $\psi^2 H^2 D$ and $\psi^4$, where each gauge field strength $X$ has been rescaled by a gauge coupling $g$, and the chirality-flip operators 
$\psi^2 H^3$ and $\psi^2 XH$, which change chirality by one unit, have been rescaled by an additional Yukawa coupling $y$. The dimension-six Lagrangian can be rewritten in terms of the  rescaled operators and their corresponding coefficients $\widehat C_i$,
\begin{align}
\lsix &= \sum_i C_i Q_i =\sum_i \widehat C_i \widehat Q_i \ .
\end{align}
The RG equations for the original and rescaled operator coefficients are given by
\begin{align}
\dot C_i &= \gamma_{ij}\ C_j , &
\dot {\widehat C}_i &= {\widehat \gamma}_{ij}\ {\widehat C}_j ,
\end{align}
where the one-loop anomalous dimension matrices $\gamma_{ij}$ and ${\widehat \gamma}_{ij}$ are related to each other by the rescaling factors and their derivatives.  In Ref.~\cite{Jenkins:2013zja}, we showed that the anomalous dimension matrix ${\widehat \gamma}$ for the rescaled operators has entries proportional to
\begin{align}
\widehat \gamma &\propto \left( \frac{\lambda}{16\pi^2}\right)^{n_\lambda} \left( \frac{y^2}{16\pi^2}\right)^{n_y}
\left( \frac{g^2}{16\pi^2}\right)^{n_g}, & N = n_\lambda + n_y + n_g
\end{align}
where $N$, the perturbative order of the anomalous dimension, is defined as the sum of the number of factors $n_\lambda$ of the Higgs 
self-coupling $\lambda$, the number of factors $n_y$ of $y^2$, and the number of factors $n_g$ of $g^2$. For the rescaled dimension-six operators, $N$ ranges from 0 to 4.  In Ref.~\cite{Jenkins:2013sda}, we derived a general formula for the perturbative order $N$ of the anomalous dimension matrix $\widehat \gamma_{ij}$,
\begin{align}
N &= 1 + w_i - w_j,
\label{nda}
\end{align}
where $w_i$ is the NDA weight of the operators $\widehat Q_i$ in the $i^{\rm th}$ class~\cite{Jenkins:2013sda}. The class 2 operator $\widehat Q_H$ has NDA weight $w_2=2$;  the operators in classes $\{3,5,7,8\}$ have NDA weight $1$; the operators in classes $\{4,6\}$ have NDA weight $0$; and the class 1 operators have NDA weight $w_1 = -1$.  Using Eq.~(\ref{nda}), the possible coupling constant dependences of $\widehat \gamma_{ij}$ are obtained.  Our previous work calculated all anomalous dimensions with nontrivial $n_\lambda$ and $n_y$ with $n_g=0$.  The present work completes the calculation of all terms with $n_g \ne 0$. 

Although the coupling constant dependence of the anomalous dimension matrix is simplest for the NDA rescaled operators, the RGE in  Refs.~\cite{Grojean:2013kd,Jenkins:2013zja,Jenkins:2013wua} and in this work are quoted in terms of the original unrescaled operators $Q_i$ of Refs.~\cite{Buchmuller:1985jz,Grzadkowski:2010es}. The possible entries of $\gamma_{ij}$ were classified in Ref.~\cite{Jenkins:2013zja} by studying all possible one-loop diagrams including EOM terms. The classification is a bit subtle. The non-zero entries arise directly from diagrams which contribute to a given term, but also indirectly via EOM.  For example, the $H^4 D^2 - H^4 D^2$ entry of the anomalous dimension matrix is computed from graphs with one insertion of a $H^4 D^2$ operator, $Q_{H \Box}$ or $Q_{HD}$, with 4 external $H$ lines. These graphs contribute to the $\gamma_{33}$ submatrix for the running of the coefficients $C_{H \Box}$ and $C_{HD}$. The graphs contributing to $\gamma_{33}$ also require a counterterm proportional to the EOM operator $E_{H \Box}$ of Ref.~\cite{Jenkins:2013zja}. This operator can be eliminated in favor of other operators such as the $\psi^2 H^3$ operators in the standard basis. Thus, the $\gamma_{33}$ graphs also contribute to the $\gamma_{53}$ submatrix via the EOM, even though they do not have any external fermion lines.

The NDA weights $w_i$ for the NDA rescaled operators $\widehat Q_i$ of the eight operator classes, and the coupling constant dependence of the allowed anomalous dimensions $\widehat \gamma_{ij}$ are shown in Table~\ref{tab:anom}, with the operators ordered according to decreasing NDA weight.  Now that the entire matrix has been computed, we can compare with the classification of Ref.~\cite{Jenkins:2013zja}. The cross-hatched entries in the table are anomalous dimension entries which could exist based on the allowed diagrams, but which  vanish by explicit computation. These entries vanish because  the relevant diagram vanishes, has no infinite part despite being naively divergent, or, in some interesting cases, by cancellation between different contributions such as a direct contribution to $ \gamma_{ij}$ and an indirect contribution obtained by using the EOM. These cancellations are discussed further in Sec.~\ref{sec:cancel}.

The diagonal blocks in Table~\ref{tab:anom} have $N=1$ since $w_i=w_j$.  Blocks one below the diagonal have $N=0$, whereas blocks one above the diagonal have $N=2$, etc.  When $N$ is less than $0$, $\gamma$ vanishes, and we find that this is always the case. However, there are many additional anomalous dimensions which vanish.  Indeed, almost all of the $N=0$ entries vanish.  The notable exception of a $N=0$ submatrix which does not vanish is $ \gamma_{68}$ which mixes class 8 four-fermion operators $\psi^4$ into the class 6 dipole operators  $\psi^2 X H$ in violation of the general ``no tree-loop mixing'' claim of Refs.~\cite{Elias-Miro:2013gya,Elias-Miro:2013mua,pomarol}.  Other examples which violate no tree-loop mixing exist~\cite{Bauer:1997gs}. ``Tree-loop'' classification \cite{Arzt:1994gp} of terms in an EFT Lagrangian has limited usefulness, and does not apply in general when the UV theory generating the dimension-six operators is itself an EFT, or is a strongly interacting theory.  Attempts to broaden this classification scheme in a very general manner relied critically on the assumption of minimal coupling.  However, in Ref.~\cite{Jenkins:2013fya}, we showed that the concept of minimal coupling is ill defined in general.

\subsection{Checks of the calculation}\label{sec:checks}

The calculations in this paper are done in background field with gauge fixing parameter $\xi$, and cancellation of  $\xi$-dependence provides a check on the results. The gauge dependence only cancels for gauge-invariant interactions, i.e.\ if the relations
\begin{align}
\hyp_q &= \hyp_d + \hyp_h, & \hyp_q &= \hyp_u - \hyp_h, & \hyp_l &= \hyp_e + \hyp_h,
\label{hreln}
\end{align}
are satisfied.  Although the expressions for the anomalous dimensions have been written in terms of all six hypercharges, $\hyp_i$ cannot be thought of as varying independently, but must satisfy the constraints Eq.~(\ref{hreln}). A check of the results that follows from custodial $SU(2)$ symmetry is discussed at the end of Sec.~\ref{sec:EWPD}.

The  SM Yukawa couplings
\begin{align}
\mathcal{L}_{\text{Yukawa}} &=- \biggl[ H^{\dagger j} \overline d_r\, [Y_d]_{rs}\, q_{js} + \widetilde H^{\dagger j} \overline u_r\, [Y_u]_{rs}\, q_{js} + H^{\dagger j} \overline e_r\, [Y_e]_{rs}\,  l_{js} + \hbox{h.c.}\biggr],
\end{align}
where $r,s$ are flavor indices and $j$ is an $SU(2)$ index,
are only gauge invariant because the $\bf{2}$ of $SU(2)$ is self-conjugate, so that $H_j$ and $\widetilde H_j=\epsilon_{jk}H^{\dagger\, k}$ belong to the same  $SU(2)$ representation. The $SU(2)$ group cannot be generalized to a $SU(N)$ group. While some of the $SU(2)$ group theory factors have been written as Casimirs such as $c_{A,2}$ and $c_{F,2}$, the results are only valid when they take on their $SU(2)$ values $c_{A,2}=2$ and $c_{F,2}=3/4$.

The $SU(3)$ results are written for an $SU(N_c)$ theory. Anomaly cancellation does not hold for the $SU(N_c)^2 \times U(1)_Y$ anomaly for arbitrary $N_c$, but  the results can still be useful in other contexts for the $SU(N_c)$ anomalous dimensions. The $SU(3)$ Fierz identity
\begin{align}
T^A_{\alpha \beta}\ T^A_{\lambda \sigma} &= \frac12 \delta_{\alpha \sigma} \delta_{\lambda \beta}-\frac{1}{2N_c} \delta_{\alpha \beta} \delta_{\lambda \sigma}
\end{align}
has been used to rearrange color indices and put operators into standard form. This identity is valid for the fundamental representation of $SU(N_c)$, but is not valid for arbitrary representations. Thus, the quadratic Casimir $c_{F,3}$ is equivalent to $(N_c^2-1)/(2N_c)$, and the fermions must be in $SU(N_c)$ fundamental or anti-fundamental representations.

\subsection{Cancellations}\label{sec:cancel}

The one-loop anomalous dimension matrix does not contain all possible terms that can arise from the allowed one-loop graphs and the EOM. In a few cases, the entries vanish because the graph has no divergent part. An example from Ref.~\cite{Jenkins:2013zja} is the $y^4$ contribution to $\gamma_{27}$, or 
$H^6-\psi^2H^2D$ mixing. 

\begin{figure}
\begin{tikzpicture}[
decoration={
	markings,
	mark=at position 0.55 with {\arrow[scale=1.5]{stealth'}};
}]

\draw[decorate,decoration={snake}] (0,0) circle (1);

\filldraw (-0.1,0.9) rectangle (0.1,1.1);

\filldraw (270:1) circle (0.075);

% top
\draw[decorate,decoration={snake}] (90:1) -- +(90:1.25) ;

% 2
\draw[decorate,decoration={snake}] (270:1) -- +(270:1.25);

\end{tikzpicture}
\hspace{1cm}
%S2X1-X3 ++++++++++++++++++++++++++++++++++++++++++++++++
% 1 : 2
% phi^2 X : X^3 ---------------------------------------------------------------------
\begin{tikzpicture}[
decoration={
	markings,
	mark=at position 0.55 with {\arrow[scale=1.5]{stealth'}};
}]

\draw[decorate,decoration={snake}] (0,0) circle (1);

\filldraw (-0.1,0.9) rectangle (0.1,1.1);

\filldraw (270:1) circle (0.075);

% top
\draw[decorate,decoration={snake}] (90:1) -- +(90:1.25) ;

% 2
\draw[dashed] (270:1)+(240:1.25) -- (270:1);
\draw[dashed] (270:1) -- +(300:1.25);

\end{tikzpicture}
\hspace{0.5cm}
%
%
%S2X1-X3 ++++++++++++++++++++++++++++++++++++++++++++++++
% 1 : 1 : 1
% phi^2 X : X^3 ---------------------------------------------------------------------
\begin{tikzpicture}[
decoration={
	markings,
	mark=at position 0.55 with {\arrow[scale=1.5]{stealth'}};
}]

\draw[decorate,decoration={snake}] (-30:1) arc (-30:210:1);
\draw[dashed] (210:1) arc (210:330:1) ;

\filldraw (-0.1,0.9) rectangle (0.1,1.1);

\filldraw (210:1) circle (0.075);
\filldraw (330:1) circle (0.075);

% top
\draw[decorate,decoration={snake}] (90:1) -- +(90:1.25) ;

% 1
\draw[dashed] (210:1)+(210:1.25) -- (210:1) ;

% 1
\draw[dashed] (330:1) -- +(330:1.25);

\draw (0,-1.5) node [align=center] {\phantom{a}};

\end{tikzpicture}
\hspace{0.5cm}
\begin{tikzpicture}[
decoration={
	markings,
	mark=at position 0.55 with {\arrow[scale=1.5]{stealth'}};
}]

\draw[decorate,decoration={snake}] (-30:1) arc (-30:210:1);
\draw (210:1) arc (210:330:1) ;

\filldraw (-0.1,0.9) rectangle (0.1,1.1);

\filldraw (210:1) circle (0.075);
\filldraw (330:1) circle (0.075);

% top
\draw[decorate,decoration={snake}] (90:1) -- +(90:1.25) ;

% 1
\draw (210:1) -- +(210:1.25);

% 1
\draw (330:1) -- +(330:1.25);

\draw (0,-1.5) node [align=center] {\phantom{a}};

\end{tikzpicture}

\caption{\label{fig:cancel} 
Graphs with insertions of the $X^3$ operator which cancel after using the equations of motion.}
\end{figure}
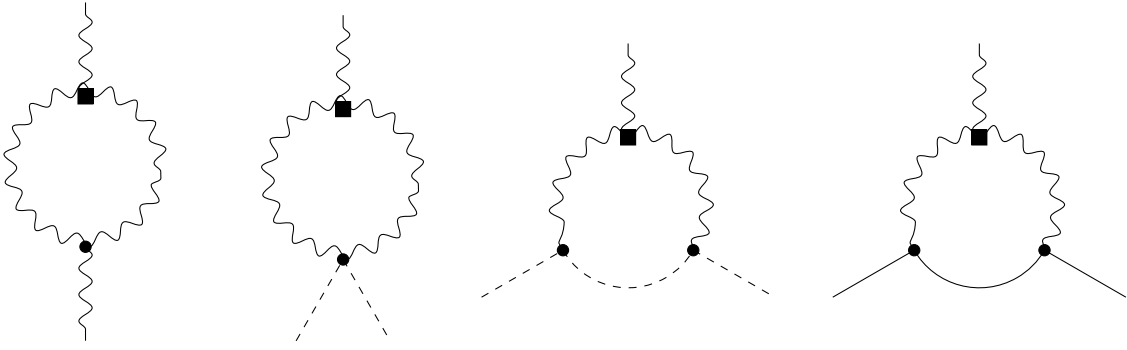

There also are a few cases with interesting non-trivial cancellations which arise when  different contributions to the same anomalous dimension are added together after using the equations of motion. An example is the contribution of insertions of the $CP$-even operators $X^3$ to the anomalous dimension from the graphs shown in Fig.~\ref{fig:cancel}. 
The divergent part of the first graph is proportional to
\begin{align}
A_1 &= -c_{A,2} g_2 C_W D^\mu W^I_{\mu \lambda} D_\nu W^{I\, \nu \lambda} - c_{A,3} g_3 C_G D^\mu G^A_{\mu \lambda} D_\nu G^{A\, \nu \lambda}.
\label{a1}
\end{align}
The divergent part of the sum of the second and third graphs is proportional to
\begin{align}
A_2 &= -i g_2^2 c_{A,2} C_W D_\mu H^\dagger \tau^I D_\nu H W^I_{\mu \nu}.
\label{a2}
\end{align}
There is no gluon term, since gluons do not couple to the Higgs field. The divergent part of the fourth graph is proportional to
\begin{align}
A_3 &=  g_2^2 c_{A,2} C_W D^\mu W^I_{\mu \nu} j^{I\, \nu}_\psi + g_3^2 c_{A,3} C_G D^\mu G^A_{\mu \nu} j^{A\, \nu}_\psi ,
\label{a3}
\end{align}
where
\begin{align}
j_\psi^{I\,\mu} &= \sum_{\psi=q,l} \overline \psi  \, \gamma^\mu \frac12 \tau^I  \,\psi , &
j_\psi^{A\,\mu} &= \sum_{\psi=q,u,d} \overline \psi  \, \gamma^\mu T^A \, \psi ,
\end{align}
are the $SU(2)$ and $SU(3)$ fermion currents, respectively.
The operator Eq.~(\ref{a2}) is equal to $g_2 c_{A,2} C_W P_{HW}$, where $P_{HW}$ is given in Eq.~(\ref{a}).  Integrating by parts, and writing the commutator of two covariant derivatives as a field-strength tensor gives the identity
\begin{align}
P_{HW} &= -i g_2 D_\mu H^\dagger \tau^I D_\nu H\ W^{I \mu \nu} \nn
&= 
g_2 j^{I\,\nu}_H\,  D^\mu W^I_{\mu \, \nu}
-\frac14 g_2^2 H^\dagger H W^I_{\mu \nu} W^{I \mu \nu} -\frac12 g_1 g_2 \hyp_h H^\dagger \tau^I H W^I_{\mu \nu} B_{ \mu \nu} ,
\end{align}
where
\begin{align}
j_H^{I\,\mu} &= \frac{i}{2} \, (H^\dagger \,  \tau^I \, \overleftrightarrow{D}^\mu H),
\end{align}
is the Higgs doublet $SU(2)$ current. The total is
\begin{align}
A_1 + A_2 + A_3 &= -g_2 c_{A,2} C_W D^\mu W^I_{\mu \lambda} \left[ D_\nu W^{I\, \nu \lambda} - g_2 j_\psi^{I\,\lambda}-
g_2 j_H^{I\,\lambda} \right]- c_{A,3} g_3 C_G D^\mu G^A_{\mu \lambda}\left[ D_\nu G^{A\, \nu \lambda} - g_2 j_\psi^{A\,\lambda} \right] \nn
&-g_2 c_{A,2} C_W \left[\frac14 g_2^2 H^\dagger H W^I_{\mu \nu} W^{I \mu \nu}+\frac12 g_1 g_2 \hyp_h H^\dagger \tau^I H\, W^I_{\mu \nu} B_{ \mu \nu} \right] \,.
\label{totalsum}
\end{align}
Using the gauge field equations of motion
\begin{align}
D_\mu W^{I\,\mu \nu} &= g_2\left( j_H^{I\,\mu} + j_\psi^{I\,\mu} \right),  &
D_\mu G^{A\,\mu \nu} &= g_3  j_\psi^{A\,\mu} ,
\label{2.18}
\end{align}
only the second line survives,
\begin{align}
A_1 + A_2 + A_3 &= -g_2 c_{A,2} C_W \left[\frac14 g_2^2 H^\dagger H W^I_{\mu \nu} W^{I \mu \nu}+\frac12 g_1 g_2 \hyp_h H^\dagger \tau^I H\, W^I_{\mu \nu} B_{ \mu \nu} \right] \,.
\label{rest}
\end{align}
The gluon term $C_G$ cancels completely and most of the $C_W$ term cancels. There is a residual contribution from Eq.~(\ref{rest}) to the anomalous dimension of $C_{HW}$ and $C_{HWB}$, the coefficients of the $X^2 H^2$ Higgs-gauge boson operators. The graphs in Fig.~\ref{fig:cancel} contribute to the running of $C_{HW}$ and $C_{HWB}$ even though none of the diagrams have two external gauge bosons and two external Higgs lines, the field content of $X^2 H^2$ operators. The cancellation of $C_G$ and $C_W$ terms in various anomalous dimensions is the reason for the absence of several terms in the last column of Table~\ref{tab:anom}.

The $C_{\widetilde W}$ and $C_{\widetilde G}$ contributions to the anomalous dimension arise from the same graphs as in Fig.~\ref{fig:cancel}, with the insertions of the 
$CP$-odd operators $ \widetilde XXX$.  In this case, one obtains Eqs.~(\ref{a1}) and~(\ref{a3}) with $D^\mu W^I_{\mu \nu}$ and $D^\mu G^A_{\mu \nu}$ replaced by $D^\mu \widetilde W^I_{\mu \nu}$ and $D^\mu \widetilde G^A_{\mu \nu}$, respectively, and Eq~(\ref{a2}) with $W^I_{\mu \nu}$ replaced by $\widetilde W^I_{\mu \nu}$. The equations of motion for $\widetilde X$ are $D^\mu \widetilde X_{\mu \nu}=0$, rather than Eq.~(\ref{2.18}), so naively there can be a difference between the $C_{\widetilde W,\widetilde G}$ and $C_{W,G}$ contributions to the anomalous dimension. However, the total sum $A_1+A_2+A_3$ is
\begin{align}
& -g_2 c_{A,2} C_{\widetilde W} D^\mu \widetilde W^I_{\mu \lambda} \left[ D_\nu W^{I\, \nu \lambda} -g_2 j_\psi^{I\,\lambda}-
g_2 j_H^{I\,\lambda} \right]- c_{A,3} g_3 C_{\widetilde G} D^\mu \widetilde G^A_{\mu \lambda}\left[ D_\nu G^{A\, \nu \lambda} - g_3 j_\psi^{A\,\lambda} \right] \nn
&-g_2 c_{A,2} C_{\widetilde W} \left[\frac14 g_2^2 H^\dagger H \widetilde W^I_{\mu \nu}  W^{I \mu \nu}+\frac12 g_1 g_2 \hyp_h H^\dagger \tau^I H\, \widetilde W^I_{\mu \nu} B_{ \mu \nu} \right] \, ,
\label{2.21}
\end{align}
instead of Eq.~(\ref{totalsum}). The first lines in both Eq.~(\ref{totalsum}) and Eq.~(\ref{2.21}), which would have produced a difference in the $C_{\widetilde W,\widetilde G}$ and $C_{W,G}$ contributions, are proportional to the gauge field equations of motion (\ref{2.18}) and vanish.  Thus, the contributions to the anomalous dimension from the $CP$-odd coefficients $C_{\widetilde W,\widetilde G}$ are the same as the contributions from the $CP$-even coefficients $C_{W,G}$.

Another interesting cancellation occurs in the contribution of the $\psi^2 X H$ dipole operators. The coefficients $C_{eW}$, etc.\ of these operators will be denoted generically by $C_{\psi X}$, where $\psi=e,u,d$. The dipole operators contribute to the running of $\psi^2 H^3$ coefficients $C_{\psi H}$, such as $C_{eH}$, and to the running of $\psi^2 H^2 D$ coefficients $C_{H \psi}$, such as $C_{He}$. The anomalous dimension for the running of $C_{\psi H}$ gets multiple contributions from $C_{\psi X}$ and $C_{\psi X}^*$ which arise from graphs with insertions of the $\psi^2 X H$ dipole operators and their hermitian conjugates. As above, the multiple contributions arise from using the EOM to bring all divergences to the canonical basis. The total contribution of $C_{\psi X}^*$ to the running of $C_{\psi H}$ cancels after using the hypercharge constraints Eq.~(\ref{hreln}), even though individual contributions do not vanish. The contribution of $C_{\psi X}$ to the running of $C_{\psi H}$ does not cancel. The total contribution of both $C_{\psi X}$ and $C_{\psi X}^*$ to the running of the $\psi^2 H^2 D$ coefficients $C_{H \psi}$ exactly cancels, which is why there is no $g^2 y^2$ entry in the anomalous dimension 
$\gamma_{76}$ from $\psi^2 H^2 D$-$\psi^2 XH$ mixing in Table~\ref{tab:anom}.

The contributions of the dipole operators and the gauge operators with $X$ and $\widetilde X$ are related by factors of $i$.  This simple factor follows from the complex self-duality of $\sigma^{\mu \nu}P_R$.  There is no $C_{\psi X}^*$ contribution to the running $\dot C_{\psi X}$, or to the runnings $\dot C_{quqd}^{(1)}$, $\dot C_{quqd}^{(8)}$ and $\dot C_{lequ}^{(3)}$, which are the $\psi^4$ operators to which the dipole operators contribute.

The examples above indicate that the RG contribution of the dipole operators respects holomorphy in $C_{\psi X}$.

\subsection{Previous work}\label{sec:prev}

Several of the gauge coupling terms of the one-loop anomalous dimension matrix have been calculated before.  However, we emphasize that with the results reported in this work, we have determined  the complete one-loop anomalous dimension matrix for dimension-six operators of the SM EFT for the first time.

Previous calculations of individual elements of the anomalous dimension matrix include the following works.\footnote{Due to the  number of operators renormalized, and the fragmentary literature on the subject, we apologize in advance to authors whose works are overlooked in this discussion.} The anomalous dimension of $Q_{G}$ and $Q_{\tilde{G}}$ were determined in Refs.~\cite{Morozov:1985ef,Braaten:1990gq,Boyd:1990bx}.  We agree with this result. Ref.~\cite{Morozov:1985ef} computed the anomalous dimension of  dimension-five and dimension-six operators in QCD.  Parts of our calculation in which the Higgs field can be treated as an external constant field agree with these results. The renormalization of four-fermion operators has been studied for many years in the context of the low-energy theory of weak interactions, and provides a check on the $\psi^4-\psi^4$ anomalous dimension. The complete one-loop RGE of the operators in class $4$ was calculated for the first time in Ref.~\cite{Grojean:2013kd}. Previously, some individual terms in this running result were calculated in  Refs. \cite{Hagiwara:1993ck, Hagiwara:1993qt, Alam:1997nk, Han:2004az}, and these terms are consistent with our calculation.  Ref.~\cite{Elias-Miro:2013gya} calculated the mixing of dipole operators $Q_{uG}$, $Q_{uW}$ and $Q_{uB}$ with the combination of Wilson coefficients $Q_{HW}$, $Q_{HB}$ and $Q_{HWB}$ that corresponds to $h \rightarrow \gamma \, \gamma$, see Section \ref{sec:gg}, which corresponds to a set of entries in $\gamma_{46}$.  We agree with these results.  Ref.~\cite{Zhang:2013xya} reports the running of the operators $Q_{uH}$ and $Q_{uG}$ due to the QCD coupling, which corresponds to entries in 
$\gamma_{55}$ and $\gamma_{66}$.  We agree with the results of this paper.

The papers mentioned in the previous paragraph allow a relatively direct comparison between results computed in the same operator basis.  Many other results in the literature are reported in a different basis, making a comparison difficult. Ref.~\cite{Elias-Miro:2013mua} presents a few terms in the anomalous dimension matrix without flavor indices (i.e.\ for $n_g=1$), and only including the top Yukawa coupling.  The exact translation between such partial results and this work requires that a complete non-redundant operator basis be defined, which often is not the case.  Ref.~\cite{Elias-Miro:2013mua} does not define such a mapping to allow us to compare our results to the terms reported, see the next Section for more discussion on this point. Nevertheless, some other classic past results in Refs.~\cite{Altarelli:1974exa,Shifman:1976ge,Gilman:1979bc,Floratos:1977au,Floratos:1978ny,
Gilman:1979ud,Gilman:1982ap,Bijnens:1983ye,Grinstein:1990tj,Bardeen:1978yd,Buras:1990fn,Buras:1992tc,Buchalla:1989we,
Ciuchini:1993vr} overlap with some of the results presented here, as do some more recent works~\cite{Arzt:1992wz,Degrassi:2005zd,Brod:2013cka,Gao:2012qpa,Borzumati:1999qt,Buras:2000if,Mebane:2013cra,Mebane:2013zga}.

\section{SILH operators}\label{sec:silh}

A minimal basis of dimension-six operators is obtained by removing all redundant operators using the SM EOM.  This paper uses the dimension-six operators $Q_i$ of Ref.~\cite{Grzadkowski:2010es} which has no redundancies.  It is a well-established result in quantum field theory that operators which vanish by the classical equations of motion do not contribute to $S$-matrix elements even at the quantum level~\cite{Politzer:1980me}, and so EOM can be used to simplify the effective Lagrangian.  Formally, the redundant operators can be eliminated by a change of variables in the functional integral.  It is clearly a nuisance to use a redundant operator basis.

Including redundant operators introduces extra parameters in the Lagrangian which can be eliminated by field redefintions, and do not contribute to any measurable quantity~\cite{Politzer:1980me}.  This redundancy is not always obvious, since intermediate steps and partial results can depend on the redundant parameters.  It is only when the complete $S$-matrix element is carefully computed that one sees that certain combinations of parameters drop out due to the EOM.  Redundant operators have led to enormous confusion in the literature over many decades, for example, this was a source of significant confusion in the early days of heavy quark effective theory.  For this reason, when choosing a basis, it is advantageous to not introduce redundant parameters.

Recently, some authors \cite{Contino:2013kra,Pomarol:2013zra,Elias-Miro:2013mua} have advocated using the ``SILH-basis.''  The definition of this operator basis varies in the papers, and the original SILH paper~\cite{Giudice:2007fh} does not define a complete basis.  We will discuss the version presented in 
Ref.~\cite{Contino:2013kra}.  The basis of Refs.~\cite{Buchmuller:1985jz,Grzadkowski:2010es} contains nine $CP$-even operators made out of only gauge and Higgs fields,
\begin{align}
&\, Q_G,\ Q_W,\ Q_H,\ Q_{H\Box},\ Q_{HD},\ Q_{HG},\ Q_{HW},\ Q_{HB},\ Q_{HWB}.
\end{align}
The SILH basis defined in Ref.~\cite{Contino:2013kra} contains 14 $CP$-even operators made out of only gauge and Higgs fields with the operator coefficients
\begin{align}
& \bar{c}_H,\ \bar{c}_T,\ \bar{c}_6,\ \bar{c}_W,\ \bar{c}_B,\ \bar{c}_{HW},\ \bar{c}_{HB},\ \bar{c}_{\gamma},\ \bar{c}_{g},\ \bar{c}_{3W},\ \bar{c}_{3G},\ \bar{c}_{2W},\ \bar{c}_{2B},\ \bar{c}_{2G}. 
\end{align}
The six operators $Q_G, Q_W, Q_H, Q_{H\Box}, Q_{HG},Q_{HB}$ coincide with the operators corresponding to $\bar{c}_{3G},\bar{c}_{3W},\bar{c}_6,\bar{c}_H, \bar{c}_{g},\bar{c}_{\gamma}$, up to simple rescalings by couplings.  In Ref.~\cite{Contino:2013kra}, it is argued that the three operators corresponding to 
$\bar{c}_{2W}$, $\bar{c}_{2B}$ and $\bar{c}_{2G}$ can be removed by the SM EOM in favor of other operators retained in the SILH operator basis.  This removal leaves five flavor-singlet operators\footnote{The SILH basis operators are denoted by $P_{i}$ to avoid confusion with similarly labelled operators $Q_{i}$ in the standard basis.}
\begin{align}
\mathcal{P}_{HW} &=  -i \, g_2 \, (D^\mu H)^\dagger \, \tau^I \, (D^\nu H) \,  W^I_{\mu \, \nu}, & \hspace{1cm}
\mathcal{P}_{HB} &=  -i \, g_1 \, (D^\mu H)^\dagger \,  (D^\nu H) \,  B_{\mu \, \nu},\nn
\mathcal{P}_{W} &=  -\frac{i \, g_2}{2} \, (H^\dagger \,  \tau^I \, \overleftrightarrow{D}^\mu H) \,  (D^\nu W^I_{\mu \, \nu}), & \hspace{1cm}
\mathcal{P}_{B} &=  -\frac{i \, g_1}{2} \, (H^\dagger \, \overleftrightarrow{D}^\mu H) \,  (D^\nu B_{\mu \, \nu}), \nn
\mathcal{P}_{T} &=  (H^\dagger \, \overleftrightarrow{D}^\mu H) \,  (H^\dagger \, \overleftrightarrow{D}^\mu H), 
\label{a}
\end{align}
in the  SILH basis, instead of the three operators
\begin{align}
Q_{HW} &= H^\dagger H \, W^{I}_{\mu\nu} \,  W_{I}^{\mu\nu}, &
Q_{HWB} &= H^\dagger \, \tau_I \, H \, W^{I}_{\mu\nu} \,  B^{\mu\nu}, &
Q_{HD} &= (H^\dagger D^\mu H)^\star \, (H^\dagger D_\mu H),
\label{b}
\end{align}
in the standard basis.  

Since Eq.~(\ref{a}) has five operators, and Eq.~(\ref{b}) has only three operators, two additional operators from the standard $Q_i$ basis can be eliminated if the operators in Eq.~(\ref{a}) are used instead of those in Eq.~(\ref{b}).
The five $P_i$ operators can be written in terms of the standard basis $Q_i$ using the equations of motion, and the conversion is given in Appendix~\ref{sec:convert}.
The relations involve non-bosonic $Q_i$ operators, 
a fact that is used in Ref.~\cite{Contino:2013kra} to remove the lepton-Higgs operators $Q_{\substack{H l  }}^{(1)}$ and $Q_{\substack{H l  }}^{(3)}$ together with $Q_{HW}$, $Q_{HWB}$ and $Q_{HD}$ in favor of the 5 $P_i$ operators of Eq.~(\ref{a}). 
However, only the flavor-singlet combinations  
\begin{align}
Q_{\substack{H l \\ pp}}^{(1)},\qquad Q_{\substack{H l \\ pp}}^{(3)}\,,
\label{remove1}
\end{align}
enter the relations in Eq.~(\ref{pops}).
One can modify  the singlet part
of the coefficients of  
 $Q_{\substack{H l  }}^{(1)}$ and $Q_{\substack{H l  }}^{(3)}$ by the shift
 \begin{align}
C_{\substack{H l \\ rs}}^{(1,3)} \to C_{\substack{H l \\ rs}}^{(1,3)} + a^{(1,3)}  \delta_{rs},%\qquad C_{\substack{H l \\ rs}}^{(3)} \to C_{\substack{H l \\ rs}}^{(3)} + a^{(3)}  \delta_{rs},
\label{modify}
\end{align}
and absorb the change in the $P_i$ operator coefficients. The constants $a^{(1,3)}$ can be chosen to eliminate the trace Eq.~(\ref{remove1}), or to set  the electron operator
$C_{\substack{H l \\ ee}}^{(1)}=0$, etc.
However, the coefficients of the flavor non-singlet parts
\begin{align}
C_{\substack{H l \\ rs}}^{(1)}-\frac{1}{n_g} \delta_{rs} C_{\substack{H l \\ pp}}^{(1)} ,\qquad C_{\substack{H l \\ rs}}^{(3)}-\frac{1}{n_g} \delta_{rs} C_{\substack{H l \\ pp}}^{(3)}
\label{remove}
\end{align}
\emph{cannot} be removed, and must be retained.
Removal of the flavor-singlet portions of $C_{\substack{H l \\ }}^{(1)}$ and $C_{\substack{H l \\ }}^{(3)}$ makes the treatment of BSM flavor violation in the SILH basis cumbersome. Furthermore, a careful and consistent treatment of EOM effects is necessary in all calculations using the ``SILH-basis,'' otherwise the basis remains redundant.

The lepton-Higgs operators $Q_{\substack{H l \\ }}^{(1)}$ and $Q_{\substack{H l \\ }}^{(3)}$ can be removed completely if one assumes completely unbroken $U(3)^5$ flavor symmetry of the UV theory, so that the coefficients of these operators are unit matrices in flavor space.  This assumption was implicitly adopted in the initial work of Ref.~\cite{Grojean:2006nn} that identified this field redefintion, and it is also adopted in Refs.~\cite{Contino:2013kra,Pomarol:2013zra,Elias-Miro:2013mua}.  
This assumption is stronger than assuming MFV, which only says that the coefficients of the lepton operators is a function of $Y_e^\dagger Y_e$, not that it is proportional to the unit matrix.  Ref.~\cite{Elias-Miro:2013mua} computes a few of the anomalous dimensions in the case of a $ U(3)^5$ flavor-symmetric BSM sector,  in an attempt to circumvent this difficulty. While the assumption of flavor-symmetric BSM physics can be adopted, it limits the applicability of the EFT.  One of the important features of the SM EFT is that it can be used to \emph{test} MFV, but this is only possible if MFV is not put in by hand.  Many SILH basis results cannot be used to test MFV in a straightforward manner, since stronger assumptions than MFV have already been built into the formalism.

In reducing the SILH operators to the operator basis of Ref.~\cite{Grzadkowski:2010es}, the EOM relations in Appendix~\ref{sec:convert} also include the SM dimension-four operator $(H^\dagger H)^2$, which is the usual $\lambda (H^\dagger H)^2$ Higgs interaction term.  
This means that the connection of the two bases also involves the redefinition of SM parameters. Explicitly, the RGE for the SM  parameters also have contributions from dimension-six operators, as pointed out in Ref.~\cite{Jenkins:2013zja}. These effects are not taken into account in Ref.~\cite{Elias-Miro:2013mua} preventing a comparison of our results with Ref.~\cite{Elias-Miro:2013mua}.\footnote{For an example of this effect, see Sec.~\ref{sec5.5}, Eq.~\ref{eqDYb}.}

Also note that Ref.~\cite{Elias-Miro:2013mua} advocates retaining redundant operators in intermediate steps of the analysis. Retaining redundant operators in partial results for an anomalous dimension matrix introduces spurious
gauge and scheme dependence, see the discussion in Ref.~\cite{Jenkins:2013zja}. It is not defined in Ref.~\cite{Elias-Miro:2013mua} 
how the partial results for the anomalous dimension matrix presented there can be converted to the full results valid for any BSM flavour structure. This is another reason we cannot compare our results with the partial calculation in Ref.~\cite{Elias-Miro:2013mua}.

\section{Minimal Flavor Violation}\label{sec:mfv}

The SM EFT provides a way to test the hypothesis of MFV in new physics. The SM has a $U(3)^5$ symmetry in the limit of vanishing Yukawa couplings under which
\begin{align}
q &\to U_q q,  & l &\to U_l l, & u &\to U_u u, & d &\to U_d d, & e &\to U_e e.
\end{align}
The MFV hypothesis~\cite{Chivukula:1987py,DAmbrosio:2002ex} is that the only source of flavor violation is the Yukawa matrices, so that the full theory is flavor invariant if the Yukawa matrices transform as
\begin{align}
Y_u &\to U_u Y_u U_q^\dagger, & Y_d &\to U_d Y_d U_q^\dagger, & Y_e &\to U_e Y_e U_l^\dagger\,.
\end{align}

If the new physics respects MFV, then the SM EFT derived from it also does. This assumption severely restricts the dimension-six coefficients. The coefficients of the flavor invariant operators in classes 1--4 can only depend on the flavor invariants\footnote{In this section, $f$ denotes an arbitrary function, and all the $f$s do not have to be the same.  Some $U(1)$s are anomalous, and one also can have dependence on certain combinations of $\det Y_{u,d,e}$ 
and the $\theta$ angles~\cite{Feldmann:2009dc,Jenkins:2009dy,Alonso:2011yg}.}
\begin{align}
\tr f(Y_e^\dagger Y_e),\qquad \tr f(Y_d^\dagger Y_d, Y_u^\dagger Y_u)\,,
\label{inv}
\end{align}
In an EFT setup, the dependence on such invariants can be absorbed into an effective coefficient.

The $\psi^2 H^3$  operators have coefficients
\begin{align}
C_{\substack{dH \\ rs}} &= \left[ f(Y_d^\dagger Y_d, Y_u^\dagger Y_u)\, Y_d^\dagger \right]_{rs},  &
C_{\substack{uH \\ rs}} &= \left[ f(Y_d^\dagger Y_d, Y_u^\dagger Y_u)\, Y_u^\dagger \right]_{rs}, &
C_{\substack{eH \\ rs}} &= \left[ f(Y_e^\dagger Y_e)\, Y_e^\dagger \right]_{rs}, 
\label{5.4}
\end{align}
where it is implicit that the above functions also can depend on the invariants of Eq.~(\ref{inv}).  For example, the quark functions can depend on the lepton invariant 
$\tr f(Y_e^\dagger Y_e)$ and vice-versa.  Analogous formulae to Eq.~(\ref{5.4}) hold for the 
$\psi^2 X H$ dipole operators $\{ C_{eW}, C_{eB} \}$, $\{ C_{uG}, C_{uW}, C_{uB} \}$ and $\{ C_{dG}, C_{dW}, C_{dB} \}$, respectively.

The $\psi^2 H^2 D$ operators have coefficients
\begin{align}
C_{\substack{Hq \\ rs}}^{(1,3)} &= \left[ f(Y_d^\dagger Y_d, Y_u^\dagger Y_u) \right]_{rs},  & C_{\substack{Hl \\ rs}}^{(1,3)} &= \left[ f(Y_e^\dagger Y_e) \right]_{rs},  \nn
C_{\substack{Hu \\ rs}} &= a \delta_{rs}+\left[Y_u\, f(Y_d^\dagger Y_d, Y_u^\dagger Y_u)\, Y_u^\dagger\right]_{rs},  &
C_{\substack{Hd \\ rs}} &= a \delta_{rs}+\left[Y_d\, f(Y_d^\dagger Y_d, Y_u^\dagger Y_u)\, Y_d^\dagger\right]_{rs},  \nn
C_{\substack{He \\ rs}} &= a \delta_{rs}+\left[Y_e\, f(Y_e^\dagger Y_e)\, Y_e^\dagger\right]_{rs},  &
C_{\substack{Hud \\ rs}} &= \left[Y_u\, f(Y_d^\dagger Y_d, Y_u^\dagger Y_u)\, Y_d^\dagger\right]_{rs}  .
\label{5.5}
\end{align}
Again, dependence of the above functions of the invariants of Eq.~(\ref{inv}) is implicit.

Similar expressions hold for the $\psi^4$ operators, with coefficients in flavor space which are products of the cases considered above. As is well-known, one can make $U(3)^5$ rotations to bring the Yukawa matrices into the form
\begin{align}
Y_e &\to \text{diag}(m_e,m_\mu,m_\tau), &
Y_d &\to  \text{diag}(m_d,m_s,m_b) ,  &
Y_u &\to  \text{diag}(m_u,m_c,m_t) K,\label{YukExpl}
\end{align}
where $K$ is the CKM matrix.
At this stage, the masslessness of neutrinos allows for the diagonalization of $Y_e$ and the absence of flavor violation in the lepton sector. The introduction of neutrino masses can be accomplished in the model-independent spirit of this paper via the $d=5$ Weinberg Operator. This operator is naturally suppressed by a scale higher than $\Lambda$ since it violates lepton number. Assuming this hierarchy of scales, the RGEs of $d=5$ and $d=6$ operators are independent and the inclusion of neutrino masses is orthogonal and does not affect the results presented here.

Since MFV is implemented as a symmetry which is respected by the SM Lagrangian, the RG evolution of $\lsix$ maintains MFV if the coefficients at scale $\Lambda$
satisfy the MFV hypothesis.  In this case, the flavor structure of $\lsix$ is the same as corresponding amplitudes computed from loop graphs in the SM.  However, it is important to emphasize that the assumption of MFV does \emph{not} imply that the coefficients of $\psi^2 H^2 D$ and $\psi^4$ operators are proportional to the unit matrix, which is a stronger assumption that requires that the functions $f$ have a perturbative expansion in $Y$ with small coefficients. In view of Eq.~(\ref{YukExpl}), this expansion in powers of Yukawa matrices can be justified for off-diagonal elements inducing flavor violation, as customary, but not for the diagonal entry of the third generation, see Ref.~\cite{Kagan:2009bn} for some discussion on this point.

One of the important applications of the SM EFT is to test the hypothesis of MFV in BSM physics in a model-independent way. Interestingly, the full SM RGE transfers flavor violation in one set of operators  to other operator sectors.
 Testing the consistency of MFV in low-energy measurements, 
taking into account the full SM EFT, is important for increasing our understanding of the flavor structure of new physics. A quick look at the
anomalous dimensions in Refs.~\cite{Jenkins:2013zja,Jenkins:2013wua} and Appendix~\ref{app:rge}  should convince the reader that any flavor ansatz not based on a symmetry will not be preserved by the RGE.

\section{Phenomenology}\label{pheno}

In this section, we outline the generalization of the analysis of observables measured at the electroweak scale from the SM to the SM EFT, and how the full one-loop RGE for the dimension-six Wilson coefficients measured at a low scale $\sim v$ can be used to obtain the Wilson coefficients at the high scale $\Lambda$.  An important point we emphasize is that if constraints at the scale $v$ are to be mapped to a high scale BSM theory, then {\it all} corrections of the order $v^2/(16 \, \pi^2 \Lambda^2)$ in the SM EFT have to be included in the analysis.  Otherwise, the analysis is inconsistent.

Our aim is not to perform a precision analysis, but to simply outline some issues that a precision Higgs and electroweak phenomenology program should take into account, and how the one-loop RGE result aids in this program.  Some aspects of how the SM EFT modifies SM phenomenology have been discussed previously 
in~Refs.~\cite{Corbett:2012ja,Contino:2013kra,Elias-Miro:2013mua,Almeida:2013jfa} and other works. However, many aspects of how the SM EFT affects precision predictions have  not been discussed in detail before, and we outline some of them below.

The Lagrangian of the SM EFT is
\begin{align}
\mathcal{L} = \mathcal{L}_{\rm SM} + \lsix + \ldots
\end{align}
where the $\ldots$ denote operators of dimension greater than six suppressed by additional powers of $\Lambda$. The dimension-six terms $\lsix$ can be treated perturbatively, i.e.\ we only need to include these to first order, since second-order contributions from $\lsix$ are as important as first-order contributions from $\mathcal{L}^{(8)}$, etc. The SM Lagrangian is
\begin{align}
\mathcal{L}_{\rm SM} &= -\frac14 G_{\mu \nu}^A G^{A\mu \nu}-\frac14 W_{\mu \nu}^I W^{I \mu \nu} -\frac14 B_{\mu \nu} B^{\mu \nu}
+ (D_\mu H^\dagger)(D^\mu H)
+\sum_{\psi=q,u,d,l,e} \overline \psi\, i \slashed{D} \, \psi\nn
&-\lambda \left(H^\dagger H -\frac12 v^2\right)^2- \biggl[ H^{\dagger j} \overline d\, Y_d\, q_{j} + \widetilde H^{\dagger j} \overline u\, Y_u\, q_{j} + H^{\dagger j} \overline e\, Y_e\,  l_{j} + \hbox{h.c.}\biggr],
\label{sm}
\end{align}
and $\lsix$ is defined in Eq.~(\ref{lsix}).  We start by discussing the modification of the SM parameters at tree-level due to $\lsix$.

\subsection{Higgs mass and self-couplings}\label{sec:Hmass}

The dimension-six Lagrangian of the SM EFT alters the definition of SM parameters at tree level in a number of ways.  The operator  $Q_H$ changes the shape of the scalar doublet potential at order $v^2/\Lambda^2$ to
\begin{align}
V(H) &= \lambda \left(H^\dagger H -\frac12 v^2\right)^2 - C_H \left( H^\dagger H \right)^3,
\label{pot}
\end{align}
yielding the new minimum
\begin{align}
\langle H^\dagger H \rangle &= \frac{v^2}{2} \left( 1+ \frac{3 C_H v^2}{4 \lambda} \right) \equiv \frac12 v_T^2,
\end{align}
on expanding the exact solution $(\lambda- \sqrt{\lambda^2-3 C_H \lambda v^2})/(3 C_H)$ to first order in $C_H$. The shift in the vacuum expectation value (VEV) is proportional to $C_H v^2$, which is of order $v^2/\Lambda^2$.

The scalar field can be written in unitary gauge as
\begin{align}
H &= \frac{1}{\sqrt 2} \left(\begin{array}{c}
0 \\
 \left[ 1+ \ckin \right]  h + v_T
 \end{array}\right),
 \label{Hvev}
\end{align}
where
\begin{align}\label{chkindef}
\ckin &\equiv \left(C_{H\Box}-\frac14 C_{HD}\right)v^2, &
v_T &\equiv \left( 1+ \frac{3 C_H v^2}{8 \lambda} \right) v.
\end{align}
The coefficient of $h$ in Eq.~(\ref{Hvev}) is no longer unity, in order for the Higgs boson kinetic term to be properly normalized when the dimension-six operators are included.  The kinetic terms 
\begin{align}
{\cal L} &= (D_\mu H^\dagger)(D^\mu H) + C_{H \Box} \left( H^\dagger H \right) \Box \left( H^\dagger H \right) + C_{HD} \left( H^\dagger D^\mu H \right)^* \left( H^\dagger D_\mu H \right),
\end{align}
and the potential in Eq.~(\ref{pot}) yield\footnote{One can always replace $v$ by $v_T$ in terms that depend on the $\lsix$ coefficients, since the change is order $1/\Lambda^4$.}
\begin{align}
{\cal L} &= {1 \over 2} \left( \partial_\mu h \right)^2 -\frac{\ckin}{v_T^2} \left[ h^2 (\partial_\mu h)^2+2v h (\partial_\mu h)^2\right] - \lambda v_T^2 \left(1-\frac{3 C_H v^2}{2 \lambda} +2 \ckin   \right)h^2 \nn
&-\lambda v_T \left(1-\frac{5 C_H v^2}{2 \lambda} +3 \ckin  \right)h^3-\frac14 \lambda  \left(1-\frac{15 C_H v^2}{2 \lambda} +4 \ckin \right)h^4+\frac34 C_H v h^5+\frac18 C_H h^6,
\end{align}
for the $h$ self-interactions. The Higgs boson mass is
\begin{align}
m_H^2 &= 2 \lambda v_T^2 \left(1-\frac{3 C_H v^2}{2 \lambda} + 2 \ckin \right)\,.
\end{align}

\subsection{Yukawa couplings}\label{sec:yuk}

The definition of the fermion mass matrices and the Yukawa matrices are modified by the presence of $\psi^2 H^3$ operators.  The Lagrangian terms in the unbroken theory
\begin{align}
{\cal L} &= - \biggl[ H^{\dagger j} \overline d_r\, \left[Y_d \right]_{rs}\, q_{j s} + \widetilde H^{\dagger j} \overline u_r\, \left[ Y_u \right]_{rs} \, q_{j s} 
+ H^{\dagger j} \overline e_r\, \left[Y_e\right]_{rs}\,  l_{j s} + \hbox{h.c.}\biggr] \nn
&+ \left[ C^*_{\substack{dH \\ sr}} \left( H^\dagger H \right) H^{\dagger j}  \overline d_r q_{j s} + C^*_{\substack{uH \\ sr}} \left( H^\dagger H \right) \tilde H^{\dagger j}  \overline u_r q_{j s} + C^*_{\substack{eH \\ sr}} \left( H^\dagger H \right) H^{\dagger j}  \overline e_r l_{j s} + \hbox{h.c.} \right],
\end{align}
yield the fermion mass matrices
\begin{align}
\left[ M_\psi \right]_{rs} &= \frac{v_T}{\sqrt 2} \left( \left[Y_\psi \right]_{rs}   - \frac12 v^2 C^*_{\substack{\psi H \\ sr}}   \right), \qquad \psi=u,d,e
\end{align}
in the broken theory.  The  coupling matrices of the $h$ boson to the fermions $\mathcal{L}=- h\ \overline u \, \mathcal{Y}\, q + \ldots$ are
\begin{align}
\left[ {\cal Y}_\psi \right]_{rs} &= \frac{1}{\sqrt 2}  \left[ Y_\psi \right]_{rs}\left[ 1+ \ckin \right]  - \frac3{2 \sqrt 2} v^2 C^*_{\substack{\psi H \\ sr}} 
\nn
& = \frac{1}{v_T}\left[ M_\psi \right]_{rs} \left[ 1+ \ckin  \right]   - \frac{v^2}{\sqrt 2} C^*_{\substack{\psi H \\ sr}} , 
\qquad \psi=u,d,e
\label{5.12}
\end{align}
and are not simply proportional to the fermion mass matrices, as is the case in the SM.  In general, the fermion mass matrices and Yukawa matrices will not be simultaneously diagonalizable (these parameters have different RGEs), so that the couplings of the Higgs boson to the fermions will not be diagonal in flavor due to terms of order $v^2/\Lambda^2$.

\subsection{$G_F$}\label{sec:GFredef}

The value of the VEV in the SM is obtained from the measurement of  $G_F$ in $\mu$ decay, $\mu^- \rightarrow e^- + \bar{\nu}_e + \nu_\mu$. Define the local effective interaction for muon decay as
\begin{align}
\mathcal{L}_{G_F} =  -\frac{4\mathcal{G}_F}{\sqrt{2}} \, \left(\bar{\nu}_\mu \, \gamma^\mu P_L \mu \right) \left(\bar{e} \, \gamma_\mu P_L \nu_e\right).
\end{align}
The parameter $\mathcal{G}_F$ is fixed by measuring the muon lifetime. In the SM EFT,\footnote{$e$ and $\mu$ are generation indices 1 and 2, and are not summed over.}
\begin{align}
-\frac{4\mathcal{G}_F}{\sqrt{2}} &=  -\frac{2}{v_T^2} +  \left(C_{\substack{ll \\ \mu ee \mu}} +  C_{\substack{ll \\ e \mu\mu e}}\right) - 2 \left(C^{(3)}_{\substack{Hl \\ ee }} +  C^{(3)}_{\substack{Hl \\ \mu\mu }}\right).
\label{gfermi}
\end{align}
The $C_{ll}$ terms are from the four-lepton interaction in $\lsix$, and the $C_{Hl}^{(3)}$ terms are from $W$ exchange, where one $W \overline l \nu$ vertex is from the $Q_{Hl}^{(3)}$ operator, and the other is the usual SM vertex. There are contributions to $\mu$ decay from $C_{\substack{ll \\ \mu e r s}}$, and $C_{\substack{ll \\ r s\mu e }}$ with $r \not = e, s \not=\mu$, as well as from $(\overline L L)(\overline R R)$ currents, but these do not interfere with the SM amplitude, and their contributions to the muon lifetime are higher order in $1/\Lambda$.

Similar expressions hold for other weak decay processes, and $\mathcal{G}_F$ in $\tau$ decay, or in quark decays, can differ from $\mu$ decay due to the $C_{ll}$ and $C_{Hl}^{(3)}$ terms.

\subsection{Gauge boson masses and couplings}\label{sec:Gmass} 

The definition of the gauge fields and the gauge couplings are affected by the dimension-six terms.  The relevant dimension-six Lagrangian terms are
\begin{align}
\lsix &=  C_{HG} H^\dagger H G_{\mu \nu}^A G^{A\mu \nu} + C_{HW} H^\dagger H W_{\mu \nu}^I W^{I \mu \nu}  + C_{HB} H^\dagger H B_{\mu \nu} B^{\mu \nu} + C_{HWB} H^\dagger \tau^I H W^I_{\mu \nu} B^{\mu \nu} \nn
&+ C_G f^{ABC} G^{A \nu}_\mu G^{B \rho}_\nu G^{C \mu}_\rho + C_W \epsilon^{IJK} W^{I \nu}_\mu W^{J \rho}_\nu W^{K \mu}_\rho \ .
\end{align}
In the broken theory, the $X^2 H^2$ operators contribute to the gauge kinetic energies,
\begin{align}
{\cal L}_{SM} +  \lsix &= - \frac{1}{2} W^+_{\mu \nu} \, W_-^{\mu \nu} - \frac{1}{4} W^3_{\mu \nu} \, W_3^{\mu \nu} - \frac{1}{4} B_{\mu \nu} \, B^{\mu \nu} -
\frac{1}{4} G_{\mu \nu}^A \, G^{A\mu \nu} +
 \frac12 v_T^2 \, C_{HG} \,G_{\mu \nu}^A \, G^{A\mu \nu}, \nn
&\, \hspace{0.4cm}+ \frac12 v_T^2 \, C_{HW}  W_{\mu \nu}^I W^{I \mu \nu}+ \frac12 v_T^2 \, C_{HB}  B_{\mu \nu} B^{\mu \nu} -\frac{1}{2} v_T^2 \, C_{HWB} W_{\mu \nu}^3 B^{\mu \nu},
\label{5.16}
\end{align}
so the gauge fields in the Lagrangian are not canonically normalized, and the last term in Eq.~(\ref{5.16}) leads to kinetic mixing between $W^3$ and $B$. The mass terms for the gauge bosons from $\mathcal{L}_{\rm SM}$ and $\lsix$ are
\begin{align}
{\cal L} &= \frac14 g_2^2 v_T^2  W_\mu^+ W^{-\,\mu} + \frac18 v_T^2  (g_2 W^3_\mu -g_1 B_\mu)^2  + \frac1{16} v_T^4 C_{H D}(g_2 W^3_\mu -g_1 B_\mu)^2\,.
\end{align}

The gauge fields need to be redefined, so that the kinetic terms are properly normalized and diagonal. The first step is to redefine the gauge fields
\begin{align}
G_\mu^A &= \mathcal{G}_\mu^A \left(1 + C_{HG} v_T^2 \right), &
W^I_\mu  &=  \mathcal{W}^I_\mu \left(1 + C_{HW} v_T^2 \right), &
B_\mu  &=  \mathcal{B}_\mu \left(1 + C_{HB} v_T^2 \right).
\label{5.16a}
\end{align}
The modified coupling constants are
\begin{align}
\gcg &= g_3 \left(1 + C_{HG} \, v_T^2 \right), & \gcw &= g_2 \left(1 + C_{HW} \, v_T^2 \right), & \gcb &= g_1 \left(1 + C_{HB} \, v_T^2 \right),
\label{5.16b}
\end{align}
so that the products $g_3 G_\mu^A=\gcg \mathcal{G}_\mu^A$, etc.\ are unchanged. This takes care of the gluon terms. The electroweak  terms are
\begin{align}
{\cal L} &=  - \frac{1}{2} \mathcal{W}^+_{\mu \nu} \, \mathcal{W}_-^{\mu \nu} - \frac{1}{4} \mathcal{W}^3_{\mu \nu} \, \mathcal{W}_3^{\mu \nu} - \frac{1}{4} \mathcal{B}_{\mu \nu} \, \mathcal{B}^{\mu \nu}  - \frac{1}{2} \left(v_T^2 C_{HWB}\right)
\mathcal{W}_{\mu \nu}^3 \mathcal{B}^{\mu \nu} +\frac14 \gcw^2 v_T^2  \mathcal{W}_\mu^+ \mathcal{W}^{-\,\mu} \nn
&+ \frac18 v_T^2  (\gcw \mathcal{W}^3_\mu - \gcb \mathcal{B}_\mu)^2  + \frac1{16} v_T^4 C_{H D}( \gcw \mathcal{W}^3_\mu - \gcb \mathcal{B}_\mu)^2.
\label{5.20}
\end{align}
The mass eigenstate basis is given by~\cite{Grinstein:1991cd}
\begin{align}
\left[ \begin{array}{cc}  \mathcal{W}_\mu^3 \\ \mathcal{B}_\mu \end{array} \right]
&=  
\left[ \begin{array}{cc}  1   &  -  \frac{1}{2} \,  v_T^2 \,  C_{HWB} \\
- \frac{1}{2} \,  v_T^2 \,  C_{HWB} & 1 \end{array} \right] \, \left[ \begin{array}{cc} \cos \tc  &  \sin \tc \\
-\sin \tc & \cos \tc \end{array} \right] \left[ \begin{array}{cc}  \mathcal{Z}_\mu \\ \mathcal{A}_\mu \end{array} \right], 
\end{align}
where the rotation angle is
\begin{align}
\tan \tc &= \frac{\gcb}{\gcw} + \frac{v_T^2}{2}  \,  C_{HWB} \, \left[1 -  \frac{\gcb^2}{\gcw^2}\right],
\end{align}
so that
\begin{align}
\sin \tc &= \frac{\gcb}{\sqrt{\gcb^2+\gcw^2}}\left[1 + \frac{v_T^2}{2}  \,   \, \frac{\gcw}{\gcb}\ \frac{\gcw^2-\gcb^2}{\gcw^2+\gcb^2} C_{HWB} \right], \nn
\cos \tc &= \frac{\gcw}{\sqrt{\gcb^2+\gcw^2}}\left[1 - \frac{v_T^2}{2}  \,   \, \frac{\gcb}{\gcw}\ \frac{\gcw^2-\gcb^2}{\gcw^2+\gcb^2} C_{HWB} \right].
\end{align}

The photon is massless, as it must be by gauge invariance, since $U(1)_Q$ is unbroken. The $W$ and $Z$ masses are
\begin{align}
M_W^2 &= \frac{\gcw^2 v_T^2}{4} , \nn
M_Z^2 &= \frac{v_T^2}{4}(\gcb^2+\gcw^2)+\frac{1}{8}v_T^4 C_{HD} (\gcb^2+\gcw^2)+\frac{1}{2}v_T^4 \gcb\gcw C_{HWB}.
\end{align}

The covariant derivative is
\begin{align}
D_{\mu} = \partial_\mu + i \, \frac{\gcw}{\sqrt{2}} \left[\mathcal{W}_{\mu}^+ T^+ + \mathcal{W}_{\mu}^- T^- \right] + i \,
\gcZ \left[T_3 - \sc^2 Q \right] \mathcal{Z}_\mu + 
i \, \ec \, Q \, \mathcal{A}_\mu,
\end{align}
where $Q=T_3+Y$,  and the effective couplings are given by
\begin{align}
\ec &= \frac{\gcb \gcw}{\sqrt{\gcw^2+\gcb^2}} \left[ 1 - \frac{\gcb \gcw}{\gcw^2+\gcb^2} v_T^2 C_{HWB} \right] = \gcw \, \sin \tc - \frac{1}{2} \, \cos \tc  \, \gcw \, v_T^2 \, C_{HWB}, \nn
\gcZ &= \sqrt{\gcw^2 + \gcb^2} + \frac{\gcb \gcw}{\sqrt{\gcw^2 + \gcb^2} } v_T^2  C_{HWB}=  \frac{\ec}{\sin \tc \cos \tc}  \left[1 +  \frac{\gcb^2+\gcw^2}{2 \gcb \gcw} v_T^2C_{HWB}\right] ,\nn
\sc^2 &= \sin^2 \tc=  \frac{\gcb^2}{{\gcw^2 + \gcb^2}} + \frac{\gcb \gcw (\gcw^2-\gcb^2)}{(\gcb^2+\gcw^2)^2}  v_T^2 C_{HWB} .
\end{align}
The $\rho$ parameter, defined as the ratio of charged and neutral currents at low energies~\cite{Ross:1975fq}, is
\begin{align}
\rho \equiv \frac{\gcw^2 M_Z^2}{\gcZ^2 M_W^2} = 1  +  \frac{1}{2} v_T^2 \, C_{HD}. 
\end{align}
Measurements of the $W$ and $Z$ masses and couplings, and the photon coupling fix $\gcb$, $\gcw$, $v_T$, $C_{HWB}$ and $C_{HD}$.
The couplings of the gauge bosons to fermions are also modified, and a recent discussion can be found in Ref.~\cite{Buchalla:2013mpa}.

\subsection{RGE for $C_H, C_{HD}, C_{H\Box}$}\label{sec5.5}

The discussion in Sections~\ref{sec:Hmass}--\ref{sec:Gmass} studied the impact of higher dimensional operators on the measured SM parameters at tree level. The coefficients $C_{HD}$, etc.\ of the higher dimension operators that enter the expressions are renormalized at the low scale, and are related to the parameters at the high scale $\Lambda$ by the RGE. As mentioned earlier, the RGE contributions are the same as the $\log \Lambda/m_H$ enhanced contributions from the finite parts of the one-loop diagrams.

The RGE for $C_{H}$, $C_{HD}$ and $C_{H\Box}$ which enter the Higgs and gauge Lagrangian are
\begin{align}
\dot C_H &= \left(108 \lambda + 6 Y(S) -\frac92 g_1^2-\frac{27}{2}g_2^2\right) C_H -12 g_1^2 \hyp_h^2 \left(4 g_1^2 \hyp_h^2 +g_2^2-4\lambda\right)C_{HB}
\nn
&
-3 g_2^2\left(4 g_1^2 \hyp_h^2+3 g_2^2-12 \lambda\right)C_{HW}  -6 g_1 g_2 \hyp_h \left(4 g_1^2 \hyp_h^2+g_2^2-4\lambda\right)C_{HWB}
\nn
& 
- \frac34 \left( (4 \hyp_h^2 g_1^2+g_2^2)^2 + 8(g_2^2-4 g_1^2 \hyp_h^2)\lambda - 64 \lambda^2 \right) C_{HD} \nn
& +\frac{40}{3}(g_2^2\lambda-12\lambda^2) C_{H\Box} + \frac{16 g_2^2 \lambda}{3} C_{\substack{H l \\ t t}}^{(3)}
+ 16 g_2^2 \lambda  C_{\substack{H q \\ t t}}^{(3)} +8 \lambda(\eta_1+\eta_2) \nn
& -4
\left( [Y_e Y_e^\dagger Y_e]_{wv} C_{\substack{ eH \\ vw}} + 3 [Y_d Y_d^\dagger Y_d]_{wv} C_{\substack{ dH \\ vw}} + 3 [Y_u Y_u^\dagger Y_u]_{wv} C_{\substack{ uH \\ vw}} +  h.c. \right),
\label{cH}
\end{align}
\begin{align}
\dot C_{H\Box} &=  \left(-\frac{16}{3} \, \hyp_h^2 \, g_1^2 - 4 \, g_2^2 + 24 \lambda + 4 Y(S) \right) C_{H \Box} + 2 g_2^2 \,  C_{\substack{H l \\ tt}}^{(3)} + 2  g_2^2 N_c \,  C_{\substack{H q \\ tt}}^{(3)} + \frac{20}{3}\, g_1^2 \, \hyp_h^2 \, C_{HD} \nn
&+ \frac{4 \, g_1^2 \, \hyp_h}{3} \left(N_c \,  \hyp_d \, C_{\substack{Hd \\ tt}} + \hyp_e \, C_{\substack{He \\ tt}} + 2 \hyp_l \, C_{\substack{H l \\ tt}}^{(1)} + 2 N_c \hyp_q C_{\substack{H q \\ tt}}^{(1)}
+ N_c \hyp_u C_{\substack{H u \\ tt}}^{(1)} \right)   - 2 \, \eta_3,
\label{cHBox}
\end{align}
\begin{align}
\dot C_{HD} &=  \left(-\frac{10}{3} \, \hyp_h^2 \, g_1^2  +\frac92 \, g_2^2 + 12 \lambda + 4 Y(S) \right) C_{H D} + \frac{80}{3}\, g_1^2 \, \hyp_h^2 \, C_{H\Box}  \nn
&+ \frac{16 \, g_1^2 \, \hyp_h}{3} \left(N_c \,  \hyp_d \, C_{\substack{Hd \\ tt}} + \hyp_e \, C_{\substack{He \\ tt}} + 2 \hyp_l \, C_{\substack{H l \\ tt}}^{(1)} + 2 N_c \hyp_q C_{\substack{H q \\ tt}}^{(1)}
+ N_c \hyp_u C_{\substack{H u \\ tt}}^{(1)} \right) - 2 \, \eta_4,
\label{cHD}
\end{align}
where $\eta_{1,2,3,4}$ are defined in our previous paper Ref.~\cite{Jenkins:2013wua}. The precision electroweak parameter $T$ is $C_{HD}$, so these RGE are also used in Sec.~\ref{sec:EWPD}.  Note that the dimension-six operator coefficients from the operators in parentheses on the second lines of Eqs.~(\ref{cHBox}) and~(\ref{cHD}) drop out of the running of the combination $\left(C_{H \Box} -  C_{HD} /4\right)$ appearing in $c_{H,{\rm kin}}$. The RGE for $C_{HWB}$ is given in Sec.~\ref{sec:EWPD}.

The RGE in Eqs.~(\ref{cH})--(\ref{cHD}) depend on other coefficients in $\lsix$. If the scale $\Lambda$ is a few TeV, the RGE can be integrated perturbatively, so that
\begin{align}
C(\mu) &\approx C(\Lambda)- \frac{1}{16\pi^2} \gamma_C \ln \frac{\Lambda}{\mu} + \ldots &\text{where} &&  \dot C &= \gamma_C\,,
\label{5.37}
\end{align}
and the $\ldots$ are part of the leading-log series $\gamma_C^2 \ln^2 \Lambda/\mu$ given by exact integration of the RGE. The $\ln \Lambda/\mu$ terms in Eq.~(\ref{5.37}) must be the same as the $\ln \mu$ terms in the finite parts of the one-loop graphs. Thus the anomalous dimensions are another way of computing the $\ln \Lambda/\mu$ enhanced terms in the finite parts of the one-loop graphs.

\subsection{$h \to f \overline f$ }\label{bbbar}

The decay of the Higgs boson into fermions is another important test of the symmetry breaking structure of the SM. Define the effective coupling $\mathcal{Y}_b$ of the $b$ quark to the Higgs by  $\mathcal{L}_{\rm Yuk} = - \mathcal{Y}_b \, h \, \, \bar{b} \, b$. The decay width is given by
\begin{align}
\Gamma(h \rightarrow \bar{b} \, b) =  \frac{\mathcal{Y}_b^2 \, m_H \, N_c}{8 \, \pi} \left(1 - 4 \frac{m_b^2}{m_H^2} \right)^{3/2},
\end{align}
where all parameters are renormalized at $\mu \sim m_H$.
 
 In the SM, the effective coupling of the $b$ quark to the Higgs field can be predicted very accurately.  The $b$-quark mass can be determined very precisely from global studies of $\bar{B} \rightarrow X_c \, \ell \bar{\nu}$ and $X_s \gamma$ \cite{Bauer:2004ve},  and then used to determine the $b$-quark Yukawa coupling at the scale $m_H$ using the SM RGE. The relation $\mathcal{Y}_b=\sqrt 2 m_b/v$ between the Higgs coupling and quark mass is modified in the SM EFT, and is given by Eq.~(\ref{5.12}), with $\mathcal{Y}_b=[\mathcal{Y}_d]_{bb}$, and the relation between $v$ and $G_F$ is modified as in Eq.~(\ref{gfermi}) due to tree level effects from $\lsix$.

The scaling of parameters from $m_b$ to $m_H$ is also modified. The dimension-six operator contribution to the one-loop running of the effective coupling of the SM Higgs to fermions is given in Ref.~\cite{Jenkins:2013zja}.  We repeat the result for the down quarks here for the sake of completeness.\footnote{Note that the usual one loop running of the SM parameters summarized in Ref.~\cite{Machacek:1983tz,Machacek:1983fi,Machacek:1984zw} should be added to this result for the full scale dependence of these effective couplings in the SM EFT.} The running of the $Y_d$  is modified by the terms
\begin{align}
\mu \frac{\rd}{\rd \mu}  [Y_d]_{rs} &=  \frac{m_H^2 }{16 \pi^2} \biggl[3 C_{\substack{dH \\ sr}}^*  - C_{H \Box} [Y_d]_{rs} + \frac12 C_{H D} [Y_d]_{rs} + [Y_d]_{rt} \left( C^{(1)}_{\substack{H q \\ ts}} +3  C^{(3)}_{\substack{H q \\ ts}}\right)  - C_{\substack{H d \\ rt}} [Y_d]_{ts} \nn
&- [Y_u]_{ts} C^*_{\substack{H ud \\ tr}}  -2 \left( C^{(1)}_{\substack{qd \\ psrt}} + c_{F,3} C^{(8)}_{\substack{qd \\ psrt}} \right) [Y_d]_{tp}  
+C_{\substack{ledq \\ ptrs}} {[Y_e]_{tp}}  + N_c C^{(1)*}_{\substack{quqd \\ ptsr}} [Y_u]^*_{tp}\nn
&+ \frac12 \left( C^{(1)*}_{\substack{quqd \\ sptr}} + c_{F,3} C^{(8)*}_{\substack{quqd \\ sptr}} \right){ [Y_u]^*_{pt} }\biggr] \,.
\label{mix:fermion}
\end{align}
These terms are of order $v^2/\Lambda^2$, and are just as important as the 
running of the $C_{\psi H}$ and $\ckin$ contributions in Eq.~(\ref{5.12}), 
%$v^2/\Lambda^2$ terms in Eq.~(\ref{5.12}), 
and must be included for a consistent calculation.

The net effect of including the RGE in Eq.~(\ref{5.12}) and Eq.~(\ref{mix:fermion}) is to introduce a shift of the form
\begin{align}\label{eqDYb}
\Gamma(h \rightarrow \bar{b} \, b) =  \frac{(\mathcal{Y}_b + \Delta \mathcal{Y}_b)^2 \, m_H \, N_c}{8 \, \pi} \left(1 - 4 \frac{m_b^2}{m_H^2} \right)^{3/2},
\end{align}
where the running effects induced by new physics are included in $\Delta \mathcal{Y}_b$:
\begin{align}
\Delta \mathcal{Y}_b &=  \frac{m_H^2 }{16 \pi^2}  \, \log \left(\frac{m_H}{m_b} \right) C_1 +  \frac{m_H^2 }{16 \pi^2}  \, \log \left(\frac{m_H}{\Lambda} \right) C_2.
\label{C12}
\end{align}
The expression for $C_1$ is obtained by setting $r = s = 3$ in the expression in square brackets on the r.h.s.\ of  Eq.~(\ref{mix:fermion}).
 The expression for $C_2$ is 
\begin{equation}
C_2=\frac {1} {2\sqrt{2}\lambda} [Y_d]_{33}\left(\dot C_{H\Box}-\frac 14 \dot C_{HD}\right)-\frac{3}{4\lambda}  \dot C_{\substack{d H \\ 33}}^*
\end{equation}
where the anomalous dimensions $\dot C_{H\Box}$, $\dot C_{HD}$,  and $\dot C_{dH}$ are given in Eqs.~(\ref{cHBox}),~(\ref{cHD}) and Sec.~\ref{Ap-psiH} respectively.
 Note that as we are considering $\Lambda \sim {\rm TeV}$, the log enhancement is modest and of about the same size for running from $m_b$ to $m_H$ and from $\Lambda$ to  $m_H$.  The $\log(m_H/m_b)$ contribution in Eq.~(\ref{C12}), and analogous terms in other amplitudes,  have been neglected in Ref.~\cite{Elias-Miro:2013mua}, and need to be included for a consistent calculation including $1/\Lambda^2$ RGE effects.

The discussion above also applies to Higgs decays into other fermions, such as $c \overline c$ and $\tau^+ \tau^-$. Using
newly developed charm tagging techniques~\cite{ATLAS-CONF-2013-068},  it may be possible to measure deviations in $\Gamma(h \rightarrow \bar{c} \, c)$ at the LHC (see the discussion in Ref. \cite{Delaunay:2013pja}).

There are also flavor-changing Higgs-fermion couplings from $\lsix$, which contribute to flavor-changing Higgs decays, such as $h \to b \overline s$.  These do not interfere with the SM Higgs amplitude, which is flavor diagonal, so the flavor-changing decay rates are order $1/\Lambda^4$. Nevertheless, as the running of $C_{eH}$, $C_{dH}$ and $C_{uH}$ is not the same as the running of the SM Yukawa couplings, searches for Higgs flavor violation is well-motivated. For some recent work on this subject, see Refs.~\cite{Blankenburg:2012ex,Harnik:2012pb}.

\subsection{$h \to WW$ and $h \to ZZ$}

The $h \to WW$ and $h \to ZZ$ amplitudes receive direct contributions from $\lsix$. The relevant $CP$-even Lagrangian terms are
\begin{align}
\mathcal{L} &= (D_\mu H)^\dagger (D^\mu H) - \frac{1}{4} \left(W^I_{\mu\nu} W^{I \mu\nu} + B_{\mu\nu} B^{\mu\nu} \right), \nn
&\hspace{0.4cm} + C_{HW} \, Q_{HW} + C_{HB} \, Q_{HB} + C_{HWB} Q_{HWB} + C_{HD} \, Q_{HD},
\label{gaugebosontree}
\end{align}
which lead to the interactions
\begin{align}
\mathcal{L}&=\frac14 \gcw^2 v_T     h \left[(\mathcal{W}^1_\mu)^2+(\mathcal{W}^2_\mu)^2\right]\left[1 + \ckin  \right] +  C_{HW}  v_T  h \left[(\mathcal{W}^1_{\mu\nu})^2+(\mathcal{W}^2_{\mu \nu})^2\right]
\label{5.41}
\end{align}
for the $W$, and
\begin{align}
\mathcal{L}&=\frac14 (\gcw^2+\gcb^2)v_T h (\mathcal{Z}_\mu)^2 \left[1 + \ckin +v_T^2 C_{HD} \right] +\frac12  \gcb \gcw v_T^3 h (\mathcal{Z}_\mu)^2 C_{HWB} \nn
& +  v_T  h (\mathcal{Z}_{\mu\nu})^2 \left( \frac{\gcw^2 C_{HW}+ \gcb^2 C_{HB} +\gcb \gcw C_{HWB}}{\gcw^2+\gcb^2} 
\right)
\label{5.42}
\end{align}
for the $Z$.

A ratio of deviations in the SM gauge boson coupling to the Higgs, reported in  \cite{ATLAS-CONF-2013-034}, is defined as
\begin{align}
\lambda_{WZ} 
&\equiv \frac{\Gamma(h \rightarrow WW)}{\Gamma(h \rightarrow WW)_{SM}} \frac{\Gamma(h \rightarrow ZZ)_{SM}}{\Gamma(h \rightarrow {ZZ})} 
\end{align}
From Eqs.~(\ref{5.41},\ref{5.42}), we see that $\ckin$ cancels out in $\lambda_{WZ}$, but there are corrections from the Higgs-gauge operators $C_{HW}$, $C_{HB}$ and $C_{HWB}$. This correction depends on the off-shellness of the $W$ and $Z$, since it is proportional to the field-strength tensors, and thus momentum-dependent. In the SM EFT, the ratio $\lambda_{WZ}$ depends on the $\lsix$ parameters $C_{HW}$, $C_{HB}$ and which are not custodial $SU(2)$ violating, as well as $C_{HWB}$ and $C_{HD}$ which are custodial $SU(2)$ violating.
The couplings of the gauge bosons to fermions are also modified. For a recent discussion on these corrections in this basis see Ref.~\cite{Buchalla:2013mpa}.

\subsection{$gg \to h$}\label{sec:ggh}

The Higgs-gluon operators $Q_{HG}$ and $Q_{H \widetilde G}$ contribute to the Higgs production rate via gluon fusion. 
The $\lsix$ contribution to $gg \to h$ is important because the SM amplitude starts at one loop order, with no tree-level contribution. 
A similar enhancement of $\lsix$ corrections occurs for $h \to \gamma \gamma$ and $h \to \gamma Z$ discussed in the next two sections.

Define $\mathscr{C}_{gg}$ and $\widetilde {\mathscr{C}}_{gg}$ by rescaling $C_{HG}$ and $C_{H \widetilde G}$ by $g_3$,
\begin{align}
C_{HG} &=g_3^2 \mathscr{C}_{gg} & C_{H\widetilde G} &= g_3^2\widetilde {\mathscr{C}}_{gg}\,.
\end{align}
The scaling by $g_3$ simplifies the RGE, and makes contact with the notation of Refs.~\cite{Manohar:2006gz,Grojean:2013kd} which uses
\begin{align}
\mathscr{C}_{gg} &=-\frac{c_G}{2\Lambda^2} & \widetilde {\mathscr{C}}_{gg} &=-\frac{\tilde c_G}{2\Lambda^2}
\end{align}
since a factor of $-1/(2\Lambda^2)$ was included in the normalization of the operators. The other advantage of the rescaling is that the field and coupling constant renormalizations Eq.~(\ref{5.16a}) and~(\ref{5.16b}) cancel out.

The change in $gg \to h$ relative to the SM is given by~\cite{Manohar:2006gz}
\begin{align}
\label{glue}
{ \sigma(gg \to h) \over \sigma^{\text{SM}}(gg \to h)} \simeq { \Gamma(h\to gg) \over \Gamma^{\text{SM}}(h \to gg)}
&\simeq \abs{1+\frac{16 \pi^2 v^2 \mathscr{C}_{gg} }{I^g} }^2+\abs{ \frac{16 \pi^2 v^2  \widetilde {\mathscr{C}}_{gg} } {I^g} }^2
\end{align}
where $I^g \approx 0.37$ is the numerical value of a Feynman parameter integral\cite{Bergstrom:1985hp,Manohar:2006gz}. 
We have neglected corrections from $\ckin$ and the Yukawa couplings Eq.~(\ref{5.12}) which are $v^2/\Lambda^2$, but not enhanced by $16 \pi^2$.  If $\mathscr{C}_{gg}$ from BSM physics is loop suppressed as in the SM,  then these terms must be included.

The complete one loop RGE of  $\mathscr{C}_{gg}$ and $\widetilde {\mathscr{C}}_{gg}$ are relatively simple,
\begin{align}
\dot {\mathscr{C}}_{gg}  &=  \left( 12 \lambda +2 Y(S) -\frac32 g_1^2 -\frac92 g_2^2 \right) \mathscr{C}_{gg} 
-2\left( [Y_d]_{wv} \mathscr{C}_{\substack{dG \\ v w}} + [Y_u]_{wv} \mathscr{C}_{\substack{uG \\ v w}} + h.c. \right) \nn
\dot {\widetilde {\mathscr{C}}}_{gg}  &=  \left( 12 \lambda +2 Y(S) -\frac32 g_1^2 -\frac92 g_2^2 \right) \widetilde {\mathscr{C}}_{gg}
+ 2\left(i  [Y_d]_{wv} \mathscr{C}_{\substack{dG \\ v w}} +i [Y_u]_{wv} \mathscr{C}_{\substack{uG \\ v w}} + h.c. \right)
\label{5.34}
\end{align}
where
\begin{align}
{C}_{\substack{dG \\ v w}} &= g_3 \mathscr{C}_{\substack{dG \\ v w}}, & {C}_{\substack{uG \\ v w}} & =g_3 \mathscr{C}_{\substack{uG \\ v w}},
\end{align}
are rescaled coefficients of the color magnetic dipole operators, and
\begin{align}
Y(S) &=\text{Tr}\left[N_c Y_u^\dagger Y_u + N_c Y_d^\dagger Y_d + Y_e^\dagger Y_e\right] .
\end{align}
The Higgs-gluon contributions in the first term of  Eq.~(\ref{5.34}) were computed in Ref.~\cite{Grojean:2013kd}. The only new contribution from the full RGE is the second term from the color dipole operators which was also calculated in Ref. \cite{Elias-Miro:2013gya} for the special case of no flavor indices, and with only a non-zero top quark Yukawa coupling.

\subsection{$h \to \gamma \gamma$}\label{sec:gg}

A very important process is $h \to \gamma \gamma$, which played a key role in the discovery of the SM scalar. Again, it is convenient to define
\begin{align}
\mathscr{C}_{\gamma \gamma} &= \frac{1}{g_2^2}C_{HW}+\frac{1}{g_1^2}C_{HB}-\frac{1}{g_1g_2}C_{HWB},
\end{align}
in terms of which our previously defined coefficients~\cite{Manohar:2006gz,Grojean:2013kd} are
\begin{align}
 \mathscr{C}_{\gamma \gamma} &= -\frac{c_{\gamma\gamma}}{2\Lambda^2},  &  \widetilde {\mathscr{C}}_{\gamma \gamma} &=-\frac{\tilde c_{\gamma \gamma}}{2\Lambda^2}.
\end{align}
The $h \to \gamma \gamma$ rate is
\begin{align}
\label{photon}
{ \Gamma(h\to \gamma \gamma) \over \Gamma^{\text{SM}}(h \to \gamma \gamma)}
&\simeq \abs{1+\frac{8 \pi^2 v^2 \mathscr{C}_{\gamma \gamma} }{I^\gamma} }^2+\abs{ \frac{8 \pi^2 v^2  \widetilde {\mathscr{C}}_{\gamma \gamma} } {I^\gamma} }^2
\end{align}
where $I^\gamma \approx -1.65$ is a Feynman parameter integral~\cite{Bergstrom:1985hp,Manohar:2006gz}. Again, as in the gluon case, we are dropping other$v^2/\Lambda^2$ terms that must be included if $\mathscr{C}_{\gamma \gamma}$ from BSM physics is loop suppressed.

The effective amplitude is
\begin{align}
\mathscr{C}_{\gamma \gamma} e^2 F_{\mu \nu} F^{\mu \nu} h v\label{gaga}
\end{align}
where
\begin{align}
g_1 &= \frac{e}{\cos \theta_W} & g_2&=\frac{e}{\sin \theta_W}
\end{align}
are the definitions of $e$ and $\theta_W$ without a bar. These differ from the coupling constants in Eq.~(\ref{5.16b}) (with a bar) at order $1/\Lambda^2$.

The complete one-loop RGE is
\begin{align}
\dot {\mathscr{C}}_{\gamma \gamma} &= \left( 12 \lambda -\frac32 g_1^2 - \frac92 g_2^2 +2 Y(S) \right) \mathscr{C}_{\gamma \gamma} + 
\left( 8 \lambda -6 g_2^2 \right) \frac{C_{HWB}}{g_1g_2} \nn
&-18 g_2 C_W + \bigl(4  \mathscr{C}_{\substack{d \gamma \\ r s}} [Y_d]_{sr} + 4 \mathscr{C}_{\substack{e \gamma \\ r s}} [Y_e]_{sr} -8 \mathscr{C}_{\substack{u \gamma \\ r s}} [Y_u]_{sr} + h.c. \bigr), \nn
{\dot {\widetilde {\mathscr{C}}}}_{\gamma \gamma} &= \left( 12 \lambda -\frac32 g_1^2 - \frac92 g_2^2 +2 Y(S) \right) \widetilde{\mathscr{C}}_{\gamma \gamma} + 
\left( 8 \lambda -6 g_2^2 \right) \frac{C_{H\widetilde WB}}{g_1g_2} \nn
&-18 g_2 C_{\widetilde W} + \bigl(-4i  \mathscr{C}_{\substack{d \gamma \\ r s}} [Y_d]_{sr} -4i \mathscr{C}_{\substack{e \gamma \\ r s}} [Y_e]_{sr} +8i \mathscr{C}_{\substack{u \gamma \\ r s}} [Y_u]_{sr} + h.c. \bigr).
\end{align}
The first line of each equation is the contribution from the $8 \times 8$ submatrix of Higgs-gauge operators computed in Ref.~\cite{Grojean:2013kd}. The second line gives the additional terms including all 59 operators. There are contributions from the triple-gauge operators
\begin{align}
Q_W &= \epsilon^{IJK} W_\mu^{I\nu} W_\nu^{J\rho} W_\rho^{K\mu}, & Q_{\widetilde W} &= \epsilon^{IJK} W_\mu^{I\nu} W_\nu^{J\rho}, 
{\widetilde W}_\rho^{K\mu}
\end{align}
and the dipole operator coefficients defined in Sec.~\ref{sec:mag}.

This result is the first {\it truly} complete one-loop result of the RGE running of $ \mathscr{C}_{\gamma \gamma}$.

\subsection{$h \to \gamma\, Z$}\label{sec:gZ}

The measurement of $h \to \gamma\, Z$ at LHC has not yet reached the sensitivity required to observe the SM rate \cite{ATLAS-CONF-2013-009,Chatrchyan:2013vaa}. Nevertheless, this process is interesting in several BSM scenarios because a suppression of BSM effects in  $h \rightarrow \gamma \, \gamma, gg$ due to a pseudo-Goldstone
Higgs does not necessarily imply a suppression of BSM effects in $h \rightarrow \gamma \, Z$ (for a recent discussion see \cite{Azatov:2013ura}). We define the effective Wilson coefficient in this case 
to be
\begin{align}
\mathscr{C}_{\gamma Z} &= \frac{1}{g_1 g_2}C_{HW}  - \frac{1}{g_1 g_2}C_{HB} -\left(\frac{1}{2 g_1^2}-\frac{1}{2 g_2^2}\right) C_{HWB}
\end{align}
so that the modification of the decay rate is
\begin{align}
\label{gammaZ}
{ \Gamma(h\to \gamma Z) \over \Gamma^{\text{SM}}(h \to \gamma Z)}
&\simeq \abs{1+\frac{8 \pi^2 v^2 \mathscr{C}_{\gamma Z} }{I^Z} }^2+\abs{ \frac{8 \pi^2 v^2  \widetilde {\mathscr{C}}_{\gamma Z} } {I^Z} }^2
\end{align}
$I^Z \approx -2.87$~\cite{Bergstrom:1985hp,Manohar:2006gz}, again neglecting $v^2/\Lambda^2$ terms due to $\ckin$, etc.,
and our previously defined coefficients are
\begin{align}
 \mathscr{C}_{\gamma Z} &= -\frac{c_{\gamma Z}}{2\Lambda^2},  &  \widetilde {\mathscr{C}}_{\gamma Z} &=-\frac{\tilde c_{\gamma Z}}{2\Lambda^2}
\end{align}
The one loop RGE results for the CP-even  term
\begin{align}
\dot {\mathscr{C}}_{\gamma Z} &= \frac12 \csc \theta_W \sec \theta_W \Bigl\{ (2 \cos 2 \theta_W+1)  [Y_d]_{wv} \mathscr{C}_{\substack{d \gamma \\ v w}} 
+  (2 \cos 2 \theta_W-1)  [Y_e]_{wv} \mathscr{C}_{\substack{e \gamma \\ v w}}
\nn
&-  (4 \cos 2 \theta_W-1)  [Y_u]_{wv} \mathscr{C}_{\substack{u \gamma \\ v w}}+h.c.\Bigr\}
+ 2 \left( [Y_d]_{wv} \mathscr{C}_{\substack{d Z \\ v w}} + [Y_e]_{wv} \mathscr{C}_{\substack{e Z \\ v w}} -2  [Y_u]_{wv} \mathscr{C}_{\substack{u Z \\ v w}}+h.c. \right)
\nn
&
+ \left(12 \lambda + 2 Y(S) -\frac{22}{3} e^2 +\frac{19}{3} e^2 \sec^2 \theta_W - \frac{20}{3}e^2 \csc^2 \theta_W \right) \mathscr{C}_{\gamma Z} 
\nn
&
+ e^2 \left(\frac{11}{3} \cos 2 \theta_W - 10\right)\csc \theta_W \sec \theta_W  \mathscr{C}_{\gamma \gamma} 
+e \left(\frac32 \sec \theta_W-\frac{33}{2} \cot \theta_W \csc \theta_W \right) C_W
\nn
&
+ \left( 6 e^2 -4 e^2  \csc^2 \theta_W +4 \lambda  \cos 2 \theta_W \right) \csc \theta_W \sec \theta_W\  \frac{C_{HWB}}{g_1 g_2}.
\end{align}
The RGE for $\dot {\widetilde {\mathscr{C}}}_{\gamma Z}$ is given by the substitution $Y_\psi \to -i Y_\psi$, $ \mathscr{C}_{\gamma Z}  \to
\widetilde{ \mathscr{C}}_{\gamma Z} $, $ \mathscr{C}_{\gamma \gamma}  \to
\widetilde{ \mathscr{C}}_{\gamma \gamma} $, $C_W \to C_{\widetilde W}$, and $C_{HWB} \to C_{H\widetilde WB}$, as for the $gg$ and $\gamma\gamma$ amplitudes.

\subsection{Electroweak precision observables}\label{sec:EWPD}

We are assuming $\Lambda$ is parametrically higher than the EW scale $v$, so the usual $S$, $T$ and $U$ parametrization~\cite{Kennedy:1988sn, Peskin:1991sw, Golden:1990ig,Holdom:1990tc} of the oblique electroweak precision data (EWPD) can be used. An operator based analysis of EWPD was first developed in Ref. \cite{Grinstein:1991cd}. The standard operator based approach identifies the $S$ parameter with the operator $Q_{HWB}$, and the $T$ parameter with the operator $Q_{HD}$,
\begin{subequations}
\begin{align}
S  &= \frac{16 \, \pi \, v^2}{g_1 \, g_2} \,  C_{H\!W\!B}, &
T &= - 2 \pi v^2 \left(\frac{1}{g_1^2}+\frac{1}{g_2^2}\right) C_{HD}\,.
\end{align}
\end{subequations}
A shift in the definition of $v$  is order $1/\Lambda^4$ for this expression, and we neglect this effect.
The $U$ parameter corresponds to the dimension-eight operator $ (H^\dagger W^{\mu \, \nu} H) (H^\dagger W_{\mu \, \nu} H)$, which we neglect.  A fit that treats $m_h = 126 \,{\rm GeV}$ as an input value \cite{Baak:2012kk} to EWPD finds $S = 0.03 \pm 0.10$ and $T = 0.05 \pm 0.12$ with a correlation coefficient between $S$ and $T$ of $0.89$. 

$S$ and $T$ depend on $C_{HWB}$ and $C_{HD}$ evaluated at the weak scale. The RG evolution of $C_{HD}$ is given in Eq.~(\ref{cHD}), and the RG evolution of $C_{HWB}$ is
\begin{align}
\dot C_{HWB} &= \left( 4 \lambda + 2 Y(S) + \frac{4}{3}g_2^2+\frac{19}{3} g_1^2 \right) C_{HWB}
+2g_1 g_2 \left( C_{HW}+C_{HB}\right) + 3g_1 g_2^2 C_W \nn
&+ g_2 \left(3 [Y_u]_{wv} C_{\substack{uB \\ v w}} - 3 [Y_d]_{wv} C_{\substack{dB \\ v w}} - [Y_e]_{wv} C_{\substack{eB \\ v w}}
+h.c. \right) \nn
&+ g_1 \left(5 [Y_u]_{wv} C_{\substack{uW \\ v w}} + [Y_d]_{wv} C_{\substack{dW \\ v w}} +3 [Y_e]_{wv} C_{\substack{eW \\ v w}}
+h.c. \right).
\end{align}

The $T$ parameter is usually interpreted as a measure of  custodial symmetry violation, whereas the $S$ parameter is considered to be sensitive to the difference between the number of left-handed and right-handed fermions. 
Interestingly, the SM EFT one loop RGE does not mix the operators $C_{HWB},C_{HD}$. However, this does not follow
from custodial symmetry. The SM violates custodial symmetry in $g_1$ interactions, and through mass splittings of the $SU(2)_L$ doublets. 
If we take the limit $Y_d \to Y_u$, $Y_e \to 0$ and $\hyp_d \to \hyp_u$, then $\hyp_h \to 0$ from Eq.~(\ref{hreln}). In this limit,
the standard model preserves custodial $SU(2)$, as does the RGE. This provides a non-trivial check of our results.

The consequences of the RGE for precision electroweak parameters was studied in Ref.~\cite{Grojean:2013kd}. The RGE allows one to compute the $\ln \Lambda/m_H$ contribution to these observables, which was computed previously in the broken theory~~\cite{Hagiwara:1993ck,Hagiwara:1993qt,Alam:1997nk}. Our computation agrees with their results for the terms they computed, but has additional effects (e.g.\ due to the top quark Yukawa) which were not in the previous results.

\subsection{Triple gauge boson couplings}\label{sec:tgc}

Another promising source of information on EW interactions are triple gauge couplings (TGC).  For some recent studies on the phenomenology of these measurements see Refs.~\cite{Eboli:2010qd,Corbett:2012ja,Corbett:2013pja,Pomarol:2013zra}.  Some of the scale dependence of the operators involved in this  process (in another basis) has been determined \cite{Hagiwara:1986vm,Hagiwara:1992eh}.  In the basis used here, only the operator $Q_W$ directly contributes to TGC measurements. (Other contributions come about indirectly due to field redefinitions.)  The full RGE of the Wilson coefficient of the operator $Q_W$ has the simple form
\begin{align}
\dot{C}_{W} &= \left(24 \,  - 3 b_{0,2} \right) g_2^2 C_W, \qquad \text{or} & \mu \frac{\rd}{\rd \mu} \left(\frac{ C_{W}}{g_2^3}\right) &= 24 g_2^2 \left(\frac{C_W}{g_2^3}\right) ,
\end{align}
where $b_{0,2}$ is the first coefficient in the $g_2$ $\beta$-function.  The triple gauge boson operators  do not mix with any other dimension-six operators. This multiplicative renormalization can be largely understood using the results of Ref.~\cite{Jenkins:2013sda}.  Consequently, TGC measurements
provide a very clean probe of this dimension-six operator.

Recently, Refs.~\cite{Isidori:2013cla,Grinstein:2013vsa} have shown that the decay spectra of the three-body decay $h \rightarrow V \, \ell^+ \, \ell^-$ are particularly rich sources of information on the possible effects of anomalous couplings of the Higgs boson, and BSM contact interactions.  The full decomposition of the modification of the $V \, \ell^+ \, \ell^-$ decay spectra in the operator basis used here was given in Ref.~\cite{Buchalla:2013mpa}, which shows that the relevant terms depend on the coefficients $C_{WB}$, $C_{HD}$, $C_{HW}$, $C_{HB}$, $C_{Hl}^1$, $C_{Hl}^3$, $C_{He}$, as well as the coefficient $\ckin$ which only modifies the total decay rate.

It has been argued that TGC measurements probe the same physics as $h \rightarrow V \, \ell^+ \, \ell^-$ decays~\cite{Pomarol:2013zra} in the SILH basis. 
This claim comes about by arbitrarily setting the operator $C_{W}$, which is present in the SILH basis, and in the analysis in Ref.~\cite{Pomarol:2013zra}, to zero.
This operator contributes to TGC measurements, but not to $h \rightarrow V \, \ell^+ \, \ell^-$ decays at tree level. It is by using this arbitrary choice that
Ref.~\cite{Pomarol:2013zra} claims a strong relationship between these experimentally measurable quantities. This makes the results in  Ref.~\cite{Pomarol:2013zra}  model-dependent, and \emph{not} general. For example, the exactly solvable model of Ref.~\cite{Manohar:2013rga} produces $C_W$ but no Higgs-lepton operators.
In the non-redundant basis of Ref.~\cite{Grzadkowski:2010es}, TGC measurements are also not related to $h \rightarrow V \, \ell^+ \, \ell^-$ decays since the combination of Wilson coefficients that contribute to the two processes is not identical.  Measurable results are basis independent, and model independent results do not arbitrarily set operators to zero, as was done in Ref.~\cite{Pomarol:2013zra}. We disagree with the conclusions of Ref.~\cite{Pomarol:2013zra} which are stated as broad, model-independent,  conclusions.

\subsection{$\mu \to e \gamma$, magnetic moments, and electric dipole moments}\label{sec:mag}

The lepton dipole operators 
\begin{align}
\mathcal{L} &= C_{\substack{e W \\ rs }}\  \overline l_{r,a} \sigma^{\mu \nu} e_s\, \tau^I_{ab} H_b \,  W^I_{\mu \nu} +C_{\substack{e B \\ rs }}\  \overline l_{r,a}H_a 
\sigma^{\mu \nu} e_s\, H_a \, B_{\mu \nu}  + h.c.
\label{dipole}
 \end{align}
contribute to radiative transitions such as $\mu \to e \gamma$ which is a remarkably clean window to physics BSM. In the broken phase, Eq.~(\ref{dipole}) gives the charged lepton operators
\begin{align}
\mathcal{L} &=\frac{ev}{\sqrt 2} \mathscr{C}_{\substack{e \gamma \\ rs }}\  \overline e_{r} \sigma^{\mu \nu} P_R e_{s}\, F_{\mu \nu} + \frac{ev}{\sqrt 2} \mathscr{C}_{\substack{e Z \\ rs }}\  \overline e_{r} \sigma^{\mu \nu} P_R e_{s}\, Z_{\mu \nu} +h.c.
\label{dipole2}
\end{align}
where $r$ and $s$ are flavor indices ($\{e_e\, ,\,e_\mu\, ,\,e_\tau \} \equiv \{e\, ,\,\mu\, ,\,\tau\}$) and
\begin{align}
\mathscr{C}_{\substack{e \gamma \\ rs }} &=  \frac{1}{g_1} C_{\substack{e B \\ rs }}  - \frac{1}{g_2} C_{\substack{e W \\ rs }} &
\mathscr{C}_{\substack{e Z \\ rs }} &= - \frac{1}{g_2} C_{\substack{e B \\ rs }}  - \frac{1}{g_1} C_{\substack{e W \\ rs }}  \nn
\mathscr{C}_{\substack{d \gamma \\ rs }} &=  \frac{1}{g_1} C_{\substack{d B \\ rs }}  - \frac{1}{g_2} C_{\substack{d W \\ rs }} &
\mathscr{C}_{\substack{d Z \\ rs }} &= - \frac{1}{g_2} C_{\substack{d B \\ rs }}  - \frac{1}{g_1} C_{\substack{d W \\ rs }} \nn
\mathscr{C}_{\substack{u \gamma \\ rs }} &=  \frac{1}{g_1} C_{\substack{u B \\ rs }}  + \frac{1}{g_2} C_{\substack{u W \\ rs }} &
\mathscr{C}_{\substack{u Z \\ rs }} &= - \frac{1}{g_2} C_{\substack{u B \\ rs }}  + \frac{1}{g_1} C_{\substack{u W \\ rs }} 
\label{2.9}
\end{align}
$C_{uW}$ has the opposite sign for $u$-type quarks in Eq.~(\ref{2.9}) because of the opposite sign for $T_{3L}$. The RGE for $\mathscr{C}_{e\gamma}$ is
\begin{align}
\dot {\mathscr{C}}_{\substack{e \gamma \\ rs }} &= 
\left\{ Y(s)+ e^2 \left(12-\frac94 \csc^2 \theta_W +\frac14 \sec^2\theta_W \right) \right\} \mathscr{C}_{\substack{e \gamma \\ rs }} \nn
&+2 \mathscr{C}_{\substack{e \gamma \\ rv }} [Y_e Y_e^\dagger]_{vs} +\left(\frac12 + 2 \cos^2\theta_W  \right)[Y_e^\dagger Y_e]_{rw} \mathscr{C}_{\substack{e \gamma \\ ws }} +e^2 \left(12 \cot 2 \theta_W \right) \mathscr{C}_{\substack{e Z \\ rs }} \nn
&- \left(2 \sin \theta_W \cos \theta_W \right)[Y_e^\dagger Y_e]_{rw} \mathscr{C}_{\substack{e Z \\ ws }} -\cot \theta_W [Y_e^\dagger]_{rs} \left( C_{HWB}+i C_{H\widetilde WB} \right) \nn
&+4 e^2 [Y_e^\dagger]_{rs}\left( \mathscr{C}_{\gamma\gamma} + i \widetilde {\mathscr{C}}_{\gamma\gamma} \right) +
e^2 \left(\cot \theta_W- 3 \tan\theta_W\right) [Y_e^\dagger]_{rs} \left( \mathscr{C}_{\gamma Z} + i \widetilde {\mathscr{C}}_{\gamma Z} \right)\nn
&+16  [Y_u]_{wv}  C^{(3)}_{\substack{lequ  \\ rsvw }} .
\end{align}
The current experimental limit~\cite{Adam:2013mnn} on $\text{BR}(\mu \to e \gamma)$ is $5.7\times 10^{-13}$ from the MEG experiment, which implies 
\begin{align}
\frac{v}{\sqrt 2\, m_e} \mathscr{C}_{\substack{e \gamma \\ \mu e }}  \lesssim\ 2.7 \times 10^{-4} \ \text{TeV}^{-2}
\label{2.11}
\end{align}
at the low energy scale $\mu \sim m_\mu$.

The lepton Yukawa couplings are diagonal in the mass eigenstate basis, so the $\mu \to e \gamma$ transition amplitude depends on  $\mathscr{C}_{e\gamma}$, 
$\mathscr{C}_{eZ}$ and $C_{lequ}^{(3)}$. The bound Eq.~(\ref{2.11}) implies
\begin{align}
\frac{m_t}{m_e}  C^{(3)}_{\substack{lequ \\ \mu e t t}}   \lesssim 1.4\times 10^{-3} \ \text{TeV}^{-2}
\end{align}
using the estimate $\ln(\Lambda/m_H)/(16 \pi^2) \sim 0.01$ for the renormalization group evolution, and assuming that this term is the only contribution to 
$ \mathscr{C}_{\substack{e \gamma \\ \mu e }}$ at low energies.

The anomalous magnetic moment of the muon is
\begin{align}
\delta a_\mu &=  \frac{4 m_\mu v}{\sqrt 2} \text{Re}\,  \mathscr{C}_{\substack{e \gamma \\ \mu \mu}}
\end{align}
which yields the limits
\begin{align}
\abs{ C_{HWB} } &\lesssim 0.6  \, \text{TeV}^{-2}, &
\abs{ \mathscr{C}_{\gamma \gamma}} &\lesssim 4  \, \text{TeV}^{-2}, &
\abs{\frac{m_t}{m_\mu} \text{Re}\, C^{(3)}_{\substack{lequ  \\ \mu \mu t t}} } &\lesssim 7\ \text{TeV}^{-2},
\end{align}
assuming that each of these is the only contribution to $ \mathscr{C}_{\substack{e \gamma \\ \mu \mu}}$.

The bound on the electric dipole moment of the electron translates to the limits
\begin{align}
\abs{ C_{H\widetilde WB} } &\lesssim 2 \times 10^{-3}  \, \text{TeV}^{-2}, &
\abs{ {\widetilde {\mathscr{C}}}_{\gamma \gamma} } &\lesssim 2 \times 10^{-2}  \, \text{TeV}^{-2}, &
\abs{ \frac{m_t}{m_e}  \text{Im}\,  C_{\substack{lequ  \\ ee tt}} } &\lesssim 3 \times 10^{-4} \ \text{TeV}^{-2},
\end{align}
using the recently measured upper bound~\cite{Baron:2013eja}, $d_e < 1.05\times 10^{-27}e\,\mbox{cm}$ from the ACME collaboration, again assuming each of these terms is the only contribution.

\section{Conclusions}\label{conclusion}

This paper completes the full calculation of the one-loop renormalization of the dimension-six Lagrangian of the SM EFT. We present all of the remaining
gauge terms in the $59 \times 59$ anomalous dimension matrix of the baryon number conserving operators. The anomalous dimension matrix of the dimension-six baryon number violating operators have been computed in Ref.~\cite{Alonso:2014zka}.

Many of the results are lengthy, but a few important cases such as $gg \to h$, $h\to \gamma \gamma$ and $h \to \gamma Z$ have simple RG equations which are given explicitly in this paper. We have computed the modification of the Higgs mass, self-interactions, and couplings to fermions and gauge bosons from $\lsix$. The dimension-six terms change the relation between the Higgs vacuum expectation value and $G_F$, and also contribute to the $\rho$ parameter. The RGE improvement of all of these  relations is now known, and will be useful for future precision studies of the SM EFT. A complete analysis of the SM EFT is a formidable task, because $\lsix$ has 2499 independent parameters.

We have also discussed how the SM EFT provides a model-independent way to test the MFV hypothesis, and how the full SM EFT RGE mixes flavor violation between the different operator sectors. A few applications of our results have been given in this paper.

\acknowledgments
This work was supported in part by DOE grant DE-SC0009919. MT thanks W.~Skiba for helpful conversations.
RA and AM thank B.~Shotwell, D.~Stone, H.-M.~Chang and C.~Murphy for useful discussions. We would also like to thank
C.~Zhang~\cite{Zhang:2014rja}, G.~Pruna and A.~Signer~\cite{Pruna:2014asa}, J.~Brod, A.~Greljo, E.~Stamou, and P.~Uttayarat~\cite{Brod:2014hsa}, and C.~Cheung and C.-H.~Shen~\cite{Cheung:2015aba}\ for pointing out typos/errors in  previous versions of the manuscript.

%
%--------------------------------------------------------------------------------------------
\begin{table}
\begin{center}
\small
\begin{minipage}[t]{4.45cm}
\renewcommand{\arraystretch}{1.5}
\begin{tabular}[t]{c|c}
\multicolumn{2}{c}{$1:X^3$} \\
\hline
$Q_G$                & $f^{ABC} G_\mu^{A\nu} G_\nu^{B\rho} G_\rho^{C\mu} $ \\
$Q_{\widetilde G}$          & $f^{ABC} \widetilde G_\mu^{A\nu} G_\nu^{B\rho} G_\rho^{C\mu} $ \\
$Q_W$                & $\epsilon^{IJK} W_\mu^{I\nu} W_\nu^{J\rho} W_\rho^{K\mu}$ \\ 
$Q_{\widetilde W}$          & $\epsilon^{IJK} \widetilde W_\mu^{I\nu} W_\nu^{J\rho} W_\rho^{K\mu}$ \\
\end{tabular}
\end{minipage}
\begin{minipage}[t]{2.7cm}
\renewcommand{\arraystretch}{1.5}
\begin{tabular}[t]{c|c}
\multicolumn{2}{c}{$2:H^6$} \\
\hline
$Q_H$       & $(H^\dag H)^3$ 
\end{tabular}
\end{minipage}
\begin{minipage}[t]{5.1cm}
\renewcommand{\arraystretch}{1.5}
\begin{tabular}[t]{c|c}
\multicolumn{2}{c}{$3:H^4 D^2$} \\
\hline
$Q_{H\Box}$ & $(H^\dag H)\Box(H^\dag H)$ \\
$Q_{H D}$   & $\ \left(H^\dag D_\mu H\right)^* \left(H^\dag D_\mu H\right)$ 
\end{tabular}
\end{minipage}
\begin{minipage}[t]{2.7cm}

\renewcommand{\arraystretch}{1.5}
\begin{tabular}[t]{c|c}
\multicolumn{2}{c}{$5: \psi^2H^3 + \hbox{h.c.}$} \\
\hline
$Q_{eH}$           & $(H^\dag H)(\bar l_p e_r H)$ \\
$Q_{uH}$          & $(H^\dag H)(\bar q_p u_r \widetilde H )$ \\
$Q_{dH}$           & $(H^\dag H)(\bar q_p d_r H)$\\
\end{tabular}
\end{minipage}

\vspace{0.25cm}

\begin{minipage}[t]{4.7cm}
\renewcommand{\arraystretch}{1.5}
\begin{tabular}[t]{c|c}
\multicolumn{2}{c}{$4:X^2H^2$} \\
\hline
$Q_{H G}$     & $H^\dag H\, G^A_{\mu\nu} G^{A\mu\nu}$ \\
$Q_{H\widetilde G}$         & $H^\dag H\, \widetilde G^A_{\mu\nu} G^{A\mu\nu}$ \\
$Q_{H W}$     & $H^\dag H\, W^I_{\mu\nu} W^{I\mu\nu}$ \\
$Q_{H\widetilde W}$         & $H^\dag H\, \widetilde W^I_{\mu\nu} W^{I\mu\nu}$ \\
$Q_{H B}$     & $ H^\dag H\, B_{\mu\nu} B^{\mu\nu}$ \\
$Q_{H\widetilde B}$         & $H^\dag H\, \widetilde B_{\mu\nu} B^{\mu\nu}$ \\
$Q_{H WB}$     & $ H^\dag \tau^I H\, W^I_{\mu\nu} B^{\mu\nu}$ \\
$Q_{H\widetilde W B}$         & $H^\dag \tau^I H\, \widetilde W^I_{\mu\nu} B^{\mu\nu}$ 
\end{tabular}
\end{minipage}
\begin{minipage}[t]{5.2cm}
\renewcommand{\arraystretch}{1.5}
\begin{tabular}[t]{c|c}
\multicolumn{2}{c}{$6:\psi^2 XH+\hbox{h.c.}$} \\
\hline
$Q_{eW}$      & $(\bar l_p \sigma^{\mu\nu} e_r) \tau^I H W_{\mu\nu}^I$ \\
$Q_{eB}$        & $(\bar l_p \sigma^{\mu\nu} e_r) H B_{\mu\nu}$ \\
$Q_{uG}$        & $(\bar q_p \sigma^{\mu\nu} T^A u_r) \widetilde H \, G_{\mu\nu}^A$ \\
$Q_{uW}$        & $(\bar q_p \sigma^{\mu\nu} u_r) \tau^I \widetilde H \, W_{\mu\nu}^I$ \\
$Q_{uB}$        & $(\bar q_p \sigma^{\mu\nu} u_r) \widetilde H \, B_{\mu\nu}$ \\
$Q_{dG}$        & $(\bar q_p \sigma^{\mu\nu} T^A d_r) H\, G_{\mu\nu}^A$ \\
$Q_{dW}$         & $(\bar q_p \sigma^{\mu\nu} d_r) \tau^I H\, W_{\mu\nu}^I$ \\
$Q_{dB}$        & $(\bar q_p \sigma^{\mu\nu} d_r) H\, B_{\mu\nu}$ 
\end{tabular}
\end{minipage}
\begin{minipage}[t]{5.4cm}
\renewcommand{\arraystretch}{1.5}
\begin{tabular}[t]{c|c}
\multicolumn{2}{c}{$7:\psi^2H^2 D$} \\
\hline
$Q_{H l}^{(1)}$      & $(H^\dag i\overleftrightarrow{D}_\mu H)(\bar l_p \gamma^\mu l_r)$\\
$Q_{H l}^{(3)}$      & $(H^\dag i\overleftrightarrow{D}^I_\mu H)(\bar l_p \tau^I \gamma^\mu l_r)$\\
$Q_{H e}$            & $(H^\dag i\overleftrightarrow{D}_\mu H)(\bar e_p \gamma^\mu e_r)$\\
$Q_{H q}^{(1)}$      & $(H^\dag i\overleftrightarrow{D}_\mu H)(\bar q_p \gamma^\mu q_r)$\\
$Q_{H q}^{(3)}$      & $(H^\dag i\overleftrightarrow{D}^I_\mu H)(\bar q_p \tau^I \gamma^\mu q_r)$\\
$Q_{H u}$            & $(H^\dag i\overleftrightarrow{D}_\mu H)(\bar u_p \gamma^\mu u_r)$\\
$Q_{H d}$            & $(H^\dag i\overleftrightarrow{D}_\mu H)(\bar d_p \gamma^\mu d_r)$\\
$Q_{H u d}$ + h.c.   & $i(\widetilde H ^\dag D_\mu H)(\bar u_p \gamma^\mu d_r)$\\
\end{tabular}
\end{minipage}

\vspace{0.25cm}

\begin{minipage}[t]{4.75cm}
\renewcommand{\arraystretch}{1.5}
\begin{tabular}[t]{c|c}
\multicolumn{2}{c}{$8:(\bar LL)(\bar LL)$} \\
\hline
$Q_{ll}$        & $(\bar l_p \gamma_\mu l_r)(\bar l_s \gamma^\mu l_t)$ \\
$Q_{qq}^{(1)}$  & $(\bar q_p \gamma_\mu q_r)(\bar q_s \gamma^\mu q_t)$ \\
$Q_{qq}^{(3)}$  & $(\bar q_p \gamma_\mu \tau^I q_r)(\bar q_s \gamma^\mu \tau^I q_t)$ \\
$Q_{lq}^{(1)}$                & $(\bar l_p \gamma_\mu l_r)(\bar q_s \gamma^\mu q_t)$ \\
$Q_{lq}^{(3)}$                & $(\bar l_p \gamma_\mu \tau^I l_r)(\bar q_s \gamma^\mu \tau^I q_t)$ 
\end{tabular}
\end{minipage}
\begin{minipage}[t]{5.25cm}
\renewcommand{\arraystretch}{1.5}
\begin{tabular}[t]{c|c}
\multicolumn{2}{c}{$8:(\bar RR)(\bar RR)$} \\
\hline
$Q_{ee}$               & $(\bar e_p \gamma_\mu e_r)(\bar e_s \gamma^\mu e_t)$ \\
$Q_{uu}$        & $(\bar u_p \gamma_\mu u_r)(\bar u_s \gamma^\mu u_t)$ \\
$Q_{dd}$        & $(\bar d_p \gamma_\mu d_r)(\bar d_s \gamma^\mu d_t)$ \\
$Q_{eu}$                      & $(\bar e_p \gamma_\mu e_r)(\bar u_s \gamma^\mu u_t)$ \\
$Q_{ed}$                      & $(\bar e_p \gamma_\mu e_r)(\bar d_s\gamma^\mu d_t)$ \\
$Q_{ud}^{(1)}$                & $(\bar u_p \gamma_\mu u_r)(\bar d_s \gamma^\mu d_t)$ \\
$Q_{ud}^{(8)}$                & $(\bar u_p \gamma_\mu T^A u_r)(\bar d_s \gamma^\mu T^A d_t)$ \\
\end{tabular}
\end{minipage}
\begin{minipage}[t]{4.75cm}
\renewcommand{\arraystretch}{1.5}
\begin{tabular}[t]{c|c}
\multicolumn{2}{c}{$8:(\bar LL)(\bar RR)$} \\
\hline
$Q_{le}$               & $(\bar l_p \gamma_\mu l_r)(\bar e_s \gamma^\mu e_t)$ \\
$Q_{lu}$               & $(\bar l_p \gamma_\mu l_r)(\bar u_s \gamma^\mu u_t)$ \\
$Q_{ld}$               & $(\bar l_p \gamma_\mu l_r)(\bar d_s \gamma^\mu d_t)$ \\
$Q_{qe}$               & $(\bar q_p \gamma_\mu q_r)(\bar e_s \gamma^\mu e_t)$ \\
$Q_{qu}^{(1)}$         & $(\bar q_p \gamma_\mu q_r)(\bar u_s \gamma^\mu u_t)$ \\ 
$Q_{qu}^{(8)}$         & $(\bar q_p \gamma_\mu T^A q_r)(\bar u_s \gamma^\mu T^A u_t)$ \\ 
$Q_{qd}^{(1)}$ & $(\bar q_p \gamma_\mu q_r)(\bar d_s \gamma^\mu d_t)$ \\
$Q_{qd}^{(8)}$ & $(\bar q_p \gamma_\mu T^A q_r)(\bar d_s \gamma^\mu T^A d_t)$\\
\end{tabular}
\end{minipage}

\vspace{0.25cm}

\begin{minipage}[t]{3.75cm}
\renewcommand{\arraystretch}{1.5}
\begin{tabular}[t]{c|c}
\multicolumn{2}{c}{$8:(\bar LR)(\bar RL)+\hbox{h.c.}$} \\
\hline
$Q_{ledq}$ & $(\bar l_p^j e_r)(\bar d_s q_{tj})$ 
\end{tabular}
\end{minipage}
\begin{minipage}[t]{5.5cm}
\renewcommand{\arraystretch}{1.5}
\begin{tabular}[t]{c|c}
\multicolumn{2}{c}{$8:(\bar LR)(\bar L R)+\hbox{h.c.}$} \\
\hline
$Q_{quqd}^{(1)}$ & $(\bar q_p^j u_r) \epsilon_{jk} (\bar q_s^k d_t)$ \\
$Q_{quqd}^{(8)}$ & $(\bar q_p^j T^A u_r) \epsilon_{jk} (\bar q_s^k T^A d_t)$ \\
$Q_{lequ}^{(1)}$ & $(\bar l_p^j e_r) \epsilon_{jk} (\bar q_s^k u_t)$ \\
$Q_{lequ}^{(3)}$ & $(\bar l_p^j \sigma_{\mu\nu} e_r) \epsilon_{jk} (\bar q_s^k \sigma^{\mu\nu} u_t)$
\end{tabular}
\end{minipage}
\end{center}
\caption{\label{op59}
The 59 independent dimension-six operators built from Standard Model fields which conserve baryon number, as given in 
Ref.~\cite{Grzadkowski:2010es}. The operators are divided into eight classes: $X^3$, $H^6$, etc. Operators with $+\hbox{h.c.}$ in the table heading also have hermitian conjugates, as does the $\psi^2H^2D$ operator $Q_{Hud}$. The subscripts $p,r,s,t$ are flavor indices.}
\end{table}
%--------------------------------------------------------------------------------------------
%
%
%
%
\begin{table}
\begin{align*}
{\footnotesize
\hspace{-1cm}
\begin{array}{rlc|ccr|ccr}
\multicolumn{1}{c}{\text{Class}} & & N_{\text{op}} & \multicolumn{3}{c|}{CP\text{-even}} &  \multicolumn{3}{c}{CP\text{-odd}} \\
& & & n_g & 1 & 3 & n_g & 1 & 3 \\
\hline
1 & &4 & 2 & 2 & 2 & 2 & 2 & 2 \\
2 & &1 &1 & 1 & 1 & 0 & 0 & 0 \\
3 & & 2 & 2 & 2 & 2 & 0 & 0 & 0 \\
4 & & 8 & 4 & 4 & 4 & 4 & 4 & 4 \\
5 & &3 & 3 n_g^2 & 3 & 27 & 3 n_g^2 & 3 & 27 \\
6 & & 8 & 8 n_g^2 & 8 & 72 & 8 n_g^2 & 8 & 72 \\
7 & & 8 & \frac12 n_g(9n_g+7) & 8 & 51 &  \frac12 n_g(9n_g-7) & 1 & 30 \\
8 & : (\overline L L)(\overline LL) & 5 & \frac14 n_g^2(7n_g^2+13) & 5 & 171 & \frac74 n_g^2(n_g-1)(n_g+1) & 0 & 126  \\
8  & : (\overline RR)(\overline RR) & 7 & \frac18 n_g(21 n_g^3+2n_g^2+31n_g+2) & 7 & 255 & \frac{1}8 n_g(21n_g+2)(n_g-1)(n_g+1) & 0 & 195 \\
8  & : (\overline LL)(\overline RR) & 8 & 4 n_g^2(n_g^2+1) & 8 & 360 & 4 n_g^2(n_g-1)(n_g+1) & 0 & 288 \\
8 & : (\overline LR)(\overline RL) & 1 & n_g^4 & 1 & 81 &  n_g^4 & 1 & 81 \\
8 & : (\overline LR)(\overline LR) & 4 & 4n_g^4 & 4 & 324 &  4n_g^4 & 4 & 324 \\
%\hline
8 & :\ \text{All} & 25 & \frac18 n_g (107 n_g^3+2n_g^2+89n_g+2) & 25 & 1191 & \frac18 n_g (107 n_g^3+2n_g^2-67n_g-2) & 5 & 1014 \\
\hline
\multicolumn{1}{c}{\text{Total}} & & 59 & \frac18 (107 n_g^4+2n_g^3+213 n_g^2+30n_g+72) & 53 & 1350 & \frac18 (107 n_g^4+2n_g^3+57 n_g^2-30n_g+48) & 23 & 1149
\end{array}
}
\end{align*}
\caption{Number of $CP$-even and $CP$-odd coefficients in $\lsix$ for $n_g$ flavors. The total number of coefficients is $(107 n_g^4+2n_g^3+135 n_g^2+60)/4$, which is 76 for $n_g=1$ and 2499 for $n_g=3$.
\label{number}}
\end{table}
%
%
%
% REARRANGED TABLE
\begin{table}
\renewcommand{\arraycolsep}{0.15cm}
\renewcommand{\arraystretch}{1.5}
\begin{align*}
\begin{array}{cc|c|cccc|cc|c}
&& H^6 & H^4 D^2  & y \psi^2 H^3 & \psi^2 H^2 D & \psi^4 & g^2 X^2 H^2 & g y \psi^2 X H & g^3 X^3  \\
\text{Class} && 2 & 3 & 5 & 7 & 8 & 4 & 6 & 1 \\
\text{NDA Weight}&& 2 & 1 & 1 & 1 & 1 & 0 & 0 & -1\\
\hline\hline
H^6 &   & \lambda, y^2, g^2 & \lambda^2, \lambda g^2, g^4 & \lambda y^2, y^4   &  \lambda y^2,  \lambda g^2, \zero{y^4} & 0 
 & \lambda g^4, g^6 & 0 & \zero{ \lambda g^6}  \\
 \hline
H^4 D^2 &   & 0 & \lambda, y^2, g^2 & \zero{y^2} & y^2,g^2 & 0  &  \zero{g^4} & \zero{y^2 g^2} &  \zero{g^6}  \\
y \psi^2 H^3 &  & 0 & \lambda, y^2, g^2  & \lambda, y^2, g^2 & \lambda, y^2, g^2 &  \lambda, y^2  & g^4  & \zero{g^2\lambda} ,g^4,g^2y^2  &  \zero{g^6} \\
\psi^2 H^2 D &   & 0 & g^2,y^2 & \zero{y^2}  & g^2, \zero{\lambda} ,y^2 & g^2, y^2 &  \zero{g^4 } &  \zero{g^2y^2} & \zero{g^6} \\
\psi^4 &  & 0 & 0 & 0  & g^2,y^2 & g^2,y^2 & 0 &  g^2y^2 & \zero{g^6} \\
\hline
g^2 X^2 H^2 &   & 0 & \zero{1} &  0 & \zero{1} & 0 & \lambda, y^2, g^2 & y^2  &  g^4  \\
g  y \psi^2 X H &  & 0 & 0  & \zero{1}  & \zero{1} & 1 &  g^2 & g^2,y^2 & g^4   \\
\hline
g^3 X^3 &   & 0 & 0  & 0  & 0 & 0 & \zero{1} & 0 & g^2 \\
\end{array}
\end{align*}
\caption{\label{tab:anom} Form of the one-loop anomalous dimension matrix $\widehat \gamma_{ij}$ for dimension-six operators 
$\widehat Q_i$ rescaled according to naive dimensional analysis. The operators are ordered by NDA weight, rather than by operator class. The possible entries allowed by the one-loop Feynman graphs are shown. The cross-hatched entries vanish.}
\end{table}

\appendix

\section{Flavor representations and parameter counting}\label{app:counting}

In this appendix, we briefly discuss the flavor representations of the operators, and the parameter counting of Table~\ref{number}. 

Operators in classes 1--4 have no flavor indices, and the counting is trivial.

Class 5 and 6 operator coefficients are $n_g \times n_g$ complex matrices $M_{rs}$ in flavor space, with $n_g^2$ complex entries. The real matrix elements give the $n_g^2$ $CP$-even parmeters and the imaginary matrix elements yield $n_g^2$ $CP$-odd entries.

Class 7 operators, other than $Q_{Hud}$ are hermitian, so their coefficients are $n_g \times n_g$ hermitian matrices $H_{rs}$ in flavor space, which can be written as $H_{rs}=S_{rs}+i A_{rs}$, where $S$ is real-symmetric and $CP$-even with $n_e=n_g(n_g+1)/2$ parameters, and $A$ is real-antisymmetric and $CP$-odd, with $n_o=n_g(n_g-1)/2$ parameters. $Q_{Hud}$, which is not hermitian, is an $n_g \times n_g$ complex matrix with $n_g^2$ $CP$-even and $n_g^2$ $CP$-odd parameters.

The four-fermion operators in Class 8 are the only non-trivial case. The $(\overline L R)(\overline R L)$ and $(\overline L R)(\overline  L R)$ operators are not hermitian, and each has $n_g^4$ $CP$-even and $n_g^4$ $CP$-odd parameters, since the operator has 4 independent flavor indices. The $(\overline L L)(\overline RR)$ operators are the product of $L$ and $R$ currents, each of which has $n_e$ $CP$-even and $n_o$ $CP$-odd components, for $n_e^2+n_o^2$ $CP$-even and $2 n_e n_o$ $CP$-odd terms. The counting for $(\overline L L)(\overline LL)$ and $(\overline R R)(\overline RR)$ operators when the currents are different, $Q_{lq}^{(1,3)}$, $Q_{eu}$, $Q_{ed}$, {$Q_{ud}^{(1,8)}$}, is the same as for the $(\overline L L)(\overline RR)$ operators. The interesting case is for $Q_{ll}$, $Q_{qq}^{(1,3)}$, $Q_{uu}$, $Q_{dd}$ where the two currents are identical, so that all four flavor indices transform under the same $SU(n_g)$ flavor group. The operators transform as the $\mathbf{1} +\mathbf{1} +   \text{adj} + \text{adj} + \overline a a  + \overline ss$ where $\text{adj}$ is the adjoint representation, $\overline a a$ is the representation $T^{[ij]}_{[kl]}$ antisymmetric in the upper and lower indices, and $\overline s s$ is the representation $T^{(ij)}_{(kl)}$ symmetric in the upper and lower indices.\footnote{The relevant group theory results can be found, for example, in Refs.~\cite{Dashen:1993jt,Dashen:1994qi}.} The $\overline a a$ representation vanishes for $n_g=3$. The singlet has one $CP$-even parameter, the adjoint has $(n_g-1)(n_g+2)/2$ $CP$-even and $n_g(n_g-1)/2$ $CP$-odd parameters, $\overline aa$ has $n_g(n_g-3)(n_g^2+n_g+2) /8$ $CP$-even and $n_g(n_g-3)(n_g-1)(n_g+2)/8$ $CP$-odd parameters, and $\overline s s$ has $n_g(n_g-1)(n_g+1)(n_g+2)/8$ $CP$-even and $ n_g(n_g-1)(n_g^2+3n_g-2)/8$ $CP$-odd parameters. The operator $Q_{ee}$ is a special case, because of the Fierz identity
\begin{align}
(\bar e_p \gamma_\mu e_r) (\bar e_s \gamma_\mu e_t) &= (\bar e_s \gamma_\mu e_r) (\bar e_p \gamma_\mu e_t) ,
\label{extra}
\end{align}
which implies that the operator must be symmetric in the two $e$ indices and in the two $\overline e$ indices. This identity does not hold for the other fermions, because they have $SU(2)$ or color indices. $Q_{ee}$ transforms as $1 + \text{adj} + \overline ss$ because of the Fierz identity.

Adding up the individual contributions gives Table~\ref{number}.

\section{Conversion of $P_i$ operators to the standard basis}\label{sec:convert}

The equations of motion can be used to express the operators $P_i$ in the standard basis.  The identifications are
\begin{align}
\mathcal{P}_{B}  &=
\frac12 \hyp_h  g_1^2 Q_{H \Box} +2g_1^2 \hyp_h Q_{H D}  + \frac12  g_1^2\left[\hyp_l Q_{\substack{H l \\ tt}}^{(1)} +\hyp_e Q_{\substack{H e \\ tt}} + \hyp_q Q_{\substack{H q \\ tt}}^{(1)}+\hyp_u Q_{\substack{H u \\ tt}}+ \hyp_d Q_{\substack{H d \\ tt}} \right],\nn
\mathcal{P}_{W} &= \frac34 g_2^2 Q_{H \Box}-\frac12 g_2^2 m_H^2 (H^\dagger H)^2 +2 g_2^2 \lambda Q_H +  \frac14  g_2^2\left[Q_{\substack{H l \\ tt}}^{(3)} + Q_{\substack{H q \\ tt}}^{(3)}\right] \nn
&+\frac12 g_2^2\left( [Y_u^\dagger]_{rs} Q_{\substack{ uH \\ rs}} + [Y_d^\dagger]_{rs} Q_{\substack{ dH \\ rs}}+ [Y_e^\dagger]_{rs} Q_{\substack{ eH \\ rs}}+h.c. \right),
\nn
\mathcal{P}_{HB}  &= \frac12 g_1^2 \hyp_h Q_{H \Box}  +2g_1^2 \hyp_h Q_{H D} - \frac12 \hyp_h g_1^2 Q_{H B} - \frac14 g_1g_2  Q_{H WB} \nn
& +\frac12  g_1^2\left[\hyp_l Q_{\substack{H l \\ tt}}^{(1)} +\hyp_e Q_{\substack{H e \\ tt}} + \hyp_q Q_{\substack{H q \\ tt}}^{(1)}+\hyp_u Q_{\substack{H u \\ tt}}+ \hyp_d Q_{\substack{H d \\ tt}} \right] ,\nn
\mathcal{P}_{HW} &= \frac34 g_2^2 Q_{H \Box} -\frac12 g_2^2 m_H^2 (H^\dagger H)^2 +2 g_2^2 \lambda Q_H -\frac14 g_2^2 Q_{H W}-\frac12 \hyp_h g_1 g_2  Q_{H WB} + \frac14  g_2^2\left[Q_{H l}^{(3)} + Q_{H q}^{(3)} \right] 
\nn
&+\frac12 g_2^2\left( [Y_u^\dagger]_{rs} Q_{\substack{ uH \\ rs}} + [Y_d^\dagger]_{rs} Q_{\substack{ dH \\ rs}}+ [Y_e^\dagger]_{rs} Q_{\substack{ eH \\ rs}}+h.c. \right) ,\nn
P_T &= - Q_{H \Box} -4 Q_{H D}.
\label{pops}
\end{align}

\section{Results}\label{app:rge}

The renormalization group equations by operator class are given below.  The complete RG equations for the dimension-six operators are given by adding Eqs.~(6.1)--(6.4) of Ref.~\cite{Jenkins:2013zja},  the equations in the appendices of Ref.~\cite{Jenkins:2013wua} and the equations given below. Eqs.~(4.3)--(4.5) of Ref.~\cite{Jenkins:2013zja} give the renormalization group evolution of SM couplings due to dimension-six operators.

The parameters $\eta_{1-5}$ are defined in the appendix of Ref.~\cite{Jenkins:2013wua}. Some equations use $\xi_B$, defined by
\begin{align}
\xi_B &= \frac43 \hyp_h \left(C_{H\Box}+C_{HD}\right) + \frac83 \left[2 \hyp_l C_{\substack{H l \\ tt}}^{(1)}+2 \hyp_q N_c C_{\substack{H q \\ tt}}^{(1)} 
 + \hyp_e C_{\substack{H e \\ tt}} + \hyp_u N_c  C_{\substack{H u \\ tt}}+ \hyp_d N_c C_{\substack{H d \\ tt}} \right]  
\label{pbpwdef}
\end{align}
The other parameters are $c_{A,2}=2$, $c_{F,2}=3/4$, $c_{A,3}=N_c$, $c_{F,3}=(N_c^2-1)/(2N_c)$ with $N_c=3$, $b_{0,1}=-1/6-20n_g/9$, $b_{0,2}=43/6-4n_g/3$ and $b_{0,3}=11-4n_g/3$.

\subsection{$X^3$}

\begin{align*}
\dot C_G &= \left( 12 c_{A,3} - 3 b_{0,3}\right) g_3^2 C_G &
\dot C_{\widetilde G} &=\left( 12 c_{A,3} - 3 b_{0,3}\right) g_3^2 C_{\widetilde G} \nn
\dot C_W &= \left( 12 c_{A,2} - 3 b_{0,2}\right) g_2^2 C_W &
\dot C_{\widetilde W} &= \left( 12 c_{A,2} - 3 b_{0,2}\right) g_2^2 C_{\widetilde W} 
\end{align*}

\subsection{$H^6$}

\begin{align*}
\dot C_H  &= \left(-\frac{27}{2}g_2^2-\frac{9}{2}g_1^2\right)C_H + \lambda \left[ \frac{40}{3} g_2^2 C_{H \Box} + \left(-6 g_2^2 + 24 g_1^2 \hyp_h^2 \right) C_{H D}  \right] 
-\frac34 \left(4 \hyp_h^2 g_1^2+g_2^2\right)^2 C_{HD} \nn
&+ 12 \lambda \left(3 g_2^2 C_{HW} + 4 g_1^2 \hyp_h^2 C_{HB} + 2 g_1 g_2 \hyp_h C_{HWB}\right) -\bigl(12 g_1^2 g_2^2 \hyp_h^2 + 9 g_2^4\bigr) C_{HW}  \nn
& - \bigl(48 g_1^4  \hyp_h^4 + 12 g_1^2 g_2^2 \hyp_h^2 \bigr) C_{HB} - \bigl(24 g_1^3 g_2 \hyp_h^3 + 6 g_1 g_2^3 \hyp_h \bigr) C_{HWB}  + \frac{16}{3} \lambda g_2^2 \left( C_{\substack{ Hl \\ t t}}^{(3)} + N_c C_{\substack{ Hq \\ t t}}^{(3)}\right)
\end{align*}

\subsection{$H^4D^2$}

\begin{align*}
\dot C_{H \Box} &=  
-\left(4 g_2^2 + \frac{16}{3} g_1^2 \hyp_h^2 \right) C_{H \Box} +\frac{20}{3} g_1^2 \hyp_h^2 C_{H D}
+2 g_2^2 \left( C_{\substack{ Hl \\ t t}}^{(3)} + N_c C_{\substack{ Hq \\ t t}}^{(3)}\right) \nn
&+\frac43 g_1^2 \hyp_h \left( N_c \hyp_u  C_{\substack{ Hu \\ t t}}+N_c \hyp_d  C_{\substack{ Hd \\ t t}} + \hyp_e  C_{\substack{ He \\ t t}} +  2 N_c \hyp_q C_{\substack{ Hq \\ t t}}^{(1)}+ 2 \hyp_l C_{\substack{ Hl \\ t t}}^{(1)}\right)  \nn
\dot C_{H D} &= \frac{80}{3} g_1^2 \hyp_h^2  C_{H \Box} + \left(\frac{9}{2} g_2^2 -  \frac{10}{3} g_1^2 \hyp_h^2 \right) C_{H D} \nn
&+\frac{16}3 g_1^2 \hyp_h \left( N_c \hyp_u  C_{\substack{ Hu \\ t t}}+N_c \hyp_d  C_{\substack{ Hd \\ t t}} + \hyp_e  C_{\substack{ He \\ t t}} +  2 N_c \hyp_q C_{\substack{ Hq \\ t t}}^{(1)}+ 2 \hyp_l C_{\substack{ Hl  \\ t t}}^{(1)}\right)  \nn
\end{align*}

\subsection{$X^2H^2$}

\begin{align*}
 \dot C_{HG} &=\bigl( -6 \hyp_h^2 g_1^2-\frac92 g_2^2-2b_{0,3}g_3^2 \bigr) C_{HG}
\end{align*}
\begin{align*}
 \dot C_{HB} &= \bigl(2 \hyp_h^2  g_1^2-\frac92 g_2^2 -2b_{0,1}g_1^2\bigr) C_{HB} +6 g_1 g_2 \hyp_h C_{HWB}  \end{align*}
\begin{align*}
 \dot C_{HW} &= -15 g_2^3 C_W+ \bigl(-6 \hyp_h^2 g_1^2-\frac52 g_2^2-2b_{0,2}g_2^2 \bigr) C_{HW}  + 2 g_1 g_2 \hyp_h C_{HWB}
 \end{align*}
\begin{align*}
 \dot C_{HWB} &= 6 g_1 g_2^2 \hyp_h C_W  +\bigl( -2 \hyp_h^2 g_1^2+\frac92 g_2^2-b_{0,1}g_1^2-b_{0,2}g_2^2\bigr) C_{HWB} + 4 g_1 g_2 \hyp_h C_{HB}  + 4 g_1 g_2 \hyp_h C_{HW}
\end{align*}
\begin{align*}
\dot C_{H\widetilde  G} &=\bigl( -6 \hyp_h^2 g_1^2-\frac92 g_2^2-2b_{0,3}g_3^2 \bigr) C_{H\widetilde G}
\end{align*}
\begin{align*}
 \dot C_{H\widetilde  B} &= \bigl(2 \hyp_h^2  g_1^2-\frac92 g_2^2 -2b_{0,1}g_1^2\bigr) C_{H\widetilde B} +6 g_1 g_2 \hyp_h C_{H\widetilde WB}
 \end{align*}
\begin{align*}
\dot C_{H\widetilde  W}&= -15 g_2^3C_{\widetilde W} + \bigl(-6 \hyp_h^2 g_1^2-\frac52 g_2^2-2b_{0,2}g_2^2 \bigr) C_{H\widetilde W}  + 2 g_1 g_2 \hyp_h C_{H\widetilde WB} 
\end{align*}
\begin{align*}
 \dot C_{H\widetilde WB} &= 6 g_1 g_2^2 \hyp_h C_{\widetilde W} + \bigl( -2 \hyp_h^2 g_1^2+\frac92 g_2^2-b_{0,1}g_1^2-b_{0,2}g_2^2\bigr) C_{H\widetilde WB} 
 + 4 g_1 g_2 \hyp_h C_{H\widetilde B}  + 4 g_1 g_2 \hyp_h C_{H\widetilde W} 
\end{align*}

\subsection{$\psi^2 H^3$}\label{Ap-psiH}

\begin{align*}
\dot  C_{\substack{e H \\ rs}} &=  [Y_e^\dagger]_{rs} \left[ \frac{10}{3} g_2^2 C_{H \Box} + \left(-\frac32 g_2^2 + 6 g_1^2 \hyp_h^2 \right) C_{H D}  \right] - \left[ 3 (3 \hyp_l^2+3\hyp_e^2-4 \hyp_l \hyp_e) g_1^2+\frac{27}{4}g_2^2 \right]C_{\substack{e H \\ rs}} \nn
 & +  3 [Y_e^\dagger]_{rs} \bigl(3 g_2^2(C_{HW} +i C_{H \widetilde W}) \nn
 &+ 4(\hyp_h^2+2 \hyp_l \hyp_e) g_1^2  (C_{HB}+i C_{H \widetilde B}) 
+ 2 g_1 g_2 \hyp_l (C_{HWB}+i C_{H \widetilde WB})\bigr) - 3 \bigl( 3g_1 \hyp_e C_{\substack{eB \\ rt}}+g_2 C_{\substack{eW \\ rt}}  \bigr) [Y_e Y_e^\dagger]_{ts}\nn 
 &  - 3 [ Y_e^\dagger Y_e]_{rv} \bigl(   2g_1( \hyp_l+\hyp_e) C_{\substack{eB \\ vs}}- g_2 C_{\substack{eW \\ vs}}  \bigr) 
-6\left(4 g_1^3 \hyp_h^2 \hyp_e+ 4 g_1^3 \hyp_h^2 \hyp_l + g_2^2 g_1 \hyp_h\right)C_{\substack{eB \\ rs}}\nn
&  -3\left(4  g_1^2 g_2 \hyp_h \hyp_e +4 g_1^2 g_2 \hyp_h \hyp_l 
 + 3 g_2^3 \right)  C_{\substack{eW \\ rs}} +\bigl(3 g_2^2 + 12 g_1^2 \hyp_l \hyp_h  \bigr)  [Y_e^\dagger]_{rt} C_{\substack{He \\ ts}} \nn
&+ 1 2 g_1^2 \hyp_e  \hyp_h   C^{(1)}_{\substack{Hl \\ rt}}[Y_e^\dagger]_{ts} 
 + 1 2 g_1^2 \hyp_e  \hyp_h C^{(3)}_{\substack{Hl \\ rt}}
[Y_e^\dagger]_{ts}+\frac{4}{3} g_2^2 [Y_e^\dagger]_{rs}  \left( C_{\substack{ Hl \\ t t}}^{(3)} + N_c C_{\substack{ Hq \\ t t}}^{(3)}\right) 
 \end{align*}
\begin{align*}
\dot  C_{\substack{u H \\ rs}} &=  [Y_u^\dagger]_{rs} \left[ \frac{10}{3} g_2^2 C_{H \Box} + \left(-\frac32 g_2^2 + 6 g_1^2 \hyp_h^2 \right) C_{H D}  \right]  - \left[ 3 (3 \hyp_q^2+3\hyp_u^2-4 \hyp_q \hyp_u) g_1^2
  +\frac{27}{4}g_2^2 + 6 c_{F,3} g_3^2 \right]C_{\substack{ uH \\ r s }} 
\nn 
&+ 3 [Y_u^\dagger]_{rs} \bigl(8g_3^2 c_{F,3} (C_{HG} +i C_{H \widetilde G})+3 g_2^2  (C_{HW}+i C_{H \widetilde W}) \nn
&+ 4(\hyp_h^2+2\hyp_q\hyp_u) g_1^2   (C_{HB}+i C_{H \widetilde B}) - 2\hyp_q g_1 g_2 (C_{HWB} +i C_{H \widetilde WB})\bigr) \nn 
&- 12 [Y_d^\dagger Y_d]_{rt} g_2 C_{\substack{uW \\ ts}}   -6 g_2 C_{\substack{dW \\ rt}}     [Y_dY_u^\dagger]_{ts} - 3 \bigl( 4g_3 c_{F,3} C_{\substack{uG \\ rt}} + g_2 C_{\substack{uW \\ rt}}+  (3\hyp_u+\hyp_d )g_1C_{\substack{uB \\ rt}} \bigr) [Y_u Y_u^\dagger]_{ts} \nn
& -3[ Y_u^\dagger Y_u]_{rv} \bigl( 4 c_{F,3}g_3 C_{\substack{uG \\ vs}} - g_2 C_{\substack{uW \\ vs}} +  2( \hyp_q+ \hyp_u)g_1 C_{\substack{uB \\ vs}} \bigr)
-6\left(4 g_1^3 \hyp_h^2 \hyp_u+ 4 g_1^3 \hyp_h^2 \hyp_q-  g_2^2 g_1 \hyp_h\right) C_{\substack{uB \\ rs}}  \nn
&+3 \left(4  g_1^2 g_2 \hyp_h \hyp_u  + 4  g_1^2 g_2 \hyp_h \hyp_q 
- 3 g_2^3  \right) C_{\substack{uW \\ rs}} -\bigl(3 g_2^2 - 12 g_1^2 \hyp_q \hyp_h  \bigr)  [Y_u^\dagger]_{rt} C_{\substack{Hu \\ ts}} \nn
&
+ 3 g_2^2   [Y_d^\dagger]_{rt} C^*_{\substack{Hud \\ st}} 
  + 1 2 g_1^2 \hyp_u  \hyp_h   C^{(1)}_{\substack{Hq \\ rt}}[Y_u^\dagger]_{ts} 
 - 1 2 g_1^2 \hyp_u  \hyp_h C^{(3)}_{\substack{Hq \\ rt}}[Y_u^\dagger]_{ts} 
+\frac{4}{3} g_2^2 [Y_u^\dagger]_{rs}  \left( C_{\substack{ Hl \\ t t}}^{(3)} + N_c C_{\substack{ Hq \\ t t}}^{(3)}\right) \nn
\end{align*}
%%%
%
\begin{align*}
\dot  C_{\substack{d H \\ rs}} &= [Y_d^\dagger]_{rs} \left[ \frac{10}{3} g_2^2 C_{H \Box} + \left(-\frac32 g_2^2 + 6 g_1^2 \hyp_h^2 \right) C_{H D}  \right] - \left[ 3 (3 \hyp_q^2+3\hyp_d^2-4 \hyp_q \hyp_d) g_1^2 + \frac{27}{4}g_2^2 + 6 c_{F,3} g_3^2 \right]C_{\substack{d H \\ rs}} \nn 
&+ 3 [Y_d^\dagger]_{rs} \bigl(8 c_{F,3} g_3^2 (C_{HG}+i C_{H \widetilde G})+3 g_2^2  (C_{HW}+i C_{H \widetilde W}) \nn
&+4(\hyp_h^2+2\hyp_q \hyp_d)g_1^2  (C_{HB}+i C_{H \widetilde B})  +2\hyp_q g_1 g_2 (C_{HWB}+i C_{H \widetilde WB})\bigr)\\
&- 12 [Y_u^\dagger Y_u]_{rt} g_2 C_{\substack{dW \\ ts}} -6 g_2 C_{\substack{uW \\ rt}}  [Y_uY_d^\dagger]_{ts} - 3 \bigl( 4 c_{F,3} g_3C_{\substack{dG \\ rt}}+g_2 C_{\substack{dW \\ rt}} + ( 3\hyp_d+ \hyp_u)g_1 C_{\substack{dB \\ rt}}  \bigr) [Y_d Y_d^\dagger]_{ts}\\&  -3  [Y_d^\dagger Y_d]_{rt} \bigl(4 c_{F,3}g_3 C_{\substack{dG \\ ts}} - g_2 C_{\substack{dW \\ ts}} 
+2\left(\hyp_q+\hyp_d\right) g_1 C_{\substack{dB \\ ts}}  \bigr) 
-6\left(4 g_1^3 \hyp_h^2 \hyp_d + 4 g_1^3 \hyp_h^2 \hyp_q+  g_2^2 g_1 \hyp_h \right)C_{\substack{dB \\ rs}}  \nn
& -3\left(4  g_1^2 g_2 \hyp_h \hyp_d+4  g_1^2 g_2 \hyp_h \hyp_q 
+3 g_2^3  \right) C_{\substack{dW \\ rs}} + \bigl(3 g_2^2 + 12 g_1^2 \hyp_q \hyp_h  \bigr)  [Y_d^\dagger]_{rt} C_{\substack{Hd \\ ts}} 
\nn
&
+ 3 g_2^2   [Y_u^\dagger]_{rt} C_{\substack{Hud \\ ts}}  + 1 2 g_1^2 \hyp_d  \hyp_h   C^{(1)}_{\substack{Hq \\ rt}}[Y_d^\dagger]_{ts} 
 + 1 2 g_1^2 \hyp_d  \hyp_h C^{(3)}_{\substack{Hq \\ rt}}[Y_d^\dagger]_{ts}
+\frac{4}{3} g_2^2 [Y_d^\dagger]_{rs}  \left( C_{\substack{ Hl \\ t t}}^{(3)} + N_c C_{\substack{ Hq \\ t t}}^{(3)}\right) \nn
\end{align*}

\subsection{$\psi^2 X H$}\label{sec:dipole}

\begin{align*}
\dot C_{\substack{ eW \\ rs}}
&=\bigl[  \left( 3 c_{F,2}-b_{0,2} \right) g_2^2 + \left( -3 \hyp_e^2+8\hyp_e \hyp_l -3 \hyp_l^2 \right) g_1^2\bigr] C_{\substack{ eW \\ rs}} + g_1 g_2 (3 \hyp_l-\hyp_e) C_{\substack{ eB \\ rs}}  \nn
&-[Y_e^\dagger]_{rs} \bigl( g_2 (C_{HW} + i C_{H \widetilde W}) +g_1 (\hyp_l+\hyp_e) (C_{HWB} + i C_{H \widetilde WB}) \bigr) 
\end{align*}
%%%
\begin{align*}
\dot C_{\substack{ eB \\ rs}}
&= \bigl[   -3 c_{F,2} g_2^2 + \left(3 \hyp_e^2+4\hyp_e \hyp_l +3 \hyp_l^2 -b_{0,1}\right) g_1^2\bigr] C_{\substack{ eB \\ rs}}+ 4 c_{F,2} g_1 g_2 (3\hyp_l-\hyp_e) C_{\substack{ eW \\ rs}} \nn
& - [Y_e^\dagger]_{rs}\bigl(2 g_1 (\hyp_l+\hyp_e)  (C_{HB} + i C_{H \widetilde B}) + \frac32 g_2 (C_{HWB} + i C_{H \widetilde WB}) \bigr) 
\end{align*}
%%%
%%%
\begin{align*}
\dot C_{\substack{ uG \\ rs}}
&=  \bigl[ \left(10 c_{F,3} - 4c_{A,3} -b_{0,3}\right) g_3^2 -3 c_{F,2} g_2^2 + \left( -3 \hyp_u^2+8\hyp_u \hyp_q -3 \hyp_q^2 \right) g_1^2\bigr] C_{\substack{ uG \\ rs}} \nn
&+ 8 c_{F,2} g_2 g_3 C_{\substack{ uW \\ rs}} +  4 g_1 g_3 (\hyp_u+\hyp_q) C_{\substack{ uB \\ rs }}
-4[Y_u^\dagger]_{rs} g_3 (C_{HG} + i C_{H \widetilde G}) + 3 g_3^2 c_{A,3} [Y_u^\dagger]_{rs} \left(C_{G} + i C_{\widetilde G} \right) 
\end{align*}
%%%
\begin{align*}
\dot  C_{\substack{ uW \\ rs}}
&=\bigl[  2 c_{F,3}  g_3^2 + \left( 3 c_{F,2}-b_{0,2} \right) g_2^2 + \left( -3 \hyp_u^2+8\hyp_u \hyp_q -3 \hyp_q^2 \right) g_1^2\bigr] C_{\substack{ uW \\ rs}} + 2 c_{F,3} g_2 g_3 C_{\substack{ uG \\ rs}} \nn
&+ g_1 g_2 (3 \hyp_q-\hyp_u) C_{\substack{ uB \\ rs}}  - [Y_u^\dagger]_{rs} \bigl( g_2 (C_{HW} + i C_{H \widetilde W})- g_1 (\hyp_q+\hyp_u)  (C_{HWB} + i C_{H \widetilde WB}) \bigr) 
\end{align*}
%%%
\begin{align*}
\dot  C_{\substack{ uB \\ rs}} 
&= \bigl[ 2 c_{F,3}  g_3^2  -3 c_{F,2} g_2^2 + \left(3 \hyp_u^2+4\hyp_u \hyp_q +3 \hyp_q^2 -b_{0,1}\right) g_1^2\bigr] C_{\substack{ uB \\ rs}} + 4 c_{F,3} g_1 g_3 \left(\hyp_u+\hyp_q\right) C_{\substack{ uG \\ rs}} \nn
& + 4 c_{F,2} g_1 g_2 (3\hyp_q-\hyp_u) C_{\substack{ uW \\ rs}} 
 - [Y_u^\dagger]_{rs}\bigl(2 g_1 (\hyp_q+\hyp_u)   (C_{HB} + i C_{H \widetilde B}) - \frac32 g_2  (C_{HWB} + i C_{H \widetilde WB}) \bigr) 
\end{align*}
%%%
%%%
\begin{align*}
\dot C_{\substack{ dG \\ rs}}
&=  \bigl[ \left(10 c_{F,3} - 4c_{A,3} -b_{0,3}\right) g_3^2 -3 c_{F,2} g_2^2 + \left( -3 \hyp_d^2+8\hyp_d \hyp_q -3 \hyp_q^2 \right) g_1^2\bigr] C_{\substack{ dG \\ rs}} \nn
&+ 8 c_{F,2} g_2 g_3 C_{\substack{ dW \\ rs}} +  4 g_1 g_3 (\hyp_d+\hyp_q) C_{\substack{ dB \\ rs }}
-4[Y_d^\dagger]_{rs} g_3(C_{HG} + i C_{H \widetilde G})+ 3 g_3^2 c_{A,3} [Y_d^\dagger]_{rs} \left(C_{G} + i C_{\widetilde G} \right) 
\end{align*}
%%%
\begin{align*}
\dot C_{\substack{ dW \\ rs}}
&=\bigl[ 2 c_{F,3}  g_3^2 + \left( 3 c_{F,2}-b_{0,2} \right) g_2^2 + \left( -3 \hyp_d^2+8\hyp_d \hyp_q -3 \hyp_q^2 \right) g_1^2\bigr] C_{\substack{ dW \\ rs}} + 2 c_{F,3} g_2 g_3 C_{\substack{ dG \\ rs}}  \nn
&+ g_1 g_2 (3 \hyp_q-\hyp_d) C_{\substack{ dB \\ rs}}  -[Y_d^\dagger]_{rs} \bigl( g_2(C_{HW} + i C_{H \widetilde W}) +g_1 (\hyp_q+\hyp_d)  (C_{HWB} + i C_{H \widetilde WB}) \bigr) 
\end{align*}
%%%
\begin{align*}
\dot C_{\substack{ dB \\ rs}}
&= \bigl[ 2 c_{F,3}  g_3^2  -3 c_{F,2} g_2^2 + \left(3 \hyp_d^2+4\hyp_d \hyp_q +3 \hyp_q^2 -b_{0,1}\right) g_1^2\bigr] C_{\substack{ dB \\ rs}} 
+ 4 c_{F,3} g_1 g_3 \left(\hyp_d+\hyp_q\right) C_{\substack{ dG \\ rs}} \nn
&+ 4 c_{F,2} g_1 g_2 (3\hyp_q-\hyp_d) C_{\substack{ dW \\ rs}} 
- [Y_d^\dagger]_{rs}\bigl(2 g_1 (\hyp_q+\hyp_d)   (C_{HB} + i C_{H \widetilde B}) + \frac32 g_2  (C_{HWB} + i C_{H \widetilde WB}) \bigr)
\end{align*}

\subsection{$\psi^2 H^2 D$}

%NEW
\begin{align*}
\dot C^{(1)}_{\substack{H l \\ rs}} &= \frac12\xi_B  g_1^2 \delta_{rs} \hyp_l + \frac43 g_1^2 \hyp_h^2  C_{\substack{Hl \\ rs}}^{(1)} +\frac{4}{3} g_1^2 N_c \hyp_d \hyp_h C_{ \substack {ld \\ r s w w } } 
+\frac{4}{3} g_1^2 \hyp_e \hyp_h C_{ \substack {le \\ r s w w } } 
+\frac{8}{3} g_1^2 \hyp_h \hyp_l C_{ \substack {ll \\ r s w w } } 
+\frac{4}{3} g_1^2 \hyp_h \hyp_l C_{ \substack {ll \\ r w w s } } 
\nn
&
+\frac{4}{3} g_1^2 \hyp_h \hyp_l C_{ \substack {ll \\ w s r w } } 
+\frac{8}{3} g_1^2 \hyp_h \hyp_l C_{ \substack {ll \\ w w r s } } 
+\frac{8}{3} g_1^2 N_c \hyp_h \hyp_q C^{(1)}_{ \substack {lq \\ r s w w } } 
+\frac{4}{3} g_1^2 N_c \hyp_h \hyp_u C_{ \substack {lu \\ r s w w } }
\end{align*}
\begin{align*}
\dot C^{(3)}_{\substack{H l \\ rs}} &=\frac16 g_2^2 C_{H \Box} \delta_{rs}+ \frac23 g_2^2  C^{(3)}_{\substack{H l \\ tt}}\delta_{rs} +  \frac23 g_2^2 N_c  C^{(3)}_{\substack{H q \\ tt}}\delta_{rs}
+\frac13 g_2^2  C_{\substack{Hl \\ rs}}^{(3)} +\frac{1}{3} g_2^2 C_{ \substack {ll \\ r w w s } } +\frac{1}{3} g_2^2 C_{ \substack {ll \\ w s r w } } 
+\frac{2}{3} g_2^2 N_c C^{(3)}_{ \substack {lq \\ r s w w } }\nn
& - 6 g_2^2  C_{\substack{Hl \\ rs}}^{(3)} 
\end{align*}
\begin{align*}
\dot C_{\substack{H e \\ rs}} &=\frac12 \xi_B  g_1^2 \delta_{rs} \hyp_e + \frac43 g_1^2 \hyp_h^2  C_{\substack{He \\ rs}} +\frac{4}{3} g_1^2 N_c \hyp_d \hyp_h C_{ \substack {ed \\ r s w w } } 
+\frac{4}{3} g_1^2 \hyp_e \hyp_h C_{ \substack {ee \\ r s w w } } 
+\frac{4}{3} g_1^2 \hyp_e \hyp_h C_{ \substack {ee \\ r w w s } } 
+\frac{4}{3} g_1^2 \hyp_e \hyp_h C_{ \substack {ee \\ w s r w } } \nn
&+\frac{4}{3} g_1^2 \hyp_e \hyp_h C_{ \substack {ee \\ w w r s } } 
+\frac{4}{3} g_1^2 N_c \hyp_h \hyp_u C_{ \substack {eu \\ r s w w } } 
+\frac{8}{3} g_1^2 \hyp_h \hyp_l C_{ \substack {le \\ w w r s } } 
+\frac{8}{3} g_1^2 N_c \hyp_h \hyp_q C_{ \substack {qe \\ w w r s } } 
\end{align*}
\begin{align*}
\dot C^{(1)}_{\substack{H q \\ rs}} &= \frac12 \xi_B  g_1^2 \delta_{rs} \hyp_q + \frac43 g_1^2 \hyp_h^2  C_{\substack{Hq \\ rs}}^{(1)} +\frac{8}{3} g_1^2 \hyp_h \hyp_l C^{(1)}_{ \substack {lq \\ w w r s } } 
+\frac{4}{3} g_1^2 N_c \hyp_d \hyp_h C^{(1)}_{ \substack {qd \\ r s w w } } 
+\frac{4}{3} g_1^2 \hyp_e \hyp_h C_{ \substack {qe \\ r s w w } } +\frac{8}{3} g_1^2 N_c \hyp_h \hyp_q C^{(1)}_{ \substack {qq \\ r s w w } } \nn
&+\frac{4}{3} g_1^2 \hyp_h \hyp_q C^{(1)}_{ \substack {qq \\ r w w s } } 
+\frac{4}{3} g_1^2 \hyp_h \hyp_q C^{(1)}_{ \substack {qq \\ w s r w } } 
+\frac{8}{3} g_1^2 N_c \hyp_h \hyp_q C^{(1)}_{ \substack {qq \\ w w r s } } 
+4 g_1^2 \hyp_h \hyp_q C^{(3)}_{ \substack {qq \\ r w w s } } +4 g_1^2 \hyp_h \hyp_q C^{(3)}_{ \substack {qq \\ w s r w } } \nn
& +\frac{4}{3} g_1^2 N_c \hyp_h \hyp_u C^{(1)}_{ \substack {qu \\ r s w w } }
\end{align*}
\begin{align*}
\dot C^{(3)}_{\substack{H q \\ rs}} &= \frac16 g_2^2 C_{H \Box} \delta_{rs}+ \frac23 g_2^2  C^{(3)}_{\substack{H l \\ tt}}\delta_{rs} +  \frac23 g_2^2 N_c  C^{(3)}_{\substack{H q \\ tt}}\delta_{rs} +\frac13 g_2^2   C_{\substack{Hq \\ rs}}^{(3)} +\frac{2}{3}g_2^2 C^{(3)}_{ \substack {lq \\ w w r s } } 
+\frac{1}{3} g_2^2 C^{(1)}_{ \substack {qq \\ r w w s } } 
+\frac{1}{3}  g_2^2 C^{(1)}_{ \substack {qq \\ w s r w } } \nn
&+\frac{2}{3} g_2^2 N_c C^{(3)}_{ \substack {qq \\ r s w w } } -\frac{1}{3} g_2^2 C^{(3)}_{ \substack {qq \\ r w w s } } 
-\frac{1}{3} g_2^2 C^{(3)}_{ \substack {qq \\ w s r w } } 
+\frac{2}{3}        g_2^2 N_c C^{(3)}_{ \substack {qq \\ w w r s } } - 6 g_2^2  C_{\substack{Hq \\ rs}}^{(3)} 
\end{align*}
\begin{align*}
\dot C_{\substack{H u \\ rs}} &=  \frac12  \xi_B g_1^2 \delta_{rs} \hyp_u + \frac43 g_1^2 \hyp_h^2  C_{\substack{Hu \\ rs}} +\frac{4}{3} g_1^2 \hyp_e \hyp_h C_{ \substack {eu \\ w w r s } } 
+\frac{8}{3} g_1^2 \hyp_h \hyp_l C_{ \substack {lu \\ w w r s } } 
+\frac{8}{3} g_1^2 N_c \hyp_h \hyp_q C^{(1)}_{ \substack {qu \\ w w r s } } 
+\frac{4}{3} g_1^2 N_c \hyp_d \hyp_h C^{(1)}_{ \substack {ud \\ r s w w } } \nn
&+\frac{4}{3} g_1^2 N_c \hyp_h \hyp_u         C_{ \substack {uu \\ r s w w } } 
+\frac{4}{3} g_1^2 \hyp_h \hyp_u C_{ \substack {uu \\ r w w s } } 
+\frac{4}{3} g_1^2 \hyp_h \hyp_u C_{ \substack {uu \\ w s r w } } 
+\frac{4}{3} g_1^2 N_c \hyp_h \hyp_u C_{ \substack {uu \\ w w r s } }
\end{align*}
\begin{align*}
\dot C_{\substack{H d \\ rs}} &= \frac12 \xi_B  g_1^2 \delta_{rs} \hyp_d +  \frac43 g_1^2 \hyp_h^2  C_{\substack{Hd \\ rs}} +\frac{4}{3} g_1^2 N_c \hyp_d \hyp_h C_{ \substack {dd \\ r s w w } } 
+\frac{4}{3} g_1^2 \hyp_d \hyp_h C_{ \substack {dd \\ r w w s } } 
+\frac{4}{3} g_1^2 \hyp_d \hyp_h C_{ \substack {dd \\ w s r w } } 
+\frac{4}{3} g_1^2 N_c \hyp_d \hyp_h C_{ \substack {dd \\ w w r s } } \nn
&+\frac{4}{3} g_1^2 \hyp_e \hyp_h C_{ \substack {ed \\ w w r s } } 
+\frac{8}{3} g_1^2 \hyp_h \hyp_l C_{ \substack {ld \\ w w r s } } 
+\frac{8}{3} g_1^2 N_c \hyp_h \hyp_q C^{(1)}_{ \substack {qd \\ w w r s } } 
+\frac{4}{3} g_1^2 N_c \hyp_h \hyp_u C^{(1)}_{ \substack {ud \\ w w r s }}
\end{align*}
\begin{align*}
\dot  C_{\substack{ Hud \\ rs}} &= - 3 g_1^2 (\hyp_u-\hyp_d)^2 C_{\substack{Hud \\ rs}}
\end{align*}

\subsection{$\psi^4$ }

\subsubsection{$(\overline L L) (\overline L L)$}\label{sec:LLLL}

\begin{align*}
\dot  C_{\substack{ ll \\ prst}}   &= \frac23 g_1^2 \hyp_h \hyp_l C^{(1)}_{ \substack {Hl \\ s t } } \delta_{p r} 
-       \frac16 g_2^2 C^{(3)}_{ \substack {Hl \\ s t } } \delta_{p r}
+ \frac13 g_2^2 C^{(3)}_{ \substack {Hl \\ s r } } \delta_{p t} 
+    \frac13   g_2^2 C^{(3)}_{ \substack {Hl \\ p t } } \delta_{r s} 
\nn
&
+ \frac23 g_1^2 \hyp_h \hyp_l C^{(1)}_{ \substack {Hl \\ p r } }   \delta_{s t} 
- \frac16 g_2^2 C^{(3)}_{ \substack {Hl \\ p r } } \delta_{s t}  
+     \frac43 g_1^2 \hyp_l^2 C_{ \substack {ll \\ p r w w } } \delta_{s t}
+        \frac43 g_1^2 \hyp_l^2 C_{ \substack {ll \\ s t w w } } \delta_{p r} 
+        \frac43 g_1^2 \hyp_l^2 C_{ \substack {ll \\ w w s t } } \delta_{p r} 
\nn
&
+        \frac43 g_1^2 \hyp_l^2 C_{ \substack {ll \\ w w p r } } \delta_{s t} 
+        \frac23 g_1^2 \hyp_l^2 C_{ \substack {ll \\ p w w r } } \delta_{s t} 
+        \frac23 g_1^2 \hyp_l^2 C_{ \substack {ll \\ s w w t } } \delta_{p r} 
+        \frac23 g_1^2 \hyp_l^2 C_{ \substack {ll \\ w r p w } } \delta_{s t}
+        \frac23 g_1^2 \hyp_l^2 C_{ \substack {ll \\ w t s w } } \delta_{p r} 
\nn
&
-        \frac16 g_2^2 C_{ \substack {ll \\ p w w r } } \delta_{s t} 
-        \frac16 g_2^2 C_{ \substack {ll \\ s w w t } } \delta_{p r}
-        \frac16 g_2^2 C_{ \substack {ll \\ w r p w } } \delta_{s t} 
-        \frac16 g_2^2 C_{ \substack {ll \\ w t s w } } \delta_{p r} 
+        \frac13 g_2^2 C_{ \substack {ll \\ s w w r } } \delta_{p t}
\nn
&
+        \frac13 g_2^2 C_{ \substack {ll \\ p w w t } } \delta_{r s} 
+         \frac13    g_2^2 C_{ \substack {ll \\ w r s w } } \delta_{p t} 
+        \frac13 g_2^2 C_{ \substack {ll \\ w t p w } } \delta_{r s} 
+        \frac43 g_1^2 N_c \hyp_l \hyp_q C^{(1)}_{ \substack {lq \\ p r w w } } \delta_{s t}
+        \frac43 g_1^2 N_c \hyp_l \hyp_q C^{(1)}_{ \substack {lq \\ s t w w } } \delta_{p r} 
\nn
&
-        \frac13 g_2^2 N_c C^{(3)}_{ \substack {lq \\ p r w w } } \delta_{s t}
-        \frac13 g_2^2 N_c C^{(3)}_{ \substack {lq \\ s t w w } } \delta_{p r} 
+        \frac23 g_2^2 N_c C^{(3)}_{ \substack {lq \\ s r w w } } \delta_{p t} 
+        \frac23 g_2^2 N_c C^{(3)}_{ \substack {lq \\ p t w w } } \delta_{r s} 
\nn
&
+        \frac23 g_1^2 N_c \hyp_l \hyp_u C_{ \substack {lu \\ p r w w } } \delta_{s t}
+        \frac23 g_1^2 N_c \hyp_l \hyp_u C_{ \substack {lu \\ s t w w } } \delta_{p r} 
+        \frac23 g_1^2 N_c \hyp_d \hyp_l C_{ \substack {ld \\ p r w w } } \delta_{s t} 
+        \frac23 g_1^2 N_c \hyp_d \hyp_l C_{ \substack {ld \\ s t w w } } \delta_{p r} 
\nn
&
+        \frac23 g_1^2 \hyp_e \hyp_l C_{ \substack {le \\ p r w w } } \delta_{s t} 
+        \frac23 g_1^2 \hyp_e \hyp_l C_{ \substack {le \\ s t w w } } \delta_{p r} 
+ 6 g_2^2 \, C_{\substack{ll \\ ptsr}} - 3\left( g_2^2 - 4 \hyp_l^2 g_1^2 \right) C_{\substack{ll \\ prst}} 
\end{align*}
\begin{align*}
\dot  C_{\substack{ qq \\ prst}}^{(1)}   &=  \frac23 g_1^2 \hyp_h \hyp_q C^{(1)}_{ \substack {Hq \\ s t } } \delta_{p r} 
+   \frac23 g_1^2 \hyp_h \hyp_q C^{(1)}_{ \substack {Hq \\ p r } } \delta_{s t} 
+       \frac43 g_1^2 \hyp_l \hyp_q C^{(1)}_{ \substack {lq \\ w w s t } } \delta_{p r} 
\nn
&
+        \frac43 g_1^2 \hyp_l \hyp_q C^{(1)}_{ \substack {lq \\ w w p r } } \delta_{s t} 
+        \frac43 g_1^2 N_c \hyp_q^2 C^{(1)}_{ \substack {qq \\ p r w w } } \delta_{s t} 
+        \frac43 g_1^2 N_c \hyp_q^2 C^{(1)}_{ \substack {qq \\ s t w w } } \delta_{p r} 
+        \frac43 g_1^2 N_c \hyp_q^2 C^{(1)}_{ \substack {qq \\ w w s t } } \delta_{p r} 
+        \frac43 g_1^2 N_c \hyp_q^2 C^{(1)}_{ \substack {qq \\ w w p r } } \delta_{s t} 
\nn
&
+        \frac23 g_1^2 \hyp_q^2 C^{(1)}_{ \substack {qq \\ p w w r } } \delta_{s t} 
+        \frac23 g_1^2 \hyp_q^2 C^{(1)}_{ \substack {qq \\ s w w t } } \delta_{p r} 
+        \frac23 g_1^2 \hyp_q^2 C^{(1)}_{ \substack {qq \\ w r p w } } \delta_{s t} 
+        \frac23 g_1^2 \hyp_q^2 C^{(1)}_{ \substack {qq \\ w t s w } } \delta_{p r}
+        \frac16 g_3^2 C^{(1)}_{ \substack {qq \\ s w w r } } \delta_{p t} 
\nn
&
+        \frac16 g_3^2 C^{(1)}_{ \substack {qq \\ p w w t } } \delta_{r s} 
+        \frac16 g_3^2 C^{(1)}_{ \substack {qq \\ w r s w } } \delta_{p t} 
+        \frac16 g_3^2 C^{(1)}_{ \substack {qq \\ w t p w } } \delta_{r s} 
-        \frac1{3N_c} g_3^2 C^{(1)}_{ \substack {qq \\ p w w r } } \delta_{s t} 
-        \frac1{3 N_c} g_3^2 C^{(1)}_{ \substack {qq \\ s w w t } } \delta_{p r} 
\nn
&
-        \frac1{3N_c} g_3^2 C^{(1)}_{ \substack {qq \\ w r p w } } \delta_{s t} 
-        \frac1{3N_c} g_3^2 C^{(1)}_{ \substack {qq \\ w t s w } } \delta_{p r} 
+        2 g_1^2 \hyp_q^2 C^{(3)}_{ \substack {qq \\ p w w r } } \delta_{s t} 
+        2 g_1^2 \hyp_q^2 C^{(3)}_{ \substack {qq \\ s w w t } } \delta_{p r} 
+        2 g_1^2 \hyp_q^2 C^{(3)}_{ \substack {qq \\ w r p w } } \delta_{s t} 
\nn
&
+        2 g_1^2 \hyp_q^2 C^{(3)}_{ \substack {qq \\ w t s w } } \delta_{p r} 
+        \frac12 g_3^2 C^{(3)}_{ \substack {qq \\ s w w r } } \delta_{p t} 
+        \frac12 g_3^2 C^{(3)}_{ \substack {qq \\ p w w t } } \delta_{r s} 
+        \frac12 g_3^2 C^{(3)}_{ \substack {qq \\ w r s w } } \delta_{p t} 
+        \frac12 g_3^2 C^{(3)}_{ \substack {qq \\ w t p w } } \delta_{r s} 
-        \frac1{N_c} g_3^2 C^{(3)}_{ \substack {qq \\ p w w r } } \delta_{s t} 
\nn
&
-        \frac1{N_c} g_3^2 C^{(3)}_{ \substack {qq \\ s w w t } } \delta_{p r} 
-        \frac1{N_c} g_3^2 C^{(3)}_{ \substack {qq \\ w r p w } } \delta_{s t} 
-        \frac1{N_c} g_3^2 C^{(3)}_{ \substack {qq \\ w t s w } } \delta_{p r} 
+        \frac23 g_1^2 N_c \hyp_q \hyp_u C^{(1)}_{ \substack {qu \\ p r w w } } \delta_{s t} 
+        \frac23 g_1^2 N_c \hyp_q \hyp_u C^{(1)}_{ \substack {qu \\ s t w w } } \delta_{p r} 
\nn
&
+        \frac23 g_1^2 N_c \hyp_d \hyp_q C^{(1)}_{ \substack {qd \\ p r w w } } \delta_{s t} 
+        \frac23 g_1^2 N_c \hyp_d \hyp_q C^{(1)}_{ \substack {qd \\ s t w w } } \delta_{p r} 
+        \frac1{12} g_3^2 C^{(8)}_{ \substack {qu \\ s r w w } } \delta_{p t} 
+        \frac1{12} g_3^2 C^{(8)}_{ \substack {qu \\ p t w w } } \delta_{r s}
-        \frac1 {6 N_c} g_3^2 C^{(8)}_{ \substack {qu \\ p r w w } } \delta_{s t} 
\nn
&
-        \frac1{6 N_c} g_3^2 C^{(8)}_{ \substack {qu \\ s t w w } } \delta_{p r} 
+        \frac1{12} g_3^2 C^{(8)}_{ \substack {qd \\ s r w w } } \delta_{p t} 
+        \frac1{12} g_3^2 C^{(8)}_{ \substack {qd \\ p t w w } } \delta_{r s} 
-        \frac1 {6N_c} g_3^2 C^{(8)}_{ \substack {qd \\ p r w w } } \delta_{s t}
-        \frac1{6N_c} g_3^2 C^{(8)}_{ \substack {qd \\ s t w w } } \delta_{p r} 
\nn
&
+        \frac23 g_1^2 \hyp_e \hyp_q C_{ \substack {qe \\ p r w w } } \delta_{s t}
+        \frac23 g_1^2 \hyp_e \hyp_q C_{ \substack {qe \\ s t w w } } \delta_{p r}
+ 3 g_3^2 C_{\substack{ qq \\ ptsr }}^{(1)} 
+ 9 g_3^2 C_{\substack{ qq \\ ptsr }}^{(3)} + 9 g_2^2 C_{\substack{ qq \\ prst }}^{(3)}   
-{6 \over N_c} \left( g_3^2 - 2 N_c \hyp_q^2 g_1^2 \right) C_{\substack{ qq \\ prst }}^{(1)}
\end{align*}
\begin{align*}
\dot  C_{\substack{ qq \\ prst}}^{(3)}   &=  \frac16 g_2^2 C^{(3)}_{ \substack {Hq \\ s t } } \delta_{p r}
+ \frac16 g_2^2 C^{(3)}_{ \substack {Hq \\ p r } } \delta_{s t}  
+       \frac13 g_2^2 C^{(3)}_{ \substack {lq \\ w w s t } } \delta_{p r} 
+        \frac13 g_2^2 C^{(3)}_{ \substack {lq \\ w w p r } } \delta_{s t}
\nn
&
+        \frac16 g_2^2 C^{(1)}_{ \substack {qq \\ s w w t } } \delta_{p r} 
+        \frac16 g_2^2 C^{(1)}_{ \substack {qq \\ w t s w } } \delta_{p r} 
+        \frac16 g_2^2 C^{(1)}_{ \substack {qq \\ p w w r } } \delta_{s t} 
+        \frac16 g_2^2 C^{(1)}_{ \substack {qq \\ w r p w } } \delta_{s t} 
+        \frac16 g_3^2 C^{(1)}_{ \substack {qq \\ p w w t } } \delta_{r s} 
\nn
&
+        \frac16 g_3^2 C^{(1)}_{ \substack {qq \\ s w w r } } \delta_{p t} 
+        \frac16 g_3^2 C^{(1)}_{ \substack {qq \\ w t p w } } \delta_{r s}
+        \frac16 g_3^2 C^{(1)}_{ \substack {qq \\ w r s w } } \delta_{p t} 
+        \frac13 g_2^2 N_c C^{(3)}_{ \substack {qq \\ p r w w } } \delta_{s t} 
+        \frac13 g_2^2 N_c C^{(3)}_{ \substack {qq \\ s t w w } } \delta_{p r}
\nn
&
+        \frac13 g_2^2 N_c C^{(3)}_{ \substack {qq \\ w w s t } } \delta_{p r} 
+        \frac13 g_2^2 N_c C^{(3)}_{ \substack {qq \\ w w p r } } \delta_{s t} 
-        \frac16 g_2^2 C^{(3)}_{ \substack {qq \\ p w w r } } \delta_{s t} 
-        \frac16 g_2^2 C^{(3)}_{ \substack {qq \\ s w w t } } \delta_{p r} 
-        \frac16 g_2^2 C^{(3)}_{ \substack {qq \\ w r p w } } \delta_{s t} 
-        \frac16 g_2^2 C^{(3)}_{ \substack {qq \\ w t s w } } \delta_{p r} 
\nn
&
+        \frac12 g_3^2 C^{(3)}_{ \substack {qq \\ p w w t } } \delta_{r s} 
+        \frac12 g_3^2 C^{(3)}_{ \substack {qq \\ s w w r } } \delta_{p t} 
+        \frac12 g_3^2 C^{(3)}_{ \substack {qq \\ w t p w } } \delta_{r s} 
+        \frac12 g_3^2 C^{(3)}_{ \substack {qq \\ w r s w } } \delta_{p t}
\nn
&
+        \frac1{12} g_3^2 C^{(8)}_{ \substack {qu \\ p t w w } } \delta_{r s} 
+        \frac1{12} g_3^2 C^{(8)}_{ \substack {qu \\ s r w w } } \delta_{p t} 
+        \frac1{12} g_3^2 C^{(8)}_{ \substack {qd \\ p t w w } } \delta_{r s} 
+        \frac1{12} g_3^2 C^{(8)}_{ \substack {qd \\ s r w w } } \delta_{p t}
\nn
&
-3 g_3^2 C_{\substack{ qq \\ ptsr }}^{(3)} -{6 \over N_c}  g_3^2 C_{\substack{ qq \\ prst }}^{(3)} 
- 6 g_2^2 C_{\substack{ qq \\ prst }}^{(3)} + 12 \hyp_q^2 g_1^2 C_{\substack{ qq \\ prst }}^{(3)}
+3  g_3^2 C_{\substack{ qq \\ ptsr }}^{(1)} + 3 g_2^2 C_{\substack{ qq \\ prst }}^{(1)}
 \end{align*} 
\begin{align*}
\dot  C_{\substack{ lq \\ prst}}^{(1)}   &=  \frac43 g_1^2 \hyp_h \hyp_l C^{(1)}_{ \substack {Hq \\ s t } } \delta_{p r} 
+   \frac43 g_1^2 \hyp_h \hyp_q C^{(1)}_{ \substack {Hl \\ p r } } \delta_{s t}   
+        \frac83 g_1^2 \hyp_l \hyp_q C_{ \substack {ll \\ p r w w } } \delta_{s t} 
+        \frac83 g_1^2 \hyp_l \hyp_q C_{ \substack {ll \\ w w p r } } \delta_{s t}
+        \frac43 g_1^2 \hyp_l \hyp_q C_{ \substack {ll \\ p w w r } } \delta_{s t} 
\nn
&
+        \frac43 g_1^2 \hyp_l \hyp_q C_{ \substack {ll \\ w r p w } } \delta_{s t} 
+        \frac83 g_1^2 N_c \hyp_q^2 C^{(1)}_{ \substack {lq \\ p r w w } } \delta_{s t} 
+        \frac83 g_1^2 \hyp_l^2 C^{(1)}_{ \substack {lq \\ w w s t } } \delta_{p r} 
+        \frac83 g_1^2 N_c \hyp_l \hyp_q C^{(1)}_{ \substack {qq \\ s t w w } } \delta_{p r}
+        \frac83 g_1^2 N_c \hyp_l \hyp_q C^{(1)}_{ \substack {qq \\ w w s t } } \delta_{p r} 
\nn
&
+        \frac43 g_1^2 \hyp_l \hyp_q C^{(1)}_{ \substack {qq \\ s w w t } } \delta_{p r} 
+        \frac43 g_1^2 \hyp_l \hyp_q C^{(1)}_{ \substack {qq \\ w t s w } } \delta_{p r} 
+        4 g_1^2 \hyp_l \hyp_q C^{(3)}_{ \substack {qq \\ s w w t } } \delta_{p r} 
+        4 g_1^2 \hyp_l \hyp_q C^{(3)}_{ \substack {qq \\ w t s w } } \delta_{p r} 
+        \frac43 g_1^2 N_c \hyp_l \hyp_u C^{(1)}_{ \substack {qu \\ s t w w } } \delta_{p r} 
\nn
&
+        \frac43 g_1^2 N_c \hyp_d \hyp_l C^{(1)}_{ \substack {qd \\ s t w w } } \delta_{p r} 
+        \frac43 g_1^2 \hyp_e \hyp_l C_{ \substack {qe \\ s t w w } } \delta_{p r} 
+        \frac43 g_1^2 N_c \hyp_q \hyp_u C_{ \substack {lu \\ p r w w } } \delta_{s t}
+        \frac43 g_1^2 N_c \hyp_d \hyp_q C_{ \substack {ld \\ p r w w } } \delta_{s t} 
\nn
&
+        \frac43 g_1^2 \hyp_e \hyp_q C_{ \substack {le \\ p r w w } } \delta_{s t} 
+ 12 \hyp_l \hyp_q g_1^2 C_{\substack{ lq \\ prst }}^{(1)} + 9 g_2^2 C_{\substack{ lq \\ prst }}^{(3)} 
\end{align*}
\begin{align*}
\dot  C_{\substack{ lq \\ prst}}^{(3)}   &=    
 \frac13 g_2^2 C^{(3)}_{ \substack {Hq \\ s t } } \delta_{p r} 
+ \frac13 g_2^2 C^{(3)}_{ \substack {Hl \\ p r } } \delta_{s t} 
+        \frac23 g_2^2 N_c C^{(3)}_{ \substack {lq \\ p r w w } } \delta_{s t} 
+        \frac23 g_2^2 C^{(3)}_{ \substack {lq \\ w w s t } } \delta_{p r} 
+        \frac13 g_2^2 C^{(1)}_{ \substack {qq \\ s w w t } } \delta_{p r} 
+        \frac13 g_2^2 C^{(1)}_{ \substack {qq \\ w t s w } } \delta_{p r} 
\nn
&
+        \frac23 g_2^2 N_c C^{(3)}_{ \substack {qq \\ s t w w } } \delta_{p r} 
+        \frac23 g_2^2 N_c C^{(3)}_{ \substack {qq \\ w w s t } } \delta_{p r} 
-        \frac13 g_2^2 C^{(3)}_{ \substack {qq \\ s w w t } } \delta_{p r} 
-        \frac13 g_2^2 C^{(3)}_{ \substack {qq \\ w t s w } } \delta_{p r} 
\nn
&
+        \frac13 g_2^2 C_{ \substack {ll \\ p w w r } } \delta_{s t} 
+        \frac13 g_2^2 C_{ \substack {ll \\ w r p w } } \delta_{s t} 
+ 3 g_2^2 C_{\substack{ lq \\ prst }}^{(1)} 
-6 \left( g_2^2 - 2 \hyp_l \hyp_q g_1^2 \right) C_{\substack{ lq \\ prst }}^{(3)}  
\end{align*}

\subsubsection{$(\overline R R) (\overline R R)$}

\begin{align*}
 \dot C_{\substack{ee\\ prst}} &=\frac13 g_1^2 \hyp_e \hyp_h C_{ \substack {He \\ s t } } \delta_{p r} 
+    \frac13 g_1^2 \hyp_e \hyp_h C_{ \substack {He \\ p r } } \delta_{s t} 
+ \frac13 g_1^2 \hyp_e \hyp_h C_{ \substack {He \\ s s } } \delta_{p t} 
+    \frac13 g_1^2 \hyp_e \hyp_h C_{ \substack {He \\ p t } } \delta_{s r} 
\nn
&
+        \frac23 g_1^2 \hyp_e \hyp_l C_{ \substack {le \\ w w s t } } \delta_{p r} 
+       \frac23 g_1^2 \hyp_e \hyp_l C_{ \substack {le \\ w w p r } } \delta_{s t} 
+        \frac23 g_1^2 \hyp_e \hyp_l C_{ \substack {le \\ w w s r } } \delta_{p t} 
+       \frac23 g_1^2 \hyp_e \hyp_l C_{ \substack {le \\ w w p t } } \delta_{s r} 
\nn
&
+        \frac23 g_1^2 N_c \hyp_e \hyp_q C_{ \substack {qe \\ w w s t } } \delta_{p r} 
+        \frac23 g_1^2 N_c \hyp_e \hyp_q C_{ \substack {qe \\ w w p r } } \delta_{s t} 
+        \frac23 g_1^2 N_c \hyp_e \hyp_q C_{ \substack {qe \\ w w s r } } \delta_{p t} 
+        \frac23 g_1^2 N_c \hyp_e \hyp_q C_{ \substack {qe \\ w w p t } } \delta_{s r} 
\nn
&
+        \frac13 g_1^2 N_c \hyp_e \hyp_u C_{ \substack {eu \\ s t w w } } \delta_{p r} 
+        \frac13 g_1^2 N_c \hyp_e \hyp_u C_{ \substack {eu \\ p r w w } } \delta_{s t} 
+        \frac13 g_1^2 N_c \hyp_e \hyp_u C_{ \substack {eu \\ s r w w } } \delta_{p t} 
+        \frac13 g_1^2 N_c \hyp_e \hyp_u C_{ \substack {eu \\ p t w w } } \delta_{s r} 
\nn
&
+        \frac13 g_1^2 N_c \hyp_d \hyp_e C_{ \substack {ed \\ s t w w } } \delta_{p r} 
+        \frac13 g_1^2 N_c \hyp_d \hyp_e C_{ \substack {ed \\ p r w w } } \delta_{s t} 
+        \frac13 g_1^2 N_c \hyp_d \hyp_e C_{ \substack {ed \\ s r w w } } \delta_{p t} 
+        \frac13 g_1^2 N_c \hyp_d \hyp_e C_{ \substack {ed \\ p t w w } } \delta_{s r} 
\nn
&
+        \frac13 g_1^2 \hyp_e^2 C_{ \substack {ee \\ s t w w } } \delta_{p r} 
+        \frac13 g_1^2 \hyp_e^2 C_{ \substack {ee \\ p r w w } } \delta_{s t} 
+        \frac13 g_1^2 \hyp_e^2 C_{ \substack {ee \\ s r w w } } \delta_{p t} 
+        \frac13 g_1^2 \hyp_e^2 C_{ \substack {ee \\ p t w w } } \delta_{s r}
\nn
&
+        \frac13 g_1^2 \hyp_e^2 C_{ \substack {ee \\ w w s t } } \delta_{p r} 
+        \frac13 g_1^2 \hyp_e^2 C_{ \substack {ee \\ w w p r } } \delta_{s t} 
+        \frac13 g_1^2 \hyp_e^2 C_{ \substack {ee \\ w w s r } } \delta_{p t} 
+        \frac13 g_1^2 \hyp_e^2 C_{ \substack {ee \\ w w p t} } \delta_{s r} 
\nn
&
+        \frac13 g_1^2 \hyp_e^2 C_{ \substack {ee \\ s w w t } } \delta_{p r} 
+        \frac13 g_1^2 \hyp_e^2 C_{ \substack {ee \\ p w w r } } \delta_{s t} 
+        \frac13 g_1^2 \hyp_e^2 C_{ \substack {ee \\ s w w r} } \delta_{p t} 
+        \frac13 g_1^2 \hyp_e^2 C_{ \substack {ee \\ p w w t } } \delta_{s r} 
\nn
&
+        \frac13 g_1^2 \hyp_e^2 C_{ \substack {ee \\ w t s w } } \delta_{p r}
+        \frac13 g_1^2 \hyp_e^2 C_{ \substack {ee \\ w r p w } } \delta_{s t} 
+        \frac13 g_1^2 \hyp_e^2 C_{ \substack {ee \\ w r s w } } \delta_{p t}
+        \frac13 g_1^2 \hyp_e^2 C_{ \substack {ee \\ w t p w } } \delta_{s r} 
\nn
&
+  12 \hyp_e^2 g_1^2 C_{\substack{ ee \\ prst }} \nn
\end{align*}

\begin{align*}
\dot  C_{\substack{ uu \\ prst}}  &=    \frac23 g_1^2 \hyp_h \hyp_u C_{ \substack {Hu \\ s t } } \delta_{p r}  
+       \frac23 g_1^2 \hyp_h \hyp_u C_{ \substack {Hu \\ p r } } \delta_{s t} 
\nn
&          
+        \frac23 g_1^2 \hyp_e \hyp_u C_{ \substack {eu \\ w w s t } } \delta_{p r} 
+        \frac23 g_1^2 \hyp_e \hyp_u C_{ \substack {eu \\ w w p r } } \delta_{s t} 
+        \frac43 g_1^2 \hyp_l \hyp_u C_{ \substack {lu \\ w w p r } } \delta_{s t} 
+        \frac43 g_1^2 \hyp_l \hyp_u C_{ \substack {lu \\ w w s t } } \delta_{p r} 
\nn
&
+        \frac43 g_1^2 N_c \hyp_q \hyp_u C^{(1)}_{ \substack {qu \\ w w s t } } \delta_{p r} 
+        \frac43 g_1^2 N_c \hyp_q \hyp_u C^{(1)}_{ \substack {qu \\ w w p r } } \delta_{s t} 
+        \frac13 g_3^2 C^{(8)}_{ \substack {qu \\ w w p t } } \delta_{r s} 
+        \frac13 g_3^2 C^{(8)}_{ \substack {qu \\ w w s r } } \delta_{p t} 
-        \frac1{3N_c} g_3^2 C^{(8)}_{ \substack {qu \\ w w s t } } \delta_{p r} 
\nn
&
-        \frac1{3N_c} g_3^2 C^{(8)}_{ \substack {qu \\ w w p r } } \delta_{s t}  
+        \frac23 g_1^2 N_c \hyp_u^2 C_{ \substack {uu \\ p r w w } } \delta_{s t} 
+        \frac23 g_1^2 N_c \hyp_u^2 C_{ \substack {uu \\ s t w w } } \delta_{p r} 
+        \frac23 g_1^2 N_c \hyp_u^2 C_{ \substack {uu \\ w w p r } } \delta_{s t}
+        \frac23 g_1^2 N_c \hyp_u^2 C_{ \substack {uu \\ w w s t } } \delta_{p r}
\nn
&
+        \frac23 g_1^2 \hyp_u^2 C_{ \substack {uu \\ p w w r } } \delta_{s t} 
+        \frac23 g_1^2 \hyp_u^2 C_{ \substack {uu \\ s w w t } } \delta_{p r} 
+        \frac23 g_1^2 \hyp_u^2 C_{ \substack {uu \\ w r p w } } \delta_{s t} 
+        \frac23 g_1^2 \hyp_u^2 C_{ \substack {uu \\ w t s w } } \delta_{p r}
+        \frac13 g_3^2 C_{ \substack {uu \\ p w w t } } \delta_{r s} 
\nn
& 
+        \frac13 g_3^2 C_{ \substack {uu \\ s w w r } } \delta_{p t} 
+        \frac13 g_3^2 C_{ \substack {uu \\ w t p w } } \delta_{r s} 
+        \frac13 g_3^2 C_{ \substack {uu \\ w r s w } } \delta_{p t}  
-        \frac1{3N_c} g_3^2 C_{ \substack {uu \\ p w w r } } \delta_{s t} 
-        \frac1{3N_c} g_3^2 C_{ \substack {uu \\ s w w t } } \delta_{p r} 
\nn
&
-        \frac1{3N_c} g_3^2 C_{ \substack {uu \\ w r p w } } \delta_{s t} 
-        \frac1{3N_c} g_3^2 C_{ \substack {uu \\ w t s w } } \delta_{p r} 
+        \frac23 g_1^2 N_c \hyp_d \hyp_u C^{(1)}_{ \substack {ud \\ p r w w } } \delta_{s t} 
+        \frac23 g_1^2 N_c \hyp_d \hyp_u C^{(1)}_{ \substack {ud \\ s t w w } } \delta_{p r} 
+        \frac16 g_3^2 C^{(8)}_{ \substack {ud \\ p t w w } } \delta_{r s} 
\nn
&
+        \frac16 g_3^2 C^{(8)}_{ \substack {ud \\ s r w w } } \delta_{p t} 
-        \frac1{6N_c} g_3^2 C^{(8)}_{ \substack {ud \\ p r w w } } \delta_{s t} 
-        \frac1{6N_c} g_3^2 C^{(8)}_{ \substack {ud \\ s t w w } } \delta_{p r} 
+  6 g_3^2 C_{\substack{ uu \\ ptsr }} - 6 \left( {1 \over N_c} g_3^2 - 2 \hyp_u^2 g_1^2 \right) C_{\substack{ uu \\ prst }} 
\end{align*}
\begin{align*}
\dot  C_{\substack{ dd \\ prst}}   &=     \frac23 g_1^2 \hyp_d \hyp_h C_{ \substack {Hd \\ s t } } \delta_{p r} 
 +       \frac23 g_1^2 \hyp_d \hyp_h C_{ \substack {Hd \\ p r } } \delta_{s t} \nn
&      
+        \frac23 g_1^2 N_c \hyp_d^2 C_{ \substack {dd \\ p r w w } } \delta_{s t} 
+        \frac23 g_1^2 N_c \hyp_d^2 C_{ \substack {dd \\ s t w w } } \delta_{p r} 
+        \frac23 g_1^2 N_c \hyp_d^2 C_{ \substack {dd \\ w w p r } } \delta_{s t} 
+        \frac23 g_1^2 N_c \hyp_d^2 C_{ \substack {dd \\ w w s t } } \delta_{p r} 
\nn
&
+        \frac23 g_1^2 \hyp_d^2 C_{ \substack {dd \\ p w w r } } \delta_{s t} 
+        \frac23 g_1^2 \hyp_d^2 C_{ \substack {dd \\ s w w t } } \delta_{p r}
+        \frac23 g_1^2 \hyp_d^2 C_{ \substack {dd \\ w t s w } } \delta_{p r} 
+        \frac23 g_1^2 \hyp_d^2 C_{ \substack {dd \\ w r p w } } \delta_{s t} 
+        \frac13 g_3^2 C_{ \substack {dd \\ p w w t } } \delta_{r s} 
\nn
& 
+        \frac13 g_3^2 C_{ \substack {dd \\ s w w r } } \delta_{p t} 
+        \frac13 g_3^2 C_{ \substack {dd \\ w t p w } } \delta_{r s} 
+        \frac13 g_3^2 C_{ \substack {dd \\ w r s w } } \delta_{p t} 
-        \frac1{3N_c} g_3^2 C_{ \substack {dd \\ p w w r } } \delta_{s t} 
-        \frac1{3N_c} g_3^2 C_{ \substack {dd \\ s w w t } } \delta_{p r} 
\nn
&
-        \frac1{3N_c} g_3^2 C_{ \substack {dd \\ w t s w } } \delta_{p r} 
-        \frac1{3N_c} g_3^2 C_{ \substack {dd \\ w r p w } } \delta_{s t} 
+        \frac43 g_1^2 \hyp_d \hyp_l C_{ \substack {ld \\ w w p r } } \delta_{s t} 
+        \frac43 g_1^2 \hyp_d \hyp_l C_{ \substack {ld \\ w w s t } } \delta_{p r}
+        \frac43 g_1^2 N_c \hyp_d \hyp_q C^{(1)}_{ \substack {qd \\ w w p r } } \delta_{s t} 
\nn
&
+        \frac43 g_1^2 N_c \hyp_d \hyp_q C^{(1)}_{ \substack {qd \\ w w s t } } \delta_{p r} 
+        \frac13 g_3^2 C^{(8)}_{ \substack {qd \\ w w s r } } \delta_{p t} 
+        \frac13 g_3^2 C^{(8)}_{ \substack {qd \\ w w p t } } \delta_{r s} 
-        \frac1{3N_c} g_3^2 C^{(8)}_{ \substack {qd \\ w w p r } } \delta_{s t} 
-        \frac1{3N_c} g_3^2 C^{(8)}_{ \substack {qd \\ w w s t } } \delta_{p r} 
\nn
&
+        \frac23 g_1^2 \hyp_d \hyp_e C_{ \substack {ed \\ w w p r } } \delta_{s t} 
+        \frac23 g_1^2 \hyp_d \hyp_e C_{ \substack {ed \\ w w s t } } \delta_{p r} 
+        \frac23 g_1^2 N_c \hyp_d \hyp_u C^{(1)}_{ \substack {ud \\ w w p r } } \delta_{s t} 
+        \frac23 g_1^2 N_c \hyp_d \hyp_u C^{(1)}_{ \substack {ud \\ w w s t } } \delta_{p r} 
+        \frac16 g_3^2 C^{(8)}_{ \substack {ud \\ w w p t } } \delta_{r s} 
\nn
&
+        \frac16 g_3^2 C^{(8)}_{ \substack {ud \\ w w s r } } \delta_{p t}
-        \frac1{6N_c} g_3^2 C^{(8)}_{ \substack {ud \\ w w p r } } \delta_{s t}
-        \frac1{6N_c} g_3^2 C^{(8)}_{ \substack {ud \\ w w s t } } \delta_{p r} 
+   6 g_3^2 C_{\substack{ dd \\ ptsr }} - 6 \left( {1 \over N_c} g_3^2 - 2 \hyp_d^2 g_1^2 \right) C_{\substack{ dd \\ prst }}  
\end{align*}
\begin{align*}
\dot C_{\substack{eu\\ prst}} &=                     \frac43 g_1^2 \hyp_e \hyp_h C_{ \substack {Hu \\ s t } } \delta_{p r}  
+       \frac43 g_1^2 \hyp_h \hyp_u C_{ \substack {He \\ p r } } \delta_{s t} 
+       \frac83 g_1^2 \hyp_l \hyp_u C_{ \substack {le \\ w w p r } } \delta_{s t} 
+        \frac83 g_1^2 \hyp_e \hyp_l C_{ \substack {lu \\ w w s t } } \delta_{p r} 
+        \frac83 g_1^2 N_c \hyp_q \hyp_u C_{ \substack {qe \\ w w p r } } \delta_{s t}
\nn
&
+        \frac83 g_1^2 N_c \hyp_e \hyp_q C^{(1)}_{ \substack {qu \\ w w s t } } \delta_{p r}
+        \frac43 g_1^2 N_c \hyp_e \hyp_u C_{ \substack {uu \\ s t w w } } \delta_{p r} 
+        \frac43 g_1^2 N_c \hyp_e \hyp_u C_{ \substack {uu \\ w w s t } } \delta_{p r}
+        \frac43 g_1^2 \hyp_e \hyp_u C_{ \substack {uu \\ s w w t } } \delta_{p r} 
\nn
&
+        \frac43 g_1^2 \hyp_e \hyp_u C_{ \substack {uu \\ w t s w } } \delta_{p r} 
+        \frac43 g_1^2 \hyp_e \hyp_u C_{ \substack {ee \\ p r w w } } \delta_{s t} 
+        \frac43 g_1^2 \hyp_e \hyp_u C_{ \substack {ee \\ w w p r } } \delta_{s t} 
+        \frac43 g_1^2 \hyp_e \hyp_u C_{ \substack {ee \\ p w w r } } \delta_{s t} 
+        \frac43 g_1^2 \hyp_e \hyp_u C_{ \substack {ee \\ w r p w } } \delta_{s t}
\nn
&
+        \frac43 g_1^2 N_c \hyp_u^2 C_{ \substack {eu \\ p r w w } } \delta_{s t} 
+        \frac43 g_1^2 \hyp_e^2 C_{ \substack {eu \\ w w s t } } \delta_{p r} 
+        \frac43 g_1^2 N_c \hyp_d \hyp_e C^{(1)}_{ \substack {ud \\ s t w w } } \delta_{p r}
+        \frac43 g_1^2 N_c \hyp_d \hyp_u C_{ \substack {ed \\ p r w w } } \delta_{s t} 
+ 12 \hyp_e \hyp_u g_1^2 C_{\substack{ eu \\ prst }} 
\end{align*}
 \begin{align*}
\dot C_{\substack{ed\\ prst}} &=                 \frac43 g_1^2 \hyp_e \hyp_h C_{ \substack {Hd \\ s t } } \delta_{p r} 
+        \frac43 g_1^2 \hyp_d \hyp_h C_{ \substack {He \\ p r } } \delta_{s t}  
+        \frac43 g_1^2 N_c \hyp_d \hyp_e C_{ \substack {dd \\ s t w w } } \delta_{p r} 
+        \frac43 g_1^2 N_c \hyp_d \hyp_e C_{ \substack {dd \\ w w s t } } \delta_{p r} 
+        \frac43 g_1^2 \hyp_d \hyp_e C_{ \substack {dd \\ s w w t } } \delta_{p r} 
\nn
&
+        \frac43 g_1^2 \hyp_d \hyp_e C_{ \substack {dd \\ w t s w } } \delta_{p r} 
+        \frac43 g_1^2 \hyp_d \hyp_e C_{ \substack {ee \\ p r w w } } \delta_{s t} 
+        \frac43 g_1^2 \hyp_d \hyp_e C_{ \substack {ee \\ w w p r } } \delta_{s t} 
+        \frac43 g_1^2 \hyp_d \hyp_e C_{ \substack {ee \\ p w w r } } \delta_{s t} 
+        \frac43 g_1^2 \hyp_d \hyp_e C_{ \substack {ee \\ w r p w } } \delta_{s t}
\nn
&
+        \frac83 g_1^2 \hyp_e \hyp_l C_{ \substack {ld \\ w w s t } } \delta_{p r} 
+        \frac83 g_1^2 \hyp_d \hyp_l C_{ \substack {le \\ w w p r } } \delta_{s t} 
+        \frac83 g_1^2 N_c \hyp_e \hyp_q C^{(1)}_{ \substack {qd \\ w w s t } } \delta_{p r} 
+        \frac83 g_1^2 N_c \hyp_d \hyp_q C_{ \substack {qe \\ w w p r } } \delta_{s t}
\nn
&
+        \frac43 g_1^2 N_c \hyp_d \hyp_u C_{ \substack {eu \\ p r w w } } \delta_{s t} 
+        \frac43 g_1^2 N_c \hyp_e \hyp_u C^{(1)}_{ \substack {ud \\ w w s t } } \delta_{p r}
+        \frac43 g_1^2 N_c \hyp_d^2 C_{ \substack {ed \\ p r w w } } \delta_{s t} 
+        \frac43 g_1^2 \hyp_e^2 C_{ \substack {ed \\ w w s t } } \delta_{p r} 
+ 12 \hyp_e \hyp_d g_1^2 C_{\substack{ ed \\ prst }} 
\end{align*}
 \begin{align*}
\dot  C_{\substack{ ud \\ prst}}^{(1)}   &=        \frac43 g_1^2 \hyp_h \hyp_u C_{ \substack {Hd \\ s t } } \delta_{p r} 
+        \frac43 g_1^2 \hyp_d \hyp_h C_{ \substack {Hu \\ p r } } \delta_{s t} 
+        \frac43 g_1^2 N_c \hyp_d \hyp_u C_{ \substack {uu \\ p r w w } } \delta_{s t} 
+        \frac43 g_1^2 N_c \hyp_d \hyp_u C_{ \substack {uu \\ w w p r } } \delta_{s t}
\nn
&
+        \frac43 g_1^2 \hyp_d \hyp_u C_{ \substack {uu \\ p w w r } } \delta_{s t} 
+        \frac43 g_1^2 \hyp_d \hyp_u C_{ \substack {uu \\ w r p w } } \delta_{s t} 
+        \frac43 g_1^2 N_c \hyp_d \hyp_u C_{ \substack {dd \\ s t w w } } \delta_{p r} 
+        \frac43 g_1^2 N_c \hyp_d \hyp_u C_{ \substack {dd \\ w w s t } } \delta_{p r}
\nn
&
+        \frac43 g_1^2 \hyp_d \hyp_u C_{ \substack {dd \\ s w w t } } \delta_{p r} 
+        \frac43 g_1^2 \hyp_d \hyp_u C_{ \substack {dd \\ w t s w } } \delta_{p r} 
+        \frac83 g_1^2 N_c \hyp_d \hyp_q C^{(1)}_{ \substack {qu \\ w w p r } } \delta_{s t}  
+        \frac83 g_1^2 N_c \hyp_q \hyp_u C^{(1)}_{ \substack {qd \\ w w s t } } \delta_{p r} 
\nn
&
+        \frac83 g_1^2 \hyp_d \hyp_l C_{ \substack {lu \\ w w p r } } \delta_{s t} 
+        \frac83 g_1^2 \hyp_l \hyp_u C_{ \substack {ld \\ w w s t } } \delta_{p r} 
+        \frac43 g_1^2 N_c \hyp_d^2 C^{(1)}_{ \substack {ud \\ p r w w } } \delta_{s t} 
+        \frac43 g_1^2 N_c \hyp_u^2 C^{(1)}_{ \substack {ud \\ w w s t } } \delta_{p r}
\nn
&
+        \frac43 g_1^2 \hyp_d \hyp_e C_{ \substack {eu \\ w w p r } } \delta_{s t} 
+        \frac43 g_1^2 \hyp_e \hyp_u C_{ \substack {ed \\ w w s t } } \delta_{p r} 
+ 12 \hyp_u \hyp_d g_1^2 C_{\substack{ ud \\ prst }}^{(1)} + 3 \left({{N_c^2 -1} \over N_c^2}\right) g_3^2 C_{\substack{ ud \\ prst }}^{(8)} 
\end{align*}
\begin{align*}
\dot  C_{\substack{ ud \\ prst}}^{(8)}   &=    \frac43 g_3^2 C_{ \substack {uu \\ p w w r } } \delta_{s t} 
+       \frac43 g_3^2 C_{ \substack {uu \\ w r p w } } \delta_{s t}
+       \frac43 g_3^2 C_{ \substack {dd \\ s w w t } } \delta_{p r} 
+        \frac43 g_3^2 C_{ \substack {dd \\ w t s w } } \delta_{p r}
+        \frac43 g_3^2 C^{(8)}_{ \substack {qu \\ w w p r } } \delta_{s t} 
\nn
&
+        \frac43 g_3^2 C^{(8)}_{ \substack {qd \\ w w s t } } \delta_{p r} 
+        \frac23 g_3^2 C^{(8)}_{ \substack {ud \\ p r w w } } \delta_{s t} 
+        \frac23 g_3^2 C^{(8)}_{ \substack {ud \\ w w s t } } \delta_{p r}
+ 12 \left( \hyp_u \hyp_d g_1^2 - {1 \over N_c} g_3^2  \right) C_{\substack{ ud \\ prst }}^{(8)}  +  12 g_3^2 C_{\substack{ ud \\ prst }}^{(1)} 
\end{align*}

\subsubsection{$(\overline L L) (\overline R R)$}

\begin{align*}
\dot C_{\substack{le \\ prst}} &=    \frac43 g_1^2 \hyp_h \hyp_l C_{ \substack {He \\ s t } } \delta_{p r}  
 +        \frac43 g_1^2 \hyp_e \hyp_h C^{(1)}_{ \substack {Hl \\ p r } } \delta_{s t}    
+        \frac83 g_1^2 \hyp_e \hyp_l C_{ \substack {ll \\ p r w w } } \delta_{s t} 
+        \frac83 g_1^2 \hyp_e \hyp_l C_{ \substack {ll \\ w w p r } } \delta_{s t} 
+        \frac43 g_1^2 \hyp_e \hyp_l C_{ \substack {ll \\ p w w r } } \delta_{s t}
\nn
&
+        \frac43 g_1^2 \hyp_e \hyp_l C_{ \substack {ll \\ w r p w } } \delta_{s t} 
+        \frac83 g_1^2 N_c \hyp_e \hyp_q C^{(1)}_{ \substack {lq \\ p r w w } } \delta_{s t} 
+        \frac83 g_1^2 N_c \hyp_l \hyp_q C_{ \substack {qe \\ w w s t } } \delta_{p r}
+        \frac43 g_1^2 \hyp_e^2 C_{ \substack {le \\ p r w w } } \delta_{s t} 
+        \frac83 g_1^2 \hyp_l^2 C_{ \substack {le \\ w w s t } } \delta_{p r} 
\nn
&
+        \frac43 g_1^2 N_c \hyp_e \hyp_u C_{ \substack {lu \\ p r w w } } \delta_{s t}
+        \frac43 g_1^2 N_c \hyp_d \hyp_e C_{ \substack {ld \\ p r w w } } \delta_{s t} 
+        \frac43 g_1^2 N_c \hyp_l \hyp_u C_{ \substack {eu \\ s t w w } } \delta_{p r} 
+        \frac43 g_1^2 N_c \hyp_d \hyp_l C_{ \substack {ed \\ s t w w } } \delta_{p r} 
\nn
&
+        \frac43 g_1^2 \hyp_e \hyp_l C_{ \substack {ee \\ s t w w } } \delta_{p r} 
+        \frac43 g_1^2 \hyp_e \hyp_l C_{ \substack {ee \\ s w w t } } \delta_{p r} 
+        \frac43 g_1^2 \hyp_e \hyp_l C_{ \substack {ee \\ w t s w } } \delta_{p r}
+        \frac43 g_1^2 \hyp_e \hyp_l C_{ \substack {ee \\ w w s t } } \delta_{p r} 
-12 \hyp_l \hyp_e g_1^2 C_{\substack{ le \\ prst }} 
\end{align*}
\begin{align*}
\dot C_{\substack{lu\\ prst}} &=          \frac43 g_1^2 \hyp_h \hyp_l C_{ \substack {Hu \\ s t } } \delta_{p r}  
 +       \frac43 g_1^2 \hyp_h \hyp_u C^{(1)}_{ \substack {Hl \\ p r } } \delta_{s t} 
 +       \frac83 g_1^2 \hyp_l \hyp_u C_{ \substack {ll \\ p r w w } } \delta_{s t} 
+        \frac83 g_1^2 \hyp_l \hyp_u C_{ \substack {ll \\ w w p r } } \delta_{s t} 
+        \frac43 g_1^2 \hyp_l \hyp_u C_{ \substack {ll \\ p w w r } } \delta_{s t}
\nn
&
+        \frac43 g_1^2 \hyp_l \hyp_u C_{ \substack {ll \\ w r p w } } \delta_{s t} 
+        \frac83 g_1^2 N_c \hyp_q \hyp_u C^{(1)}_{ \substack {lq \\ p r w w } } \delta_{s t} 
+        \frac83 g_1^2 N_c \hyp_l \hyp_q C^{(1)}_{ \substack {qu \\ w w s t } } \delta_{p r} 
+        \frac43 g_1^2 N_c \hyp_u^2 C_{ \substack {lu \\ p r w w } } \delta_{s t}
+        \frac83 g_1^2 \hyp_l^2 C_{ \substack {lu \\ w w s t } } \delta_{p r} 
\nn
&
+        \frac43 g_1^2 N_c \hyp_d \hyp_u C_{ \substack {ld \\ p r w w } } \delta_{s t} 
+        \frac43 g_1^2 \hyp_e \hyp_u C_{ \substack {le \\ p r w w } } \delta_{s t} 
+        \frac43 g_1^2 N_c \hyp_d \hyp_l C^{(1)}_{ \substack {ud \\ s t w w } } \delta_{p r}
+        \frac43 g_1^2 \hyp_e \hyp_l C_{ \substack {eu \\ w w s t } } \delta_{p r} 
\nn
&
+        \frac43 g_1^2 N_c \hyp_l \hyp_u C_{ \substack {uu \\ s t w w } } \delta_{p r} 
+        \frac43 g_1^2 N_c \hyp_l \hyp_u C_{ \substack {uu \\ w w s t } } \delta_{p r}
+        \frac43 g_1^2 \hyp_l \hyp_u C_{ \substack {uu \\ s w w t } } \delta_{p r} 
+        \frac43 g_1^2 \hyp_l \hyp_u C_{ \substack {uu \\ w t s w } } \delta_{p r} 
-12 \hyp_l \hyp_u g_1^2 C_{\substack{ lu \\ prst }} 
\end{align*}
\begin{align*}
\dot C_{\substack{ld\\ prst}} &=  \frac43 g_1^2 \hyp_h \hyp_l C_{ \substack {Hd \\ s t } } \delta_{p r}  
 +        \frac43 g_1^2 \hyp_d \hyp_h C^{(1)}_{ \substack {Hl \\ p r } } \delta_{s t}
 +        \frac83 g_1^2 \hyp_d \hyp_l C_{ \substack {ll \\ p r w w } } \delta_{s t} 
+        \frac83 g_1^2 \hyp_d \hyp_l C_{ \substack {ll \\ w w p r } } \delta_{s t} 
+        \frac43 g_1^2 \hyp_d \hyp_l C_{ \substack {ll \\ p w w r } } \delta_{s t}
\nn
&
+        \frac43 g_1^2 \hyp_d \hyp_l C_{ \substack {ll \\ w r p w } } \delta_{s t} 
+        \frac83 g_1^2 N_c \hyp_d \hyp_q C^{(1)}_{ \substack {lq \\ p r w w } } \delta_{s t} 
+        \frac83 g_1^2 N_c \hyp_l \hyp_q C^{(1)}_{ \substack {qd \\ w w s t } } \delta_{p r} 
+        \frac43 g_1^2 N_c \hyp_d^2 C_{ \substack {ld \\ p r w w } } \delta_{s t} 
+        \frac83 g_1^2 \hyp_l^2 C_{ \substack {ld \\ w w s t } } \delta_{p r} 
\nn
&
+        \frac43 g_1^2 N_c \hyp_d \hyp_u C_{ \substack {lu \\ p r w w } } \delta_{s t}
+        \frac43 g_1^2 \hyp_d \hyp_e C_{ \substack {le \\ p r w w } } \delta_{s t} 
+        \frac43 g_1^2 N_c \hyp_l \hyp_u C^{(1)}_{ \substack {ud \\ w w s t } } \delta_{p r}
+        \frac43 g_1^2 \hyp_e \hyp_l C_{ \substack {ed \\ w w s t } } \delta_{p r} 
\nn
&
+        \frac43 g_1^2 N_c \hyp_d \hyp_l C_{ \substack {dd \\ s t w w } } \delta_{p r} 
+        \frac43 g_1^2 N_c \hyp_d \hyp_l C_{ \substack {dd \\ w w s t } } \delta_{p r}
+        \frac43 g_1^2 \hyp_d \hyp_l C_{ \substack {dd \\ s w w t } } \delta_{p r} 
+        \frac43 g_1^2 \hyp_d \hyp_l C_{ \substack {dd \\ w t s w } } \delta_{p r} 
-12 \hyp_l \hyp_d g_1^2 C_{\substack{ ld \\ prst }} 
\end{align*}
\begin{align*}
\dot C_{\substack{qe\\ prst}} &=  \frac43 g_1^2 \hyp_h \hyp_q C_{ \substack {He \\ s t } } \delta_{p r}  
 +      \frac43 g_1^2 \hyp_e \hyp_h C^{(1)}_{ \substack {Hq \\ p r } } \delta_{s t}
 +       \frac83 g_1^2 N_c \hyp_e \hyp_q C^{(1)}_{ \substack {qq \\ p r w w } } \delta_{s t}
+        \frac83 g_1^2 N_c \hyp_e \hyp_q C^{(1)}_{ \substack {qq \\ w w p r } } \delta_{s t}
+        \frac43 g_1^2 \hyp_e \hyp_q C^{(1)}_{ \substack {qq \\ p w w r } } \delta_{s t} 
\nn
&
+        \frac43 g_1^2 \hyp_e \hyp_q C^{(1)}_{ \substack {qq \\ w r p w } } \delta_{s t} 
+        4 g_1^2 \hyp_e \hyp_q C^{(3)}_{ \substack {qq \\ p w w r } } \delta_{s t} 
+        4 g_1^2 \hyp_e \hyp_q C^{(3)}_{ \substack {qq \\ w r p w } } \delta_{s t}
+        \frac83 g_1^2 \hyp_e \hyp_l C^{(1)}_{ \substack {lq \\ w w p r } } \delta_{s t} 
+        \frac83 g_1^2 \hyp_l \hyp_q C_{ \substack {le \\ w w s t } } \delta_{p r} 
\nn
&
+        \frac43 g_1^2 \hyp_e^2 C_{ \substack {qe \\ p r w w } } \delta_{s t} 
+        \frac83 g_1^2 N_c \hyp_q^2 C_{ \substack {qe \\ w w s t } } \delta_{p r}
+        \frac43 g_1^2 N_c \hyp_e \hyp_u C^{(1)}_{ \substack {qu \\ p r w w } } \delta_{s t}
+        \frac43 g_1^2 N_c \hyp_d \hyp_e C^{(1)}_{ \substack {qd \\ p r w w } } \delta_{s t} 
\nn
&
+        \frac43 g_1^2 N_c \hyp_q \hyp_u C_{ \substack {eu \\ s t w w } } \delta_{p r} 
+       \frac43 g_1^2 N_c \hyp_d \hyp_q C_{ \substack {ed \\ s t w w } } \delta_{p r} 
+        \frac43 g_1^2 \hyp_e \hyp_q C_{ \substack {ee \\ s t w w } } \delta_{p r} 
+        \frac43 g_1^2 \hyp_e \hyp_q C_{ \substack {ee \\ w w s t } } \delta_{p r} 
\nn
&
+        \frac43 g_1^2 \hyp_e \hyp_q C_{ \substack {ee \\ s w w t } } \delta_{p r} 
+        \frac43 g_1^2 \hyp_e \hyp_q C_{ \substack {ee \\ w t s w } } \delta_{p r}
-12 \hyp_q \hyp_e g_1^2 C_{\substack{ qe \\ prst }} 
\end{align*}
\begin{align*}
\dot  C_{\substack{ qu \\ prst}}^{(1)}   &=  \frac43 g_1^2 \hyp_h \hyp_q C_{ \substack {Hu \\ s t } } \delta_{p r}  
+        \frac43 g_1^2 \hyp_h \hyp_u C^{(1)}_{ \substack {Hq \\ p r } } \delta_{s t} 
+       \frac83 g_1^2 N_c \hyp_q \hyp_u C^{(1)}_{ \substack {qq \\ p r w w } } \delta_{s t}
+        \frac83 g_1^2 N_c \hyp_q \hyp_u C^{(1)}_{ \substack {qq \\ w w p r } } \delta_{s t} 
+        \frac43 g_1^2 \hyp_q \hyp_u C^{(1)}_{ \substack {qq \\ p w w r } } \delta_{s t} 
\nn
&
+        \frac43 g_1^2 \hyp_q \hyp_u C^{(1)}_{ \substack {qq \\ w r p w } } \delta_{s t} 
+        4 g_1^2 \hyp_q \hyp_u C^{(3)}_{ \substack {qq \\ p w w r } } \delta_{s t}
+        4 g_1^2 \hyp_q \hyp_u C^{(3)}_{ \substack {qq \\ w r p w } } \delta_{s t} 
+        \frac83 g_1^2 \hyp_l \hyp_u C^{(1)}_{ \substack {lq \\ w w p r } } \delta_{s t} 
+        \frac43 g_1^2 \hyp_e \hyp_u C_{ \substack {qe \\ p r w w } } \delta_{s t} 
\nn
&
+        \frac43 g_1^2 N_c \hyp_d \hyp_u C^{(1)}_{ \substack {qd \\ p r w w } } \delta_{s t} 
+        \frac43 g_1^2 N_c \hyp_u^2 C^{(1)}_{ \substack {qu \\ p r w w } } \delta_{s t} 
+        \frac83 g_1^2 N_c \hyp_q^2 C^{(1)}_{ \substack {qu \\ w w s t } } \delta_{p r} 
+        \frac83 g_1^2 \hyp_l \hyp_q C_{ \substack {lu \\ w w s t } } \delta_{p r} 
+        \frac43 g_1^2 \hyp_e \hyp_q C_{ \substack {eu \\ w w s t } } \delta_{p r} 
\nn
&
+        \frac43 g_1^2 N_c \hyp_d \hyp_q C^{(1)}_{ \substack {ud \\ s t w w } } \delta_{p r}
+        \frac43 g_1^2 N_c \hyp_q \hyp_u C_{ \substack {uu \\ s t w w } } \delta_{p r} 
+        \frac43 g_1^2 N_c \hyp_q \hyp_u C_{ \substack {uu \\ w w s t } } \delta_{p r}
+        \frac43 g_1^2 \hyp_q \hyp_u C_{ \substack {uu \\ s w w t } } \delta_{p r} 
\nn
&
+        \frac43 g_1^2 \hyp_q \hyp_u C_{ \substack {uu \\ w t s w } } \delta_{p r} 
 -12 \hyp_q \hyp_u g_1^2 C_{\substack{ qu \\ prst }}^{(1)} -3 \left({{N_c^2 -1} \over N_c^2}\right) g_3^2 C_{\substack{ qu \\ prst }}^{(8)}
\end{align*}
\begin{align*}
\dot  C_{\substack{ qu \\ prst}}^{(8)}   &=   \frac43 g_3^2 C^{(1)}_{ \substack {qq \\ p w w r } } \delta_{s t} 
+        \frac43 g_3^2 C^{(1)}_{ \substack {qq \\ w r p w } } \delta_{s t}
+ 4 g_3^2 C^{(3)}_{ \substack {qq \\ p w w r } }         \delta_{s t}
+ 4 g_3^2 C^{(3)}_{ \substack {qq \\ w r p w } } \delta_{s t} 
\nn
&
+        \frac23 g_3^2 C^{(8)}_{ \substack {qu \\ p r w w } } \delta_{s t}
+        \frac23 g_3^2 C^{(8)}_{ \substack {qd \\ p r w w } } \delta_{s t} 
+       \frac43 g_3^2 C^{(8)}_{ \substack {qu \\ w w s t } } \delta_{p r} 
+        \frac23 g_3^2 C^{(8)}_{ \substack {ud \\ s t w w } } \delta_{p r} 
+        \frac43 g_3^2 C_{ \substack {uu \\ s w w t } } \delta_{p r} 
+        \frac43 g_3^2 C_{ \substack {uu \\ w t s w } } \delta_{p r}
\nn
&-\left( 12 \hyp_q \hyp_u g_1^2  + 6 \left( N_c - {2 \over N_c} \right) g_3^2 \right) C_{\substack{ qu \\ prst }}^{(8)} 
- 12 g_3^2 C_{\substack{ qu \\ prst }}^{(1)}
\end{align*}
\begin{align*}
\dot  C_{\substack{ qd \\ prst}}^{(1)}   &=  \frac43 g_1^2 \hyp_h \hyp_q C_{ \substack {Hd \\ s t } } \delta_{p r} 
+        \frac43 g_1^2 \hyp_d \hyp_h C^{(1)}_{ \substack {Hq \\ p r } } \delta_{s t}
+        \frac83 g_1^2 N_c \hyp_d \hyp_q C^{(1)}_{ \substack {qq \\ p r w w } } \delta_{s t} 
+        \frac83 g_1^2 N_c \hyp_d \hyp_q C^{(1)}_{ \substack {qq \\ w w p r } } \delta_{s t} 
+        \frac43 g_1^2 \hyp_d \hyp_q C^{(1)}_{ \substack {qq \\ p w w r } } \delta_{s t} 
\nn
&
+        \frac43 g_1^2 \hyp_d \hyp_q C^{(1)}_{ \substack {qq \\ w r p w } } \delta_{s t} 
+        4 g_1^2 \hyp_d \hyp_q C^{(3)}_{ \substack {qq \\ p w w r } } \delta_{s t} 
+        4 g_1^2 \hyp_d \hyp_q C^{(3)}_{ \substack {qq \\ w r p w } } \delta_{s t} 
+        \frac83 g_1^2 \hyp_d \hyp_l C^{(1)}_{ \substack {lq \\ w w p r } } \delta_{s t} 
+        \frac43 g_1^2 \hyp_d \hyp_e C_{ \substack {qe \\ p r w w } } \delta_{s t} 
\nn
&
+        \frac43 g_1^2 N_c \hyp_d \hyp_u C^{(1)}_{ \substack {qu \\ p r w w } } \delta_{s t}
+        \frac43 g_1^2 N_c \hyp_d^2 C^{(1)}_{ \substack {qd \\ p r w w } } \delta_{s t} 
+        \frac83 g_1^2 N_c \hyp_q^2 C^{(1)}_{ \substack {qd \\ w w s t } } \delta_{p r} 
+        \frac83 g_1^2 \hyp_l \hyp_q C_{ \substack {ld \\ w w s t } } \delta_{p r} 
+        \frac43 g_1^2 \hyp_e \hyp_q C_{ \substack {ed \\ w w s t } } \delta_{p r} 
\nn
&
+        \frac43 g_1^2 N_c \hyp_q \hyp_u C^{(1)}_{ \substack {ud \\ w w s t } } \delta_{p r}
+        \frac43 g_1^2 N_c \hyp_d \hyp_q C_{ \substack {dd \\ s t w w } } \delta_{p r} 
+        \frac43 g_1^2 \hyp_d \hyp_q C_{ \substack {dd \\ s w w t } } \delta_{p r} 
+        \frac43 g_1^2 \hyp_d \hyp_q C_{ \substack {dd \\ w t s w } } \delta_{p r} 
\nn
&
+        \frac43 g_1^2 N_c \hyp_d \hyp_q C_{ \substack {dd \\ w w s t } } \delta_{p r} 
-12 \hyp_q \hyp_d g_1^2 C_{\substack{ qd \\ prst }}^{(1)} -3 \left({{N_c^2 -1} \over N_c^2}\right) g_3^2 C_{\substack{ qd \\ prst }}^{(8)}
\end{align*}
\begin{align*}
\dot  C_{\substack{ qd \\ prst}}^{(8)}   &=    \frac43 g_3^2 C^{(1)}_{ \substack {qq \\ p w w r } } \delta_{s t} 
+        \frac43 g_3^2 C^{(1)}_{ \substack {qq \\ w r p w } } \delta_{s t} 
+ 4 g_3^2 C^{(3)}_{ \substack {qq \\ p w w r } }         \delta_{s t}  
+ 4 g_3^2 C^{(3)}_{ \substack {qq \\ w r p w } } \delta_{s t} 
\nn
&
+        \frac23 g_3^2 C^{(8)}_{ \substack {qu \\ p r w w } } \delta_{s t}
+        \frac23 g_3^2 C^{(8)}_{ \substack {qd \\ p r w w } } \delta_{s t} 
+        \frac43 g_3^2 C^{(8)}_{ \substack {qd \\ w w s t } } \delta_{p r} 
+        \frac23 g_3^2 C^{(8)}_{ \substack {ud \\ w w s t } } \delta_{p r} 
+       \frac43 g_3^2 C_{ \substack {dd \\ s w w t } } \delta_{p r} 
+        \frac43 g_3^2 C_{ \substack {dd \\ w t s w } } \delta_{p r} 
\nn
&-\left( 12 \hyp_q \hyp_d g_1^2  + 6 \left( N_c - {2 \over N_c} \right) g_3^2 \right) C_{\substack{ qd \\ prst }}^{(8)} 
- 12 g_3^2 C_{\substack{ qd \\ prst }}^{(1)} 
\end{align*}

\subsubsection{$(\overline L R) (\overline R L)$}

\begin{align*}
\dot C_{\substack{ ledq \\ prst}} &= - \left( 6 \left( \hyp_d \left( \hyp_q - \hyp_e \right) + \hyp_e \left( \hyp_e + \hyp_q \right) \right) g_1^2 + 3 \left( N_c - {1 \over N_c} \right) g_3^2 \right) C_{\substack{ ledq \\ prst}} 
\end{align*}

\subsubsection{$(\overline L R) (\overline L R)$}

\begin{align*}
%%%
\dot C_{\substack{quqd \\ prst}}^{(1)} &= 4g_1(\hyp_q+ \hyp_u) C_{\substack{dB \\ st}} [Y_u^\dagger]_{pr} - 6 g_2 C_{\substack{dW \\ st}} [Y_u^\dagger]_{pr}-\frac{8}{N_c} g_1(\hyp_q+ \hyp_u) C_{\substack{dB \\ pt}} [Y_u^\dagger]_{sr} + \frac{12}{N_c} g_2 C_{\substack{dW \\ pt}} [Y_u^\dagger]_{sr}
\nn
&
-8 \frac{N_c^2-1}{N_c^2} g_3  C_{\substack{dG \\ pt}} [Y_u^\dagger]_{sr} 
+4g_1(\hyp_q+ \hyp_d) C_{\substack{uB \\ pr}} [Y_d^\dagger]_{st} - 6 g_2 C_{\substack{uW \\ pr}} [Y_d^\dagger]_{st}
-\frac{8}{N_c} g_1(\hyp_q+ \hyp_d) C_{\substack{uB \\ sr}} [Y_d^\dagger]_{pt}
\nn
&
 + \frac{12}{N_c} g_2 C_{\substack{uW \\ sr}} [Y_d^\dagger]_{pt}
-8 \frac{N_c^2-1}{N_c^2} g_3  C_{\substack{uG \\ sr}} [Y_d^\dagger]_{pt}  \nn
&-\frac12 \left(\left( 3 \hyp_d^2 + 2 \hyp_d \hyp_u + 3 \hyp_u^2 \right) g_1^2 + 3 g_2^2  
+ 12 \left( N_c - {1 \over N_c} \right) g_3^2 \right) C_{\substack{ quqd \\ prst}}^{(1)}\nn
& - {1 \over N_c} \left(\left( \left( 3 \hyp_d^2 + 10 \hyp_d \hyp_u + 3 \hyp_u^2 \right) g_1^2 - 3 g_2^2 \right) + 8 \left( N_c - {1 \over N_c} \right) g_3^2 \right) 
C_{\substack{ quqd \\ srpt}}^{(1)}\nn
&- \frac12 \left( 1 - {1 \over N_c^2}\right) \left(\left( 3 \hyp_d^2 + 10 \hyp_d \hyp_u + 3 \hyp_u^2 \right) g_1^2 - 3 g_2^2  + 4 \left( N_c - {2 \over N_c} \right) g_3^2 \right) C_{\substack{ quqd \\ srpt}}^{(8)}\nn
&+2 \left( 1 - {1 \over N_c^2}\right) g_3^2 C_{\substack{ quqd \\ prst}}^{(8)}\nn
\end{align*}
%%%
\begin{align*}
\dot C_{\substack{quqd \\ prst}}^{(8)} &= 8 g_3  C_{\substack{dG \\ st}} [Y_u^\dagger]_{pr} 
-16  g_1(\hyp_q+ \hyp_u) C_{\substack{dB \\ pt}} [Y_u^\dagger]_{sr} +24  g_2 C_{\substack{dW \\ pt}} [Y_u^\dagger]_{sr}
+ \frac{16}{N_c} g_3  C_{\substack{dG \\ pt}} [Y_u^\dagger]_{sr}  \nn
&+ 8 g_3  C_{\substack{uG \\ pr}} [Y_d^\dagger]_{st}
-16  g_1(\hyp_q+ \hyp_d) C_{\substack{uB \\ sr}} [Y_d^\dagger]_{pt} +24  g_2 C_{\substack{uW \\ sr}} [Y_d^\dagger]_{pt}
+ \frac{16}{N_c} g_3  C_{\substack{uG \\ sr}} [Y_d^\dagger]_{pt}  \nn
&+  8 g_3^2 C^{(1)}_{\substack{quqd \\ prst}} + \left(  -2 \left( 3 \hyp_d^2 + 10 \hyp_d \hyp_u + 3 \hyp_u^2 \right) g_1^2 + 6 g_2^2  + 16 {1 \over N_c} g_3^2 \right) C^{(1)}_{\substack{quqd \\ srpt}} \nn
& +\left( \left( -\frac32 \hyp_d^2 - \hyp_d \hyp_u -\frac32 \hyp_u^2 \right) g_1^2 -\frac32 g_2^2 + 2 \left( N_c  - {1 \over N_c} \right) g_3^2 \right) 
C^{(8)}_{\substack{quqd \\ prst}} \nn
& + {1 \over N_c} \left( \left( 3 \hyp_d^2 + 10 \hyp_d \hyp_u + 3 \hyp_u^2 \right) g_1^2 - 3 g_2^2 +4 \left( -N_c -{2 \over N_c} \right) g_3^2 \right) C^{(8)}_{\substack{quqd \\ srpt}}
\nn 
\end{align*}
%%%
\begin{align*}
\dot  C_{\substack{ lequ \\ prst}}^{(1)} &=  - \left( 6 \left( \hyp_e^2 + \hyp_e\left( \hyp_u - \hyp_q \right) + \hyp_q \hyp_u \right) g_1^2 + 3\left( N_c - {1 \over N_c} \right) g_3^2 \right) C_{\substack{ lequ \\ prst}}^{(1)} \nn
&- \left( 24  \left( \hyp_q + \hyp_u \right) \left( 2 \hyp_e - \hyp_q + \hyp_u \right) g_1^2 - 18 g_2^2 \right) C_{\substack{ lequ \\ prst}}^{(3)} \nn
\end{align*}
%%%
\begin{align*}
\dot C_{\substack{lequ \\ prst}}^{(3)} &= g_1(\hyp_q+ \hyp_u) C_{\substack{eB \\ pr}} [Y_u^\dagger]_{st} - \frac32 g_2 C_{\substack{uW \\ st}} [Y_e^\dagger]_{pr}
+g_1(\hyp_l+ \hyp_e) C_{\substack{uB \\ st}} [Y_e^\dagger]_{pr} - \frac32 g_2 C_{\substack{eW \\ pr}} [Y_u^\dagger]_{st} \nn
&+ \left(\left( 2 \left( \hyp_e^2 - \hyp_e \hyp_q + \hyp_e \hyp_u - 2 \hyp_q^2 + 5 \hyp_q \hyp_u - 2 \hyp_u^2  \right) g_1^2 - 3 g_2^2 \right) + \left( N_c - {1 \over N_c} \right) g_3^2 \right) C^{(3)}_{\substack{lequ \\ prst}} \nn
&+ {1 \over 8}\left( - 4 \left( \hyp_q + \hyp_u \right) \left( 2 \hyp_e - \hyp_q + \hyp_u \right) g_1^2 + 3 g_2^2 \right) C^{(1)}_{\substack{lequ \\ prst}} \nn
\end{align*}

\bibliographystyle{JHEP}
\bibliography{RG}

\providecommand{\href}[2]{#2}\begingroup\raggedright\begin{thebibliography}{100}

\bibitem{Aad:2012gk}
{\bf ATLAS} Collaboration, G.~Aad et~al., {\it {Observation of a new particle
  in the search for the Standard Model Higgs boson with the ATLAS detector at
  the LHC}},  {\em Phys.Lett.} {\bf B716} (2012) 1--29,
  [\href{http://arxiv.org/abs/1207.7214}{{\tt arXiv:1207.7214}}].

\bibitem{Chatrchyan:2012gu}
{\bf CMS} Collaboration, S.~Chatrchyan et~al., {\it {Observation of a new boson
  at a mass of 125 GeV with the CMS experiment at the LHC}},  {\em Phys.Lett.}
  {\bf B716} (2012) 30--61, [\href{http://arxiv.org/abs/1207.7235}{{\tt
  arXiv:1207.7235}}].

\bibitem{Grinstein:2007iv}
B.~Grinstein and M.~Trott, {\it {A Higgs-Higgs bound state due to new physics
  at a TeV}},  {\em Phys.Rev.} {\bf D76} (2007) 073002,
  [\href{http://arxiv.org/abs/0704.1505}{{\tt arXiv:0704.1505}}].

\bibitem{Contino:2010mh}
R.~Contino, C.~Grojean, M.~Moretti, F.~Piccinini, and R.~Rattazzi, {\it {Strong
  Double Higgs Production at the LHC}},  {\em JHEP} {\bf 1005} (2010) 089,
  [\href{http://arxiv.org/abs/1002.1011}{{\tt arXiv:1002.1011}}].

\bibitem{Alonso:2012px}
R.~Alonso, M.~B. Gavela, L.~Merlo, S.~Rigolin, and J.~Yepes, {\it The effective
  chiral lagrangian for a light dynamical "higgs particle"},  {\em Phys.Lett.}
  {\bf B722} (2013) 330--335, [\href{http://arxiv.org/abs/1212.3305}{{\tt
  arXiv:1212.3305}}].

\bibitem{Buchalla:2013rka}
G.~Buchalla, O.~Cata, and C.~Krause, {\it Complete electroweak chiral
  lagrangian with a light higgs at nlo},
  \href{http://arxiv.org/abs/1307.5017}{{\tt arXiv:1307.5017}}.

\bibitem{Brivio:2013pma}
I.~Brivio, T.~Corbett, O.~{\'E}boli, M.~B. Gavela, J.~Gonzalez-Fraile, et~al.,
  {\it Disentangling a dynamical higgs},
  \href{http://arxiv.org/abs/1311.1823}{{\tt arXiv:1311.1823}}.

\bibitem{Buchmuller:1985jz}
W.~Buchmuller and D.~Wyler, {\it {Effective Lagrangian Analysis of New
  Interactions and Flavor Conservation}},  {\em Nucl.Phys.} {\bf B268} (1986)
  621.

\bibitem{Grzadkowski:2010es}
B.~Grzadkowski, M.~Iskrzynski, M.~Misiak, and J.~Rosiek, {\it {Dimension-Six
  Terms in the Standard Model Lagrangian}},  {\em JHEP} {\bf 1010} (2010) 085,
  [\href{http://arxiv.org/abs/1008.4884}{{\tt arXiv:1008.4884}}].

\bibitem{Jenkins:2013wua}
E.~E. Jenkins, A.~V. Manohar, and M.~Trott, {\it {Renormalization Group
  Evolution of the Standard Model Dimension Six Operators II: Yukawa
  Dependence}},  \href{http://arxiv.org/abs/1310.4838}{{\tt arXiv:1310.4838}}.

\bibitem{Jenkins:2013sda}
E.~E. Jenkins, A.~V. Manohar, and M.~Trott, {\it {Naive Dimensional Analysis
  Counting of Gauge Theory Amplitudes and Anomalous Dimensions}},
  \href{http://arxiv.org/abs/1309.0819}{{\tt arXiv:1309.0819}}.

\bibitem{Jenkins:2013zja}
E.~E. Jenkins, A.~V. Manohar, and M.~Trott, {\it {Renormalization Group
  Evolution of the Standard Model Dimension Six Operators I: Formalism and
  {$\lambda$} Dependence}},  \href{http://arxiv.org/abs/1308.2627}{{\tt
  arXiv:1308.2627}}.

\bibitem{Grojean:2013kd}
C.~Grojean, E.~E. Jenkins, A.~V. Manohar, and M.~Trott, {\it {Renormalization
  Group Scaling of Higgs Operators and $h \to \gamma \gamma$ Decay }},
  \href{http://arxiv.org/abs/1301.2588}{{\tt arXiv:1301.2588}}.

\bibitem{Chivukula:1987py}
R.~S. Chivukula and H.~Georgi, {\it {Composite Technicolor Standard Model}},
  {\em Phys.Lett.} {\bf B188} (1987) 99.

\bibitem{DAmbrosio:2002ex}
G.~D'Ambrosio, G.~Giudice, G.~Isidori, and A.~Strumia, {\it {Minimal flavor
  violation: An Effective field theory approach}},  {\em Nucl.Phys.} {\bf B645}
  (2002) 155--187, [\href{http://arxiv.org/abs/hep-ph/0207036}{{\tt
  hep-ph/0207036}}].

\bibitem{Giudice:2007fh}
G.~Giudice, C.~Grojean, A.~Pomarol, and R.~Rattazzi, {\it {The
  Strongly-Interacting Light Higgs}},  {\em JHEP} {\bf 0706} (2007) 045,
  [\href{http://arxiv.org/abs/hep-ph/0703164}{{\tt hep-ph/0703164}}].

\bibitem{Chen:2013kfa}
C.-Y. Chen, S.~Dawson, and C.~Zhang, {\it {Electroweak Effective Operators and
  Higgs Physics}},  \href{http://arxiv.org/abs/1311.3107}{{\tt
  arXiv:1311.3107}}.

\bibitem{Manohar:1983md}
A.~Manohar and H.~Georgi, {\it {Chiral Quarks and the Nonrelativistic Quark
  Model}},  {\em Nucl.Phys.} {\bf B234} (1984) 189.

\bibitem{Elias-Miro:2013gya}
J.~Elias-Miro, J.~Espinosa, E.~Masso, and A.~Pomarol, {\it {Renormalization of
  dimension-six operators relevant for the Higgs decay h to gamma gamma}},
  \href{http://arxiv.org/abs/1302.5661}{{\tt arXiv:1302.5661}}.

\bibitem{Elias-Miro:2013mua}
J.~Elias-Miro, J.~Espinosa, E.~Masso, and A.~Pomarol, {\it {Higgs windows to
  new physics through d = 6 operators: Constraints and one-loop anomalous
  dimensions}},  \href{http://arxiv.org/abs/1308.1879}{{\tt arXiv:1308.1879}}.

\bibitem{pomarol}
A.~Pomarol, {\it {Invisibles Seminar, Tue, 25/06/2013
  http://invisibles.eu/journal-club}}, .

\bibitem{Bauer:1997gs}
C.~W. Bauer and A.~V. Manohar, {\it {Renormalization group scaling of the
  {$1/m^2$} HQET Lagrangian}},  {\em Phys. Rev.} {\bf D57} (1998) 337--343,
  [\href{http://arxiv.org/abs/hep-ph/9708306}{{\tt hep-ph/9708306}}].

\bibitem{Arzt:1994gp}
C.~Arzt, M.~Einhorn, and J.~Wudka, {\it {Patterns of deviation from the
  standard model}},  {\em Nucl.Phys.} {\bf B433} (1995) 41--66,
  [\href{http://arxiv.org/abs/hep-ph/9405214}{{\tt hep-ph/9405214}}].

\bibitem{Jenkins:2013fya}
E.~E. Jenkins, A.~V. Manohar, and M.~Trott, {\it {On Gauge Invariance and
  Minimal Coupling}},  \href{http://arxiv.org/abs/1305.0017}{{\tt
  arXiv:1305.0017}}.

\bibitem{Morozov:1985ef}
A.~Y. Morozov, {\it {Mixing Matrices for scalar and vector operators of
  dimension $d \le 8$ in QCD}},  {\em Sov.J.Nucl.Phys.} {\bf 40} (1984) 505.

\bibitem{Braaten:1990gq}
E.~Braaten, C.-S. Li, and T.-C. Yuan, {\it {The Evolution of Weinberg's Gluonic
  $CP$ VIolating Operator}},  {\em Phys.Rev.Lett.} {\bf 64} (1990) 1709.

\bibitem{Boyd:1990bx}
G.~Boyd, A.~K. Gupta, S.~P. Trivedi, and M.~B. Wise, {\it {Effective
  Hamiltonian for the Electric Dipole Moment of the Neutron}},  {\em
  Phys.Lett.} {\bf B241} (1990) 584.

\bibitem{Hagiwara:1993ck}
K.~Hagiwara, S.~Ishihara, R.~Szalapski, and D.~Zeppenfeld, {\it {Low-energy
  effects of new interactions in the electroweak boson sector}},  {\em
  Phys.Rev.} {\bf D48} (1993) 2182--2203.

\bibitem{Hagiwara:1993qt}
K.~Hagiwara, R.~Szalapski, and D.~Zeppenfeld, {\it {Anomalous Higgs boson
  production and decay}},  {\em Phys.Lett.} {\bf B318} (1993) 155--162,
  [\href{http://arxiv.org/abs/hep-ph/9308347}{{\tt hep-ph/9308347}}].

\bibitem{Alam:1997nk}
S.~Alam, S.~Dawson, and R.~Szalapski, {\it {Low-energy constraints on new
  physics revisited}},  {\em Phys.Rev.} {\bf D57} (1998) 1577--1590,
  [\href{http://arxiv.org/abs/hep-ph/9706542}{{\tt hep-ph/9706542}}].

\bibitem{Han:2004az}
Z.~Han and W.~Skiba, {\it {Effective theory analysis of precision electroweak
  data}},  {\em Phys.Rev.} {\bf D71} (2005) 075009,
  [\href{http://arxiv.org/abs/hep-ph/0412166}{{\tt hep-ph/0412166}}].

\bibitem{Zhang:2013xya}
C.~Zhang and F.~Maltoni, {\it {Top-quark decay into Higgs boson and a light
  quark at next-to-leading order in QCD}},  {\em Phys.Rev.} {\bf D88} (2013)
  054005, [\href{http://arxiv.org/abs/1305.7386}{{\tt arXiv:1305.7386}}].

\bibitem{Altarelli:1974exa}
G.~Altarelli and L.~Maiani, {\it {Octet Enhancement of Nonleptonic Weak
  Interactions in Asymptotically Free Gauge Theories}},  {\em Phys.Lett.} {\bf
  B52} (1974) 351--354.

\bibitem{Shifman:1976ge}
M.~A. Shifman, A.~Vainshtein, and V.~I. Zakharov, {\it {Nonleptonic Decays of K
  Mesons and Hyperons}},  {\em Sov.Phys.JETP} {\bf 45} (1977) 670.

\bibitem{Gilman:1979bc}
F.~J. Gilman and M.~B. Wise, {\it {Effective Hamiltonian for $\Delta s = 1$
  Weak Nonleptonic Decays in the Six Quark Model}},  {\em Phys.Rev.} {\bf D20}
  (1979) 2392.

\bibitem{Floratos:1977au}
E.~Floratos, D.~Ross, and C.~T. Sachrajda, {\it {Higher Order Effects in
  Asymptotically Free Gauge Theories: The Anomalous Dimensions of Wilson
  Operators}},  {\em Nucl.Phys.} {\bf B129} (1977) 66--88.

\bibitem{Floratos:1978ny}
E.~Floratos, D.~Ross, and C.~T. Sachrajda, {\it {Higher Order Effects in
  Asymptotically Free Gauge Theories. 2. Flavor Singlet Wilson Operators and
  Coefficient Functions}},  {\em Nucl.Phys.} {\bf B152} (1979) 493.

\bibitem{Gilman:1979ud}
F.~J. Gilman and M.~B. Wise, {\it {$K \to \pi e^+ e^-$ in the Six Quark
  Model}},  {\em Phys.Rev.} {\bf D21} (1980) 3150.

\bibitem{Gilman:1982ap}
F.~J. Gilman and M.~B. Wise, {\it {$K^0 - \overline K^0$ Mixing in the Six
  Quark Model}},  {\em Phys.Rev.} {\bf D27} (1983) 1128.

\bibitem{Bijnens:1983ye}
J.~Bijnens and M.~B. Wise, {\it {Electromagnetic Contribution to
  Epsilon-prime/Epsilon}},  {\em Phys.Lett.} {\bf B137} (1984) 245.

\bibitem{Grinstein:1990tj}
B.~Grinstein, R.~P. Springer, and M.~B. Wise, {\it {Strong Interaction Effects
  in Weak Radiative $\bar{B}$ Meson Decay}},  {\em Nucl.Phys.} {\bf B339}
  (1990) 269--309.

\bibitem{Bardeen:1978yd}
W.~A. Bardeen, A.~Buras, D.~Duke, and T.~Muta, {\it {Deep Inelastic Scattering
  Beyond the Leading Order in Asymptotically Free Gauge Theories}},  {\em
  Phys.Rev.} {\bf D18} (1978) 3998.

\bibitem{Buras:1990fn}
A.~J. Buras, M.~Jamin, and P.~H. Weisz, {\it {Leading and Next-to-leading {QCD}
  Corrections to $\epsilon$ Parameter and $B^0 - \bar{B}^0$ Mixing in the
  Presence of a Heavy Top Quark}},  {\em Nucl.Phys.} {\bf B347} (1990)
  491--536.

\bibitem{Buras:1992tc}
A.~J. Buras, M.~Jamin, M.~E. Lautenbacher, and P.~H. Weisz, {\it {Two loop
  anomalous dimension matrix for $\Delta S = 1$ weak nonleptonic decays. 1.
  $O(\alpha_s^2)$}},  {\em Nucl.Phys.} {\bf B400} (1993) 37--74,
  [\href{http://arxiv.org/abs/hep-ph/9211304}{{\tt hep-ph/9211304}}].

\bibitem{Buchalla:1989we}
G.~Buchalla, A.~J. Buras, and M.~K. Harlander, {\it {The Anatomy of
  Epsilon-prime / Epsilon in the Standard Model}},  {\em Nucl.Phys.} {\bf B337}
  (1990) 313--362.

\bibitem{Ciuchini:1993vr}
M.~Ciuchini, E.~Franco, G.~Martinelli, and L.~Reina, {\it {The Delta S = 1
  effective Hamiltonian including next-to-leading order QCD and QED
  corrections}},  {\em Nucl.Phys.} {\bf B415} (1994) 403--462,
  [\href{http://arxiv.org/abs/hep-ph/9304257}{{\tt hep-ph/9304257}}].

\bibitem{Arzt:1992wz}
C.~Arzt, M.~Einhorn, and J.~Wudka, {\it {Effective Lagrangian approach to
  precision measurements: The Anomalous magnetic moment of the muon}},  {\em
  Phys.Rev.} {\bf D49} (1994) 1370--1377,
  [\href{http://arxiv.org/abs/hep-ph/9304206}{{\tt hep-ph/9304206}}].

\bibitem{Degrassi:2005zd}
G.~Degrassi, E.~Franco, S.~Marchetti, and L.~Silvestrini, {\it {QCD corrections
  to the electric dipole moment of the neutron in the MSSM}},  {\em JHEP} {\bf
  0511} (2005) 044, [\href{http://arxiv.org/abs/hep-ph/0510137}{{\tt
  hep-ph/0510137}}].

\bibitem{Brod:2013cka}
J.~Brod, U.~Haisch, and J.~Zupan, {\it {Constraints on CP-violating Higgs
  couplings to the third generation}},  {\em JHEP 1311,} {\bf 180} (2013)
  [\href{http://arxiv.org/abs/1310.1385}{{\tt arXiv:1310.1385}}].

\bibitem{Gao:2012qpa}
J.~Gao, C.~S. Li, and C.~Yuan, {\it {NLO QCD Corrections to dijet Production
  via Quark Contact Interactions}},  {\em JHEP} {\bf 1207} (2012) 037,
  [\href{http://arxiv.org/abs/1204.4773}{{\tt arXiv:1204.4773}}].

\bibitem{Borzumati:1999qt}
F.~Borzumati, C.~Greub, T.~Hurth, and D.~Wyler, {\it {Gluino contribution to
  radiative B decays: Organization of QCD corrections and leading order
  results}},  {\em Phys.Rev.} {\bf D62} (2000) 075005,
  [\href{http://arxiv.org/abs/hep-ph/9911245}{{\tt hep-ph/9911245}}].

\bibitem{Buras:2000if}
A.~J. Buras, M.~Misiak, and J.~Urban, {\it {Two loop QCD anomalous dimensions
  of flavor changing four quark operators within and beyond the standard
  model}},  {\em Nucl.Phys.} {\bf B586} (2000) 397--426,
  [\href{http://arxiv.org/abs/hep-ph/0005183}{{\tt hep-ph/0005183}}].

\bibitem{Mebane:2013cra}
H.~Mebane, N.~Greiner, C.~Zhang, and S.~Willenbrock, {\it {Effective Field
  Theory of Precision Electroweak Physics at One Loop}},  {\em Phys.Lett.} {\bf
  B724} (2013) 259--263, [\href{http://arxiv.org/abs/1304.1789}{{\tt
  arXiv:1304.1789}}].

\bibitem{Mebane:2013zga}
H.~Mebane, N.~Greiner, C.~Zhang, and S.~Willenbrock, {\it {Constraints on
  Electroweak Effective Operators at One Loop}},  {\em Phys.Rev.} {\bf D88}
  (2013) 015028, [\href{http://arxiv.org/abs/1306.3380}{{\tt
  arXiv:1306.3380}}].

\bibitem{Politzer:1980me}
H.~D. Politzer, {\it {Power Corrections at Short Distances}},  {\em Nucl.Phys.}
  {\bf B172} (1980) 349.

\bibitem{Contino:2013kra}
R.~Contino, M.~Ghezzi, C.~Grojean, M.~Muhlleitner, and M.~Spira, {\it
  {Effective Lagrangian for a light Higgs-like scalar}},
  \href{http://arxiv.org/abs/1303.3876}{{\tt arXiv:1303.3876}}.

\bibitem{Pomarol:2013zra}
A.~Pomarol and F.~Riva, {\it {Towards the Ultimate SM Fit to Close in on Higgs
  Physics}},  \href{http://arxiv.org/abs/1308.2803}{{\tt arXiv:1308.2803}}.

\bibitem{Grojean:2006nn}
C.~Grojean, W.~Skiba, and J.~Terning, {\it {Disguising the oblique
  parameters}},  {\em Phys.Rev.} {\bf D73} (2006) 075008,
  [\href{http://arxiv.org/abs/hep-ph/0602154}{{\tt hep-ph/0602154}}].

\bibitem{Feldmann:2009dc}
T.~Feldmann, M.~Jung, and T.~Mannel, {\it Sequential flavour symmetry
  breaking},  {\em Phys.Rev.} {\bf D80} (2009)
  [\href{http://arxiv.org/abs/0906.1523}{{\tt arXiv:0906.1523}}].

\bibitem{Jenkins:2009dy}
E.~E. Jenkins and A.~V. Manohar, {\it {Algebraic Structure of Lepton and Quark
  Flavor Invariants and CP Violation}},  {\em JHEP} {\bf 0910} (2009) 094,
  [\href{http://arxiv.org/abs/0907.4763}{{\tt arXiv:0907.4763}}].

\bibitem{Alonso:2011yg}
R.~Alonso, M.~B. Gavela, L.~Merlo, and S.~Rigolin, {\it On the scalar potential
  of minimal flavour violation},  {\em JHEP} {\bf 1107} (2011) 012,
  [\href{http://arxiv.org/abs/1103.2915}{{\tt arXiv:1103.2915}}].

\bibitem{Kagan:2009bn}
A.~L. Kagan, G.~Perez, T.~Volansky, and J.~Zupan, {\it {General Minimal Flavor
  Violation}},  {\em Phys.Rev.} {\bf D80} (2009) 076002,
  [\href{http://arxiv.org/abs/0903.1794}{{\tt arXiv:0903.1794}}].

\bibitem{Corbett:2012ja}
T.~Corbett, O.~Eboli, J.~Gonzalez-Fraile, and M.~Gonzalez-Garcia, {\it {Robust
  Determination of the Higgs Couplings: Power to the Data}},
  \href{http://arxiv.org/abs/1211.4580}{{\tt arXiv:1211.4580}}.

\bibitem{Almeida:2013jfa}
L.~G. Almeida, S.~J. Lee, S.~Pokorski, and J.~D. Wells, {\it {Study of the 125
  GeV Standard Model Higgs Boson Partial Widths and Branching Fractions}},
  \href{http://arxiv.org/abs/1311.6721}{{\tt arXiv:1311.6721}}.

\bibitem{Grinstein:1991cd}
B.~Grinstein and M.~B. Wise, {\it {Operator analysis for precision electroweak
  physics}},  {\em Phys.Lett.} {\bf B265} (1991) 326--334.

\bibitem{Ross:1975fq}
D.~Ross and M.~Veltman, {\it {Neutral Currents in Neutrino Experiments}},  {\em
  Nucl.Phys.} {\bf B95} (1975) 135.

\bibitem{Buchalla:2013mpa}
G.~Buchalla, O.~Cata, and G.~D'Ambrosio, {\it {Nonstandard Higgs Couplings from
  Angular Distributions in $h\to Z \ell^+\ell^-$}},
  \href{http://arxiv.org/abs/1310.2574}{{\tt arXiv:1310.2574}}.

\bibitem{Bauer:2004ve}
C.~W. Bauer, Z.~Ligeti, M.~Luke, A.~V. Manohar, and M.~Trott, {\it {Global
  analysis of inclusive B decays}},  {\em Phys.Rev.} {\bf D70} (2004) 094017,
  [\href{http://arxiv.org/abs/hep-ph/0408002}{{\tt hep-ph/0408002}}].

\bibitem{Machacek:1983tz}
M.~E. Machacek and M.~T. Vaughn, {\it {Two Loop Renormalization Group Equations
  in a General Quantum Field Theory. 1. Wave Function Renormalization}},  {\em
  Nucl.Phys.} {\bf B222} (1983) 83.

\bibitem{Machacek:1983fi}
M.~E. Machacek and M.~T. Vaughn, {\it {Two Loop Renormalization Group Equations
  in a General Quantum Field Theory. 2. Yukawa Couplings}},  {\em Nucl.Phys.}
  {\bf B236} (1984) 221.

\bibitem{Machacek:1984zw}
M.~E. Machacek and M.~T. Vaughn, {\it {Two Loop Renormalization Group Equations
  in a General Quantum Field Theory. 3. Scalar Quartic Couplings}},  {\em
  Nucl.Phys.} {\bf B249} (1985) 70.

\bibitem{ATLAS-CONF-2013-068}
{\bf ATLAS} Collaboration, {\it Atlas-conf-2013-068}, .

\bibitem{Delaunay:2013pja}
C.~Delaunay, T.~Golling, G.~Perez, and Y.~Soreq, {\it {Charming the Higgs}},
  \href{http://arxiv.org/abs/1310.7029}{{\tt arXiv:1310.7029}}.

\bibitem{Blankenburg:2012ex}
G.~Blankenburg, J.~Ellis, and G.~Isidori, {\it {Flavour-Changing Decays of a
  125 GeV Higgs-like Particle}},  {\em Phys.Lett.} {\bf B712} (2012) 386--390,
  [\href{http://arxiv.org/abs/1202.5704}{{\tt arXiv:1202.5704}}].

\bibitem{Harnik:2012pb}
R.~Harnik, J.~Kopp, and J.~Zupan, {\it {Flavor Violating Higgs Decays}},  {\em
  JHEP} {\bf 1303} (2013) 026, [\href{http://arxiv.org/abs/1209.1397}{{\tt
  arXiv:1209.1397}}].

\bibitem{ATLAS-CONF-2013-034}
{\bf ATLAS} Collaboration, {\it {ATLAS-CONF-2013-034}}, .

\bibitem{Manohar:2006gz}
A.~V. Manohar and M.~B. Wise, {\it {Modifications to the properties of the
  Higgs boson}},  {\em Phys.Lett.} {\bf B636} (2006) 107--113,
  [\href{http://arxiv.org/abs/hep-ph/0601212}{{\tt hep-ph/0601212}}].

\bibitem{Bergstrom:1985hp}
L.~Bergstrom and G.~Hulth, {\it {Induced Higgs Couplings to Neutral Bosons in
  $e^+ e^-$ Collisions}},  {\em Nucl.Phys.} {\bf B259} (1985) 137.

\bibitem{ATLAS-CONF-2013-009}
{\bf ATLAS} Collaboration, {\it Atlas-conf-2013-009}, .

\bibitem{Chatrchyan:2013vaa}
{\bf CMS} Collaboration, S.~Chatrchyan et~al., {\it {}},  {\em Phys.Lett.} {\bf
  B726} (2013) 587--609, [\href{http://arxiv.org/abs/1307.5515}{{\tt
  arXiv:1307.5515}}].

\bibitem{Azatov:2013ura}
A.~Azatov, R.~Contino, A.~Di~Iura, and J.~Galloway, {\it {New Prospects for
  Higgs Compositeness in h -> Z gamma}},  {\em Phys.Rev.} {\bf D88} (2013)
  075019, [\href{http://arxiv.org/abs/1308.2676}{{\tt arXiv:1308.2676}}].

\bibitem{Kennedy:1988sn}
D.~Kennedy and B.~Lynn, {\it {Electroweak Radiative Corrections with an
  Effective Lagrangian: Four Fermion Processes}},  {\em Nucl.Phys.} {\bf B322}
  (1989) 1.

\bibitem{Peskin:1991sw}
M.~E. Peskin and T.~Takeuchi, {\it {Estimation of oblique electroweak
  corrections}},  {\em Phys.Rev.} {\bf D46} (1992) 381--409.

\bibitem{Golden:1990ig}
M.~Golden and L.~Randall, {\it {Radiative Corrections to Electroweak Parameters
  in Technicolor Theories}},  {\em Nucl.Phys.} {\bf B361} (1991) 3--23.

\bibitem{Holdom:1990tc}
B.~Holdom and J.~Terning, {\it {Large corrections to electroweak parameters in
  technicolor theories}},  {\em Phys.Lett.} {\bf B247} (1990) 88--92.

\bibitem{Baak:2012kk}
M.~Baak, M.~Goebel, J.~Haller, A.~Hoecker, D.~Kennedy, et~al., {\it {The
  Electroweak Fit of the Standard Model after the Discovery of a New Boson at
  the LHC}},  {\em Eur.Phys.J.} {\bf C72} (2012) 2205,
  [\href{http://arxiv.org/abs/1209.2716}{{\tt arXiv:1209.2716}}].

\bibitem{Eboli:2010qd}
O.~Eboli, J.~Gonzalez-Fraile, and M.~Gonzalez-Garcia, {\it {Scrutinizing the
  ZW+W- vertex at the Large Hadron Collider at 7 TeV}},  {\em Phys.Lett.} {\bf
  B692} (2010) 20--25, [\href{http://arxiv.org/abs/1006.3562}{{\tt
  arXiv:1006.3562}}].

\bibitem{Corbett:2013pja}
T.~Corbett, O.~Eboli, J.~Gonzalez-Fraile, and M.~Gonzalez-Garcia, {\it
  {Determining Triple Gauge Boson Couplings from Higgs Data}},  {\em
  Phys.Rev.Lett.} {\bf 111} (2013) 011801,
  [\href{http://arxiv.org/abs/1304.1151}{{\tt arXiv:1304.1151}}].

\bibitem{Hagiwara:1986vm}
K.~Hagiwara, R.~Peccei, D.~Zeppenfeld, and K.~Hikasa, {\it {Probing the Weak
  Boson Sector in e+ e- W+ W-}},  {\em Nucl.Phys.} {\bf B282} (1987) 253.

\bibitem{Hagiwara:1992eh}
K.~Hagiwara, S.~Ishihara, R.~Szalapski, and D.~Zeppenfeld, {\it {Low-energy
  constraints on electroweak three gauge boson couplings}},  {\em Phys.Lett.}
  {\bf B283} (1992) 353--359.

\bibitem{Isidori:2013cla}
G.~Isidori, A.~V. Manohar, and M.~Trott, {\it {Probing the nature of the
  Higgs-like Boson via $h \to VF$ decays}},
  \href{http://arxiv.org/abs/1305.0663}{{\tt arXiv:1305.0663}}.

\bibitem{Grinstein:2013vsa}
B.~Grinstein, C.~W. Murphy, and D.~Pirtskhalava, {\it {Searching for New
  Physics in the Three-Body Decays of the Higgs-like Particle}},  {\em JHEP}
  {\bf 1310} (2013) 077, [\href{http://arxiv.org/abs/1305.6938}{{\tt
  arXiv:1305.6938}}].

\bibitem{Manohar:2013rga}
A.~V. Manohar, {\it {An Exactly Solvable Model for Dimension Six Higgs
  Operators and $h \to \gamma \gamma$}},
  \href{http://arxiv.org/abs/1305.3927}{{\tt arXiv:1305.3927}}.

\bibitem{Adam:2013mnn}
{\bf MEG Collaboration} Collaboration, J.~Adam et~al., {\it {New constraint on
  the existence of the mu to e+ gamma decay}},
  \href{http://arxiv.org/abs/1303.0754}{{\tt arXiv:1303.0754}}.

\bibitem{Baron:2013eja}
{\bf ACME} Collaboration, J.~Baron et~al., {\it {Order of Magnitude Smaller
  Limit on the Electric Dipole Moment of the Electron}},
  \href{http://arxiv.org/abs/1310.7534}{{\tt arXiv:1310.7534}}.

\bibitem{Alonso:2014zka}
R.~Alonso, H.-M. Chang, E.~E. Jenkins, A.~V. Manohar, and B.~Shotwell, {\it
  {Renormalization group evolution of dimension-six baryon number violating
  operators}},  {\em Phys. Lett.} {\bf B734} (2014) 302--307,
  [\href{http://arxiv.org/abs/1405.0486}{{\tt arXiv:1405.0486}}].

\bibitem{Zhang:2014rja}
C.~Zhang, {\it {Effective field theory approach to top-quark decay at
  next-to-leading order in QCD}},  {\em Phys.Rev.} {\bf D90} (2014), no.~1
  014008, [\href{http://arxiv.org/abs/1404.1264}{{\tt arXiv:1404.1264}}].

\bibitem{Pruna:2014asa}
G.~M. Pruna and A.~Signer, {\it {The $\mu\to e\gamma$ decay in a systematic
  effective field theory approach with dimension 6 operators}},  {\em JHEP}
  {\bf 1410} (2014) 14, [\href{http://arxiv.org/abs/1408.3565}{{\tt
  arXiv:1408.3565}}].

\bibitem{Brod:2014hsa}
J.~Brod, A.~Greljo, E.~Stamou, and P.~Uttayarat, {\it {Probing anomalous $
  t\overline{t}Z $ interactions with rare meson decays}},  {\em JHEP} {\bf
  1502} (2015) 141, [\href{http://arxiv.org/abs/1408.0792}{{\tt
  arXiv:1408.0792}}].

\bibitem{Cheung:2015aba}
C.~Cheung and C.-H. Shen, {\it {Non-renormalization Theorems without
  Supersymmetry}},  \href{http://arxiv.org/abs/1505.01844}{{\tt
  arXiv:1505.01844}}.

\bibitem{Dashen:1993jt}
R.~F. Dashen, E.~E. Jenkins, and A.~V. Manohar, {\it {The $1/N_c$ expansion for
  baryons}},  {\em Phys.Rev.} {\bf D49} (1994) 4713,
  [\href{http://arxiv.org/abs/hep-ph/9310379}{{\tt hep-ph/9310379}}].

\bibitem{Dashen:1994qi}
R.~F. Dashen, E.~E. Jenkins, and A.~V. Manohar, {\it {Spin flavor structure of
  large $N_c$ baryons}},  {\em Phys.Rev.} {\bf D51} (1995) 3697--3727,
  [\href{http://arxiv.org/abs/hep-ph/9411234}{{\tt hep-ph/9411234}}].

\end{thebibliography}\endgroup

\end{document}